\pdfoutput=1
\documentclass[12pt,a4paper]{article}
\usepackage{ifthen} 
\usepackage{subfig} 
\usepackage{pdflscape}
\usepackage{fancyhdr}
\usepackage{amssymb}
\usepackage[tbtags,reqno]{amsmath}
\usepackage{amsmath}
\usepackage{latexsym}
\usepackage{amsthm}

\usepackage{pbox}
\usepackage{xspace}
\usepackage{color}
\usepackage{colortbl}
\usepackage{lscape}
\usepackage{multirow}
\usepackage[titletoc,toc,title]{appendix}
\usepackage{afterpage}
\usepackage{longtable}

\usepackage[capitalise, sort&compress]{cleveref}

\usepackage{amssymb}
\usepackage{amsfonts}
\usepackage{upgreek}
\usepackage{hyperref}
\usepackage{cleveref}
\usepackage{float}
\newboolean{pdflatex}
\setboolean{pdflatex}{true} 

\newboolean{articletitles}
\setboolean{articletitles}{true} 

\newboolean{uprightparticles}
\setboolean{uprightparticles}{false} 

\newboolean{inbibliography}
\setboolean{inbibliography}{false} 

\usepackage[top=1in, bottom=1.25in, left=1in, right=1in]{geometry}

\usepackage{microtype}
\usepackage{lineno}  
\usepackage{xspace} 
\usepackage{caption} 

\usepackage{graphicx}  
\usepackage{color}
\usepackage{colortbl}
\graphicspath{{./figs/}} 

\usepackage{amsmath} 
\usepackage{amssymb}
\usepackage{amsfonts}
\usepackage{upgreek} 

\newcommand*\patchAmsMathEnvironmentForLineno[1]{%
\expandafter\let\csname old#1\expandafter\endcsname\csname #1\endcsname
\expandafter\let\csname oldend#1\expandafter\endcsname\csname
end#1\endcsname
 \renewenvironment{#1}%
   {\linenomath\csname old#1\endcsname}%
   {\csname oldend#1\endcsname\endlinenomath}%
}
\newcommand*\patchBothAmsMathEnvironmentsForLineno[1]{%
  \patchAmsMathEnvironmentForLineno{#1}%
  \patchAmsMathEnvironmentForLineno{#1*}%
}
\AtBeginDocument{%
\patchBothAmsMathEnvironmentsForLineno{equation}%
\patchBothAmsMathEnvironmentsForLineno{align}%
\patchBothAmsMathEnvironmentsForLineno{flalign}%
\patchBothAmsMathEnvironmentsForLineno{alignat}%
\patchBothAmsMathEnvironmentsForLineno{gather}%
\patchBothAmsMathEnvironmentsForLineno{multline}%
\patchBothAmsMathEnvironmentsForLineno{eqnarray}%
}

\usepackage{hyperref}    
\usepackage[all]{hypcap} 

\def\lhcb {\mbox{LHCb}\xspace}

\def\babar  {\mbox{BaBar}\xspace}
\def\belle  {\mbox{Belle}\xspace}

\def\dzero  {\mbox{D0}\xspace}

\ifthenelse{\boolean{uprightparticles}}%
{

 \def\Pmu         {\ensuremath{\upmu}\xspace}

 \def\Ppi         {\ensuremath{\uppi}\xspace}

 \def\Ppsi        {\ensuremath{\uppsi}\xspace}

 \def\PDelta      {\ensuremath{\Delta}\xspace}                 
 \def\PXi      {\ensuremath{\Xi}\xspace}                 
 \def\PLambda      {\ensuremath{\Lambda}\xspace}                 
 \def\PSigma      {\ensuremath{\Sigma}\xspace}                 
 \def\POmega      {\ensuremath{\Omega}\xspace}                 
 \def\PUpsilon      {\ensuremath{\Upsilon}\xspace}                 
 
 \def\PB      {\ensuremath{\mathrm{B}}\xspace}                 
                  
 \def\PD      {\ensuremath{\mathrm{D}}\xspace}

 \def\PJ      {\ensuremath{\mathrm{J}}\xspace}                 
 \def\PK      {\ensuremath{\mathrm{K}}\xspace}

 \def\Pb      {\ensuremath{\mathrm{b}}\xspace}                 
 \def\Pc      {\ensuremath{\mathrm{c}}\xspace}                 
 \def\Pd      {\ensuremath{\mathrm{d}}\xspace}

 \def\Pi      {\ensuremath{\mathrm{i}}\xspace}

 \def\Ps      {\ensuremath{\mathrm{s}}\xspace}                 
                  
 \def\Pu      {\ensuremath{\mathrm{u}}\xspace}

}
{

 \def\Pmu         {\ensuremath{\mu}\xspace}

 \def\Ppi         {\ensuremath{\pi}\xspace}

 \def\Ppsi        {\ensuremath{\psi}\xspace}                 
                  
 \mathchardef\PDelta="7101
 \mathchardef\PXi="7104
 \mathchardef\PLambda="7103
 \mathchardef\PSigma="7106
 \mathchardef\POmega="710A
 \mathchardef\PUpsilon="7107
                  
 \def\PB      {\ensuremath{B}\xspace}                 
                  
 \def\PD      {\ensuremath{D}\xspace}

 \def\PJ      {\ensuremath{J}\xspace}                 
 \def\PK      {\ensuremath{K}\xspace}

 \def\Pb      {\ensuremath{b}\xspace}                 
 \def\Pc      {\ensuremath{c}\xspace}                 
 \def\Pd      {\ensuremath{d}\xspace}

 \def\Pi      {\ensuremath{i}\xspace}

 \def\Ps      {\ensuremath{s}\xspace}                 
                  
 \def\Pu      {\ensuremath{u}\xspace}

}

\makeatletter
\ifcase \@ptsize \relax
  \newcommand{\miniscule}{\@setfontsize\miniscule{4}{5}}
\or
  \newcommand{\miniscule}{\@setfontsize\miniscule{5}{6}}
\or
  \newcommand{\miniscule}{\@setfontsize\miniscule{5}{6}}
\fi
\makeatother

\DeclareRobustCommand{\optbar}[1]{\shortstack{{\miniscule (\rule[.5ex]{1.25em}{.18mm})}
  \\ [-.7ex] $#1$}}

\def\mup        {{\ensuremath{\Pmu^+}}\xspace}
\def\mun        {{\ensuremath{\Pmu^-}}\xspace} 

\def\uquark    {{\ensuremath{\Pu}}\xspace}

\def\dquark    {{\ensuremath{\Pd}}\xspace}

\def\squark    {{\ensuremath{\Ps}}\xspace}

\def\cquark    {{\ensuremath{\Pc}}\xspace}

\def\bquark    {{\ensuremath{\Pb}}\xspace}
\def\bquarkbar {{\ensuremath{\overline \bquark}}\xspace}

\def\pion   {{\ensuremath{\Ppi}}\xspace}
\def\piz    {{\ensuremath{\pion^0}}\xspace}

\def\pip    {{\ensuremath{\pion^+}}\xspace}
\def\pim    {{\ensuremath{\pion^-}}\xspace}

\def\kaon    {{\ensuremath{\PK}}\xspace}
  \def\Kbar    {{\kern 0.2em\overline{\kern -0.2em \PK}{}}\xspace}

\def\KorKbar    {\kern 0.18em\optbar{\kern -0.18em K}{}\xspace}
\def\Kz      {{\ensuremath{\kaon^0}}\xspace}
\def\Kzb     {{\ensuremath{\Kbar{}^0}}\xspace}
\def\Kp      {{\ensuremath{\kaon^+}}\xspace}
\def\Km      {{\ensuremath{\kaon^-}}\xspace}

\def\KS      {{\ensuremath{\kaon^0_{\rm\scriptscriptstyle S}}}\xspace}

\def\Kstarz  {{\ensuremath{\kaon^{*0}}}\xspace}

  \def\Dbar    {{\kern 0.2em\overline{\kern -0.2em \PD}{}}\xspace}
\def\D       {{\ensuremath{\PD}}\xspace}

\def\DorDbar    {\kern 0.18em\optbar{\kern -0.18em D}{}\xspace}
\def\Dz      {{\ensuremath{\D^0}}\xspace}

\def\Dp      {{\ensuremath{\D^+}}\xspace}
\def\Dm      {{\ensuremath{\D^-}}\xspace}

\def\Dsp     {{\ensuremath{\D^+_\squark}}\xspace}
\def\Dsm     {{\ensuremath{\D^-_\squark}}\xspace}

\def\B       {{\ensuremath{\PB}}\xspace}
\def\Bbar    {{\ensuremath{\kern 0.18em\overline{\kern -0.18em \PB}{}}}\xspace}

\def\BorBbar    {\kern 0.18em\optbar{\kern -0.18em B}{}\xspace}
\def\Bz      {{\ensuremath{\B^0}}\xspace}
\def\Bzb     {{\ensuremath{\Bbar{}^0}}\xspace}
\def\Bu      {{\ensuremath{\B^+}}\xspace}
\def\Bub     {{\ensuremath{\B^-}}\xspace}
\def\Bp      {{\ensuremath{\Bu}}\xspace}
\def\Bm      {{\ensuremath{\Bub}}\xspace}

\def\Bd      {{\ensuremath{\B^0}}\xspace}
\def\Bs      {{\ensuremath{\B^0_\squark}}\xspace}
\def\Bsb     {{\ensuremath{\Bbar{}^0_\squark}}\xspace}

\def\Bc      {{\ensuremath{\B_\cquark^+}}\xspace}

\def\jpsi     {{\ensuremath{{\PJ\mskip -3mu/\mskip -2mu\Ppsi\mskip 2mu}}}\xspace}

  \def\Y#1S{\ensuremath{\PUpsilon{(#1S)}}\xspace}

\def\Lz          {{\ensuremath{\PLambda}}\xspace}
\def\Lbar        {{\ensuremath{\kern 0.1em\overline{\kern -0.1em\PLambda}}}\xspace}
\def\LorLbar    {\kern 0.18em\optbar{\kern -0.18em \PLambda}{}\xspace}

\def\Lb      {{\ensuremath{\Lz^0_\bquark}}\xspace}
\def\Lbbar   {{\ensuremath{\Lbar{}^0_\bquark}}\xspace}

\newcommand{\decay}[2]{\ensuremath{#1\!\to #2}\xspace}         

\def\to                 {\ensuremath{\rightarrow}\xspace}

\def\order   {{\ensuremath{\mathcal{O}}}\xspace}

\def\CP                {{\ensuremath{C\!P}}\xspace}

\newcommand{\dms}{{\ensuremath{\Delta m_{\squark}}}\xspace}
\newcommand{\dmd}{{\ensuremath{\Delta m_{\dquark}}}\xspace}

\newcommand{\DGs}{{\ensuremath{\Delta\Gamma_{\squark}}}\xspace}
\newcommand{\DGd}{{\ensuremath{\Delta\Gamma_{\dquark}}}\xspace}
\newcommand{\Gs}{{\ensuremath{\Gamma_{\squark}}}\xspace}
\newcommand{\Gd}{{\ensuremath{\Gamma_{\dquark}}}\xspace}

\def\BdToJPsiKst  {\decay{\Bd}{\jpsi\Kstarz}}

\def\AT#1     {\ensuremath{A_{\mathrm{T}}^{#1}}\xspace}           

\def\C#1      {\ensuremath{\mathcal{C}_{#1}}\xspace}                       
\def\Cp#1     {\ensuremath{\mathcal{C}_{#1}^{'}}\xspace}                    
\def\Ceff#1   {\ensuremath{\mathcal{C}_{#1}^{\mathrm{(eff)}}}\xspace}        
\def\Cpeff#1  {\ensuremath{\mathcal{C}_{#1}^{'\mathrm{(eff)}}}\xspace}       
\def\Ope#1    {\ensuremath{\mathcal{O}_{#1}}\xspace}                       
\def\Opep#1   {\ensuremath{\mathcal{O}_{#1}^{'}}\xspace}                    

\newcommand{\ket}[1]{\ensuremath{|#1\rangle}}              

\newcommand{\tev}{\ensuremath{\mathrm{\,Te\kern -0.1em V}}\xspace}
\newcommand{\gev}{\ensuremath{\mathrm{\,Ge\kern -0.1em V}}\xspace}
\newcommand{\mev}{\ensuremath{\mathrm{\,Me\kern -0.1em V}}\xspace}
\newcommand{\kev}{\ensuremath{\mathrm{\,ke\kern -0.1em V}}\xspace}
\newcommand{\ev}{\ensuremath{\mathrm{\,e\kern -0.1em V}}\xspace}
\newcommand{\gevc}{\ensuremath{{\mathrm{\,Ge\kern -0.1em V\!/}c}}\xspace}
\newcommand{\mevc}{\ensuremath{{\mathrm{\,Me\kern -0.1em V\!/}c}}\xspace}
\newcommand{\gevcc}{\ensuremath{{\mathrm{\,Ge\kern -0.1em V\!/}c^2}}\xspace}
\newcommand{\gevgevcccc}{\ensuremath{{\mathrm{\,Ge\kern -0.1em V^2\!/}c^4}}\xspace}
\newcommand{\gevgevcc}{\ensuremath{{\mathrm{\,Ge\kern -0.1em V^2\!/}c^2}}\xspace}
\newcommand{\mevcc}{\ensuremath{{\mathrm{\,Me\kern -0.1em V\!/}c^2}}\xspace}

\def\mum  {\ensuremath{{\,\upmu\rm m}}\xspace}

\def\invfb   {\ensuremath{\mbox{\,fb}^{-1}}\xspace}

\def\invps{\ensuremath{{\rm \,ps^{-1}}}\xspace}

\newcommand{\stat}{\ensuremath{\mathrm{\,(stat)}}\xspace}
\newcommand{\syst}{\ensuremath{\mathrm{\,(syst)}}\xspace}

\def\order{{\ensuremath{\cal O}}\xspace}
\newcommand{\chisq}{\ensuremath{\chi^2}\xspace}
\newcommand{\chisqndf}{\ensuremath{\chi^2/\mathrm{ndf}}\xspace}
\newcommand{\chisqip}{\ensuremath{\chi^2_{\rm IP}}\xspace}

\def\gsim{{~\raise.15em\hbox{$>$}\kern-.85em
          \lower.35em\hbox{$\sim$}~}\xspace}
\def\lsim{{~\raise.15em\hbox{$<$}\kern-.85em
          \lower.35em\hbox{$\sim$}~}\xspace}

\def\sPlot{\mbox{\em sPlot}\xspace}

\def\sqs   {\ensuremath{\protect\sqrt{s}}\xspace}

\def\ptot       {\mbox{$p$}\xspace}
\def\pt         {\mbox{$p_{\rm T}$}\xspace}

\def\evtgen     {\mbox{\textsc{EvtGen}}\xspace}

\def\geant      {\mbox{\textsc{Geant4}}\xspace}

\def\photos     {\mbox{\textsc{Photos}}\xspace}

\def\pythia     {\mbox{\textsc{Pythia}}\xspace}

\def\tell1  {TELL1\xspace}
\def\ukl1   {UKL1\xspace}

\newcommand{\ie}{\mbox{\itshape i.e.}\xspace}

\usepackage{cite} 
\usepackage{mciteplus}

\begin{document}
\renewcommand{\thefootnote}{\fnsymbol{footnote}}
\setcounter{footnote}{1}
\begin{titlepage}
\pagenumbering{roman}

\vspace*{-1.5cm}
\centerline{\large EUROPEAN ORGANIZATION FOR NUCLEAR RESEARCH (CERN)}
\vspace*{1.5cm}
\hspace*{-0.5cm}
\begin{tabular*}{\linewidth}{lc@{\extracolsep{\fill}}r}
\ifthenelse{\boolean{pdflatex}}
{\vspace*{-2.7cm}\mbox{\!\!\!\includegraphics[width=.14\textwidth]{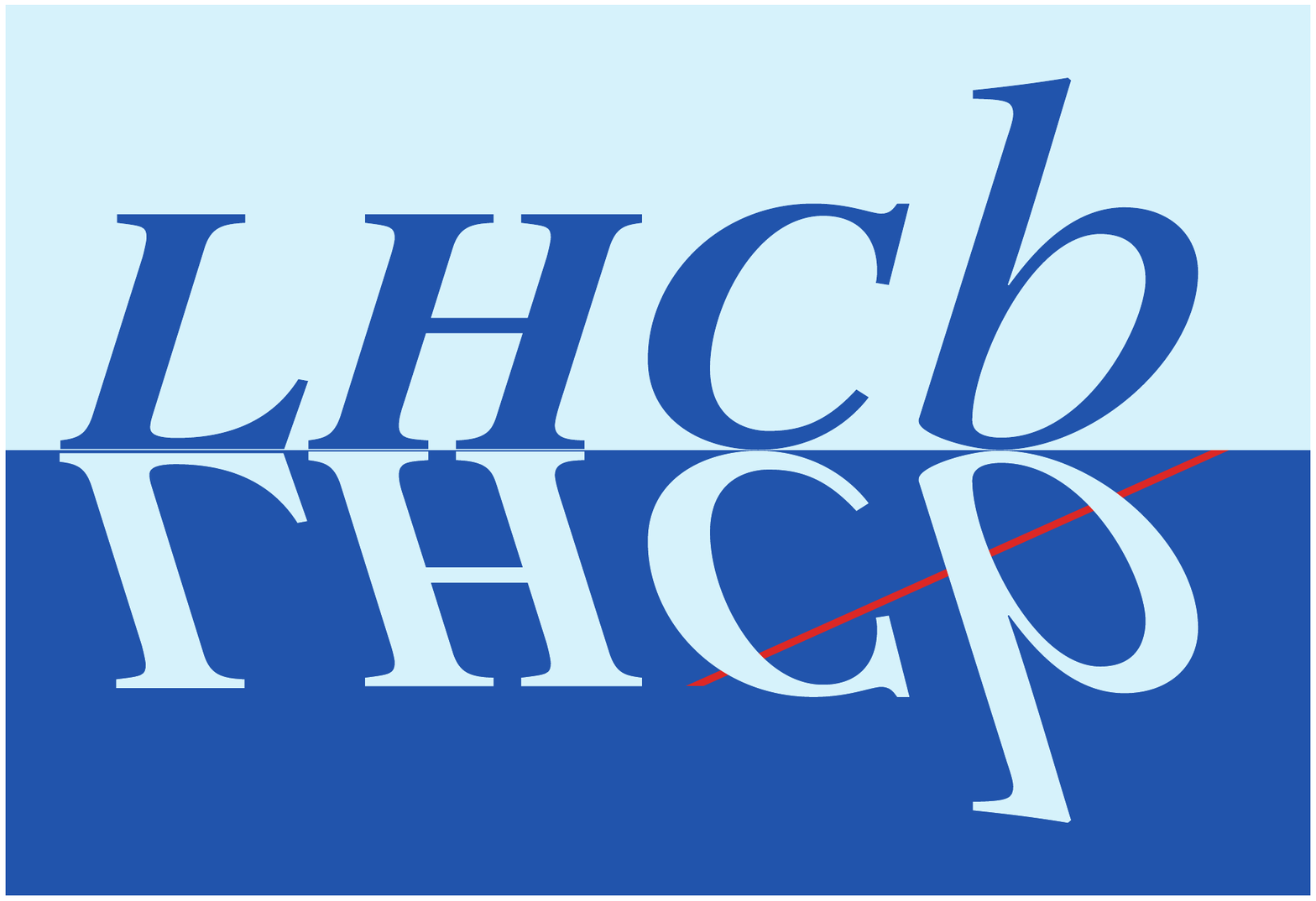}} & &}%
{\vspace*{-1.2cm}\mbox{\!\!\!\includegraphics[width=.12\textwidth]{lhcb-logo.eps}} & &}%
\\
 & & CERN-EP-2017-036 \\  
 & & LHCb-PAPER-2016-062 \\  
 & & 24 March 2017 \\ 
\end{tabular*}

\vspace*{3.0cm}

{\bf\boldmath\huge
\begin{center}
Measurement of $\Bz$, $\Bs$, $\Bp$ and $\Lb$ production asymmetries in 7 and 8~TeV proton-proton collisions
\end{center}
}
\vspace*{2.0cm}

\begin{center}
The LHCb collaboration\footnote{Authors are listed at the end of this paper.}
\end{center}

\begin{abstract}
  \noindent
The $\Bz$, $\Bs$, $\Bp$ and $\Lb$ hadron production asymmetries 
are measured using a data sample corresponding to an integrated luminosity of
3.0\invfb, collected by the LHCb experiment in proton-proton collisions at centre-of-mass
energies of 7 and 8\tev.
The measurements are performed as a function of transverse momentum
and rapidity of the $\bquark$ hadrons within the LHCb detector acceptance.
The overall production asymmetries, integrated over transverse momentum and rapidity,
are also determined. 
\end{abstract}

\vspace*{1.0cm}

\begin{center}
  Published in Phys.~Lett.~B 774 (2017) 139-158 
\end{center}

\vspace{\fill}

{\footnotesize 
\centerline{\copyright~CERN on behalf of the \lhcb collaboration, licence \href{http://creativecommons.org/licenses/by/4.0/}{CC-BY-4.0}.}}
\vspace*{2mm}

\end{titlepage}


\newpage
\setcounter{page}{2}
\mbox{~}
%
%
%
%

\cleardoublepage

\renewcommand{\thefootnote}{\arabic{footnote}}
\setcounter{footnote}{0}

\pagestyle{plain} 
\setcounter{page}{1}
\pagenumbering{arabic}
\section{Introduction}
\label{sec:Introduction}

The production rates of $\bquark$ and $\bquarkbar$ hadrons are not expected to be identical in proton-proton collisions, 
as $\bquark$ and $\bquarkbar$ quarks, produced in a hard scattering at the partonic level, might have different probabilities for coalescing with $\uquark$ or $\dquark$ valence quarks from the beam remnant. 
As a consequence, the production rates of \Bp and \Bz mesons may exceed those of \Bm and \Bzb, and $b$ baryons can be produced more abundantly than $\bar{b}$ baryons. In the case of \Bs and \Bsb the production rates depend on the values of the other production asymmetries as no valence strange quark is present within the colliding protons and \bquark and \bquarkbar quarks are predominantly produced in pairs. 

The LHCb detector, thanks to its unique geometry as a forward spectrometer, is particularly suited to measure such asymmetries, as they are expected to be enhanced at forward rapidities and small transverse momenta. Other subtle effects of quantum chromodynamics, beyond the coalescence of $\bquark$ quarks and light valence quarks, may also contribute~\cite{Chaichian:1993rh,Norrbin:2000jy,Norrbin:2000zc}.

The measurements of hadron production asymmetries are of primary importance, not only for the understanding of the production mechanisms, but also for enabling precise measurements of $\CP$ violation in \cquark and \bquark hadrons at the LHC. Indeed, observed asymmetries must be corrected for production effects to obtain the \CP asymmetries in the decays.
Simulations that model the non-perturbative fragmentation of $\bquark$ quarks in proton-proton collisions at LHC energies predict asymmetries generally up to a few percent~\cite{Sjostrand:2007gs,Bahr:2008pv}. Production asymmetries of \Bz and \Bs mesons have been measured by LHCb at a centre-of-mass energy of 7\tev, excluding values larger than a few percent~\cite{LHCb-PAPER-2014-042}. 
The LHCb collaboration has also searched for possible production asymmetries for $\Dp$ and $\Dsp$ mesons, finding the integrated \Dp production asymmetry different from zero at approximately three standard deviations~\cite{LHCb-PAPER-2012-026,LHCb-PAPER-2012-009}. In the \bquark-baryon sector, the LHCb collaboration measured the sum of the \Lb--\Lbbar production asymmetry and the \CP asymmetry in the $\Lb \to \jpsi p\Km$ decay~\cite{LHCb-PAPER-2015-032}, finding evidence for a dependence on \Lb rapidity. 

This paper reports measurements of the production asymmetries, $A_\mathrm{P}\left(\Bp\right)$, $A_\mathrm{P}\left(\Bz\right)$ and $A_\mathrm{P}\left(\Bs\right)$, measured using $\Bp \to \jpsi \Kp$, $\Bz \to \jpsi \Kstarz$ and $\Bs\to\Dsm\pip$ decays.
In addition, a measurement of $A_\mathrm{P}\left(\Lb\right)$, determined indirectly from the other asymmetries, is presented. 
Hereafter, \Kstarz is used to refer to the $K^*(892)^0$ and the inclusion of charge-conjugate decay modes is implied throughout, except when referring to the production asymmetries, which are defined as 
\begin{align}
A_\mathrm{P}\!\left(x\right)&\equiv\frac{\sigma\left(\overline{x}\right)-\sigma\left(x\right)}{\sigma\left(\overline{x}\right)+\sigma\left(x\right)}, \nonumber
\quad\mbox{with}\quad
x\in\{\Bp,\Bz,\Bs,\Lbbar\},
\end{align}
where $\sigma$ denotes the inclusive production cross-section in a given region of phase space. 
The data sample, collected by LHCb in proton-proton collisions, corresponds to an integrated luminosity of 1.0\invfb at a centre-of-mass energy of 7\tev, and 2.0\invfb at 8\tev. The measurements are performed as a function of both the component of the momentum transverse to the beam (\pt) and the rapidity ($y$) of the hadrons within the LHCb detector acceptance, and are then integrated over the ranges $0 < \pt < 30 $\gevc and/or $2.1 < y < 4.5$ for \Bp and \Bz decays, and $2 < \pt < 30 $\gevc and/or $2.1 < y < 4.5$ for \Bs and \Lb decays. The ranges in \pt are not identical due to different trigger requirements between decays with and without muons in the final states. This analysis improves the previous one performed on \Bz and \Bs production asymmetries~\cite{LHCb-PAPER-2014-042}, using a larger data sample and a finer binning scheme for investigating the dependence on \pt and $y$. In addition, new measurements of \Bp and \Lb production asymmetries have been included. Unlike in the previous analysis, the $\Bz \to \Dm\pip$ decay is not considered, as it has been found not to improve the precision on the \Bz measurement.

\section{Detector, trigger and simulation}
\label{sec:Detector}

The \lhcb detector~\cite{Alves:2008zz} is a single-arm forward spectrometer covering the \mbox{pseudorapidity} range $2 < \eta < 5$,
designed for the study of particles containing \bquark or \cquark quarks. The detector includes a high-precision tracking system consisting of a silicon-strip vertex detector surrounding the proton-proton interaction region, a large-area silicon-strip detector located upstream of a dipole magnet with a bending power of about $4{\rm\,Tm}$, and three stations of silicon-strip detectors and straw drift tubes placed downstream of the magnet.

The tracking system provides a measurement of momentum, \ptot, of charged particles with a relative uncertainty that varies from 0.5\% at low momentum to 1.0\% at 200\gevc. The minimum distance of a track to a primary vertex (PV), the impact parameter (IP), is measured with a resolution of $(15+29/\pt)\mum$, where \pt is measured
in\,\gevc.
Different types of charged hadrons are distinguished using information from two ring-imaging Cherenkov detectors.
Photons, electrons and hadrons are identified by a calorimeter system consisting of scintillating-pad and preshower detectors, an electromagnetic
calorimeter and a hadronic calorimeter. Muons are identified by a system composed of alternating layers of iron and multiwire
proportional chambers.
The trigger~\cite{LHCb-DP-2012-004} consists of a hardware stage, based on information from the calorimeter and muon systems, followed by a software stage, which applies a full event reconstruction.

For $\Bp \to \jpsi \Kp$ and $\Bz \to \jpsi \Kstarz$ decays, the data are collected by using the hardware muon trigger, which requires a single muon with large transverse momentum (from $p_\mathrm{T}>1.4$\gevc to $p_\mathrm{T}>1.8$\gevc) or a pair of muons with a large product of their transverse momenta (from $\sqrt{p_\mathrm{T,1}\,p_\mathrm{T,2}} > 1.3\gevc$ to  $\sqrt{p_\mathrm{T,1}\,p_\mathrm{T,2}} > 1.6\gevc$), depending on the data-taking conditions.
For $\Bs \to \Dsm \pip$ decays, data are collected using the hadronic hardware trigger, which requires at least one cluster in the hadronic calorimeter with a transverse energy greater than $3.5$\gev or $3.7$\gev, depending on the data-taking period.  
The output is then processed by the software trigger. 
In the case of $\Bp \to \jpsi \Kp$ and $\Bz \to \jpsi \Kstarz$ decays, $\jpsi$ mesons consistent with coming from the decay of a $\bquark$-hadron are selected by requiring that their decay products form a displaced vertex and have large IPs at the PV with respect to which the \B candidate has the smallest \chisqip.
The quantity \chisqip is defined as the difference in the vertex-fit \chisq of a given PV reconstructed with and without the particle under consideration. The $\Bs \to \Dsm \pip$ decays are selected by requiring a two- or three-track secondary vertex with a significant displacement from all PVs. At least one charged particle must have a transverse momentum $\pt > 1.7$\gevc and be inconsistent with originating from a PV. A multivariate algorithm~\cite{BBDT} is used for the identification of secondary vertices consistent with the decay of a \bquark hadron.

Simulated events are used to determine the signal selection efficiency as a function of $p_\mathrm{T}$ and $y$, and to study the modelling of the decay-time resolution, the reconstruction efficiency as function of the decay time and the shape of the invariant mass distribution of partially reconstructed background. In the simulation, proton-proton collisions are generated using \pythia~\cite{Sjostrand:2006za,Sjostrand:2007gs} with a specific \lhcb configuration~\cite{LHCb-PROC-2010-056}. Decays of hadronic particles are described by \evtgen~\cite{Lange:2001uf}, in which final-state radiation is generated using \photos~\cite{Golonka:2005pn}. The interaction of the generated particles with the detector, and its response, are implemented using the \geant toolkit~\cite{Allison:2006ve, *Agostinelli:2002hh} as described in Ref.~\cite{LHCb-PROC-2011-006}.

\section{Methodology}
\label{fitmodel}
The asymmetries $A_\mathrm{P}\left(\Bz\right)$ and $A_\mathrm{P}\left(\Bs\right)$ are measured by means of a time-dependent analysis of $\Bz \to \jpsi \Kstarz$ decays, with $\jpsi \to \mun \mup$ and $\Kstarz \to \Km\pip$, and $\Bs \to \Dsm \pip$ decays, with $\Dsm \to \Kp\Km\pim$.  The decay rate to a flavour-specific final state $f$ of a $B^0_{(s)}$ meson with average decay width $\Gamma_{\dquark (\squark)}$ can be written as
\begin{eqnarray}
S\left(t,\,\psi,\,\xi \right) & \propto  &  \left(1-\psi A_{\CP}\right)\left(1-\psi A_D\right) \label{eq:test}\\ 
&&  e^{-\Gamma_{\dquark (\squark)} t}  \left[\Omega^\xi_{+}\cosh\left(\frac{\ensuremath{\Delta\Gamma_{\dquark(\squark)}\xspace} t}{2}\right)+\psi \Omega^\xi_{-} \cos\left(\ensuremath{\Delta m_{\dquark(\squark)}\xspace} t\right) \right], \nonumber
\end{eqnarray}
where $\Delta m_{\dquark(\squark)}\xspace \equiv m_{\dquark(\squark),\,\rm H} - m_{\dquark(\squark),\,\rm L}$ and $\Delta\Gamma_{\dquark(\squark)}\xspace \equiv \Gamma_{\dquark(\squark),\,\rm L} - \Gamma_{\dquark(\squark),\,\rm H}$ are the mass and width differences of the $B^{0}_{(s)}$--$\Bbar{}^0_{(s)}$ system mass eigenstates. The subscripts H and L denote the heavy and light eigenstates, respectively. The symbol $\psi$ is the tag of the final state, which assumes the values $\psi=1$ if the final state is $f$ and $\psi=-1$ if the final state is the \CP conjugate $\bar{f}$, while $\xi$ indicates the tag of the initial flavour of the $B^0_{(s)}$ meson, which takes the values $\xi=1$ for $B^0_{(s)}$ and $\xi=-1$ for $\Bbar{}^0_{(s)}$. The terms $\Omega_{+}^{\xi}$ and  $\Omega_{-}^{\xi}$ are defined as
\begin{eqnarray}
\Omega_{\pm}^{\xi}\equiv\delta_{+1\xi}\left(1-A_\mathrm{P}\right)\left|\frac{q}{p}\right|^{1-\psi} \pm \delta_{-1\xi}\left(1+A_\mathrm{P}\right)\left|\frac{q}{p}\right|^{-1-\psi}, \nonumber
\label{eq:propertime_master1}
\end{eqnarray}
where $p$ and $q$ are complex parameters entering the definition of the two mass eigenstates of the effective Hamiltonian of the $B^0_{(s)}$ system, $\ket{B_{\mathrm{H}}}= p\ket{B^0_{(s)}} - q\ket{\Bbar{}^0_{(s)}}$ and $\ket{B_{\mathrm{L}}}= p\ket{B^0_{(s)}} + q\ket{\Bbar{}^0_{(s)}}$, and $\delta_{ij}$ is the Kronecker delta. The symbol $A_{\rm D}$ represents the detection asymmetry of the final state, defined in terms of the $f$ and $\bar{f}$ detection efficiencies, $\varepsilon$, as
\begin{eqnarray}
A_{\rm D}\equiv\frac{\varepsilon_{\bar{f}}-\varepsilon_f}{\varepsilon_{\bar{f}}+\varepsilon_f}. \nonumber
\end{eqnarray} 
The direct \CP asymmetry $A_{\CP}$ is defined as 
\begin{eqnarray}
A_{\CP} \equiv \frac{\mathcal{B}\left(\Bbar{}^0_{(s)}\rightarrow \bar{f}\right) - \mathcal{B}\left( B^0_{(s)} \rightarrow f \right) }{\mathcal{B}\left(\Bbar{}^0_{(s)}\rightarrow \bar{f}\right) + \mathcal{B}\left(B^0_{(s)}\rightarrow f \right)}\nonumber
\end{eqnarray}

where the symbol $\mathcal{B}$ stands for the branching fraction of the decay considered.

\begin{figure}[tb]
  \begin{center}
    \includegraphics[width=0.48\textwidth]{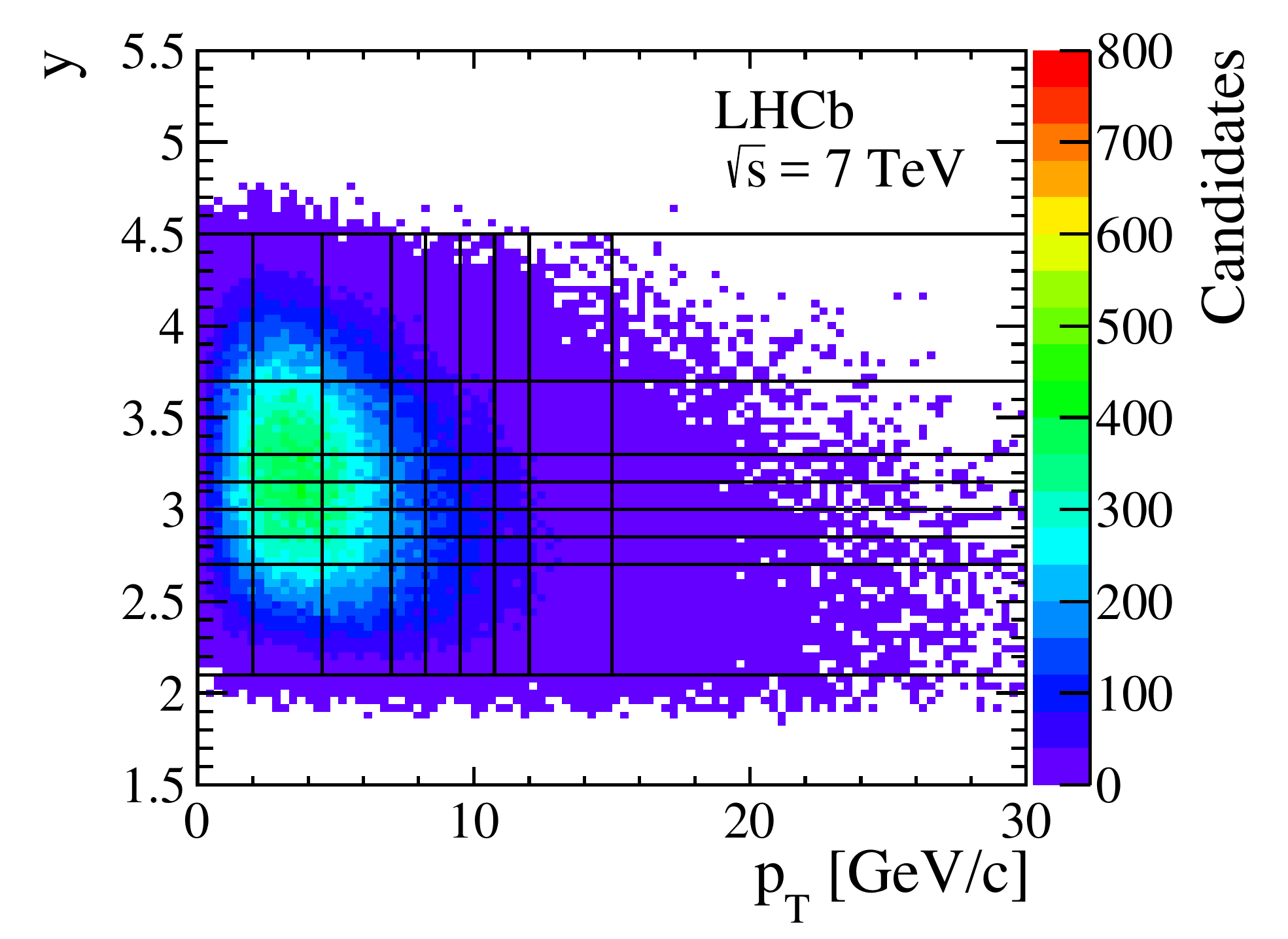}
    \includegraphics[width=0.48\textwidth]{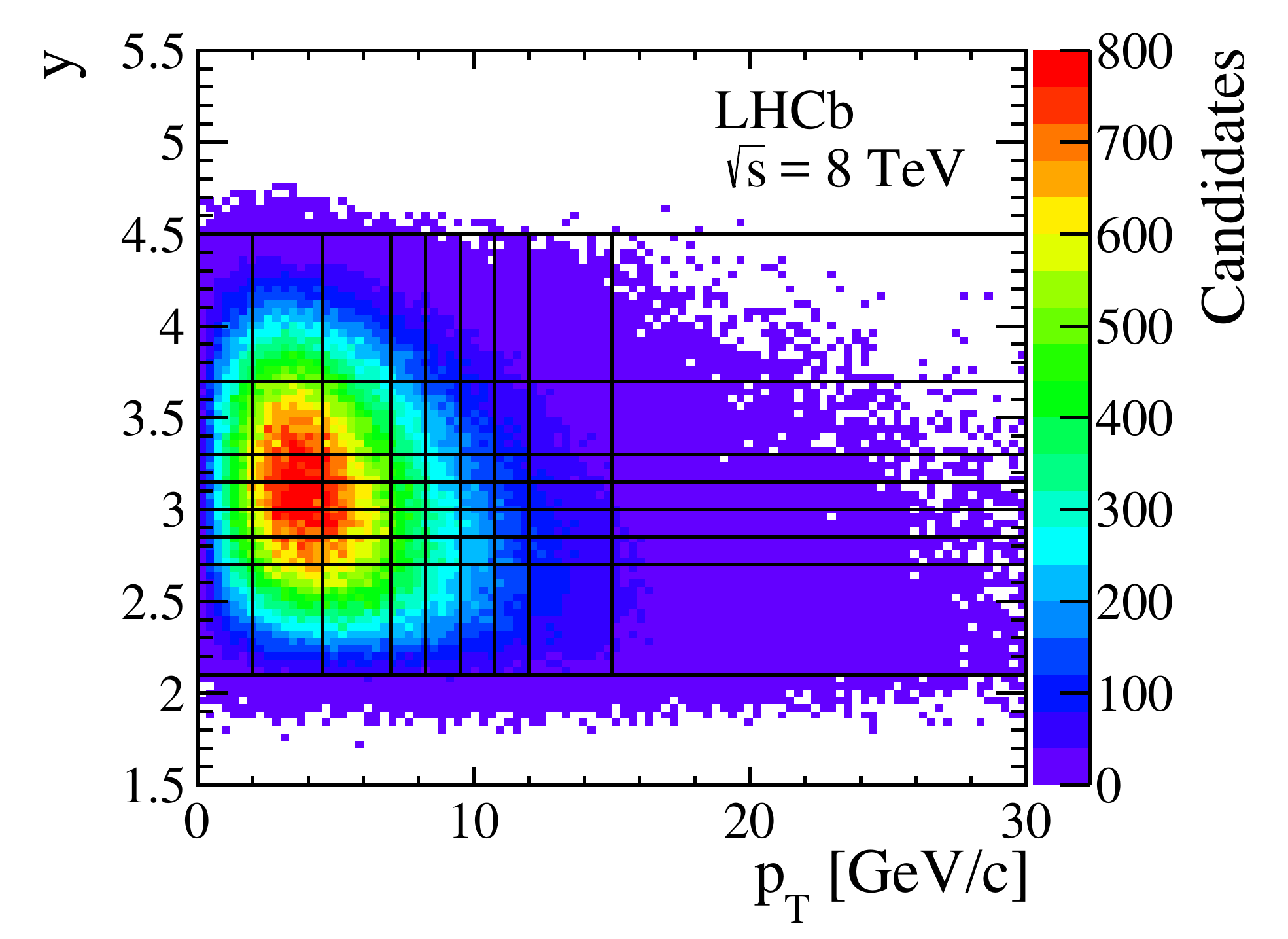}
    \includegraphics[width=0.48\linewidth]{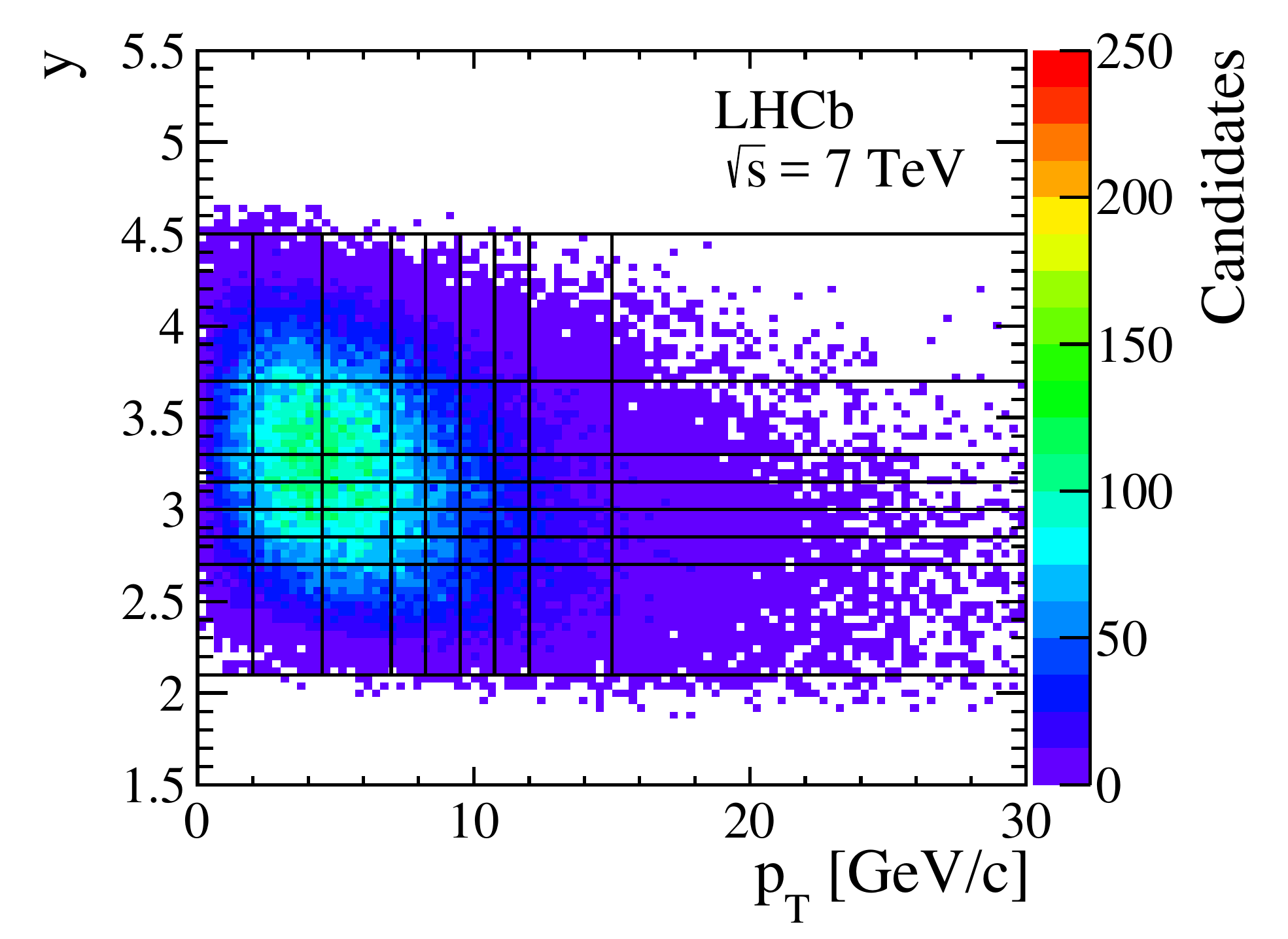}
    \includegraphics[width=0.48\linewidth]{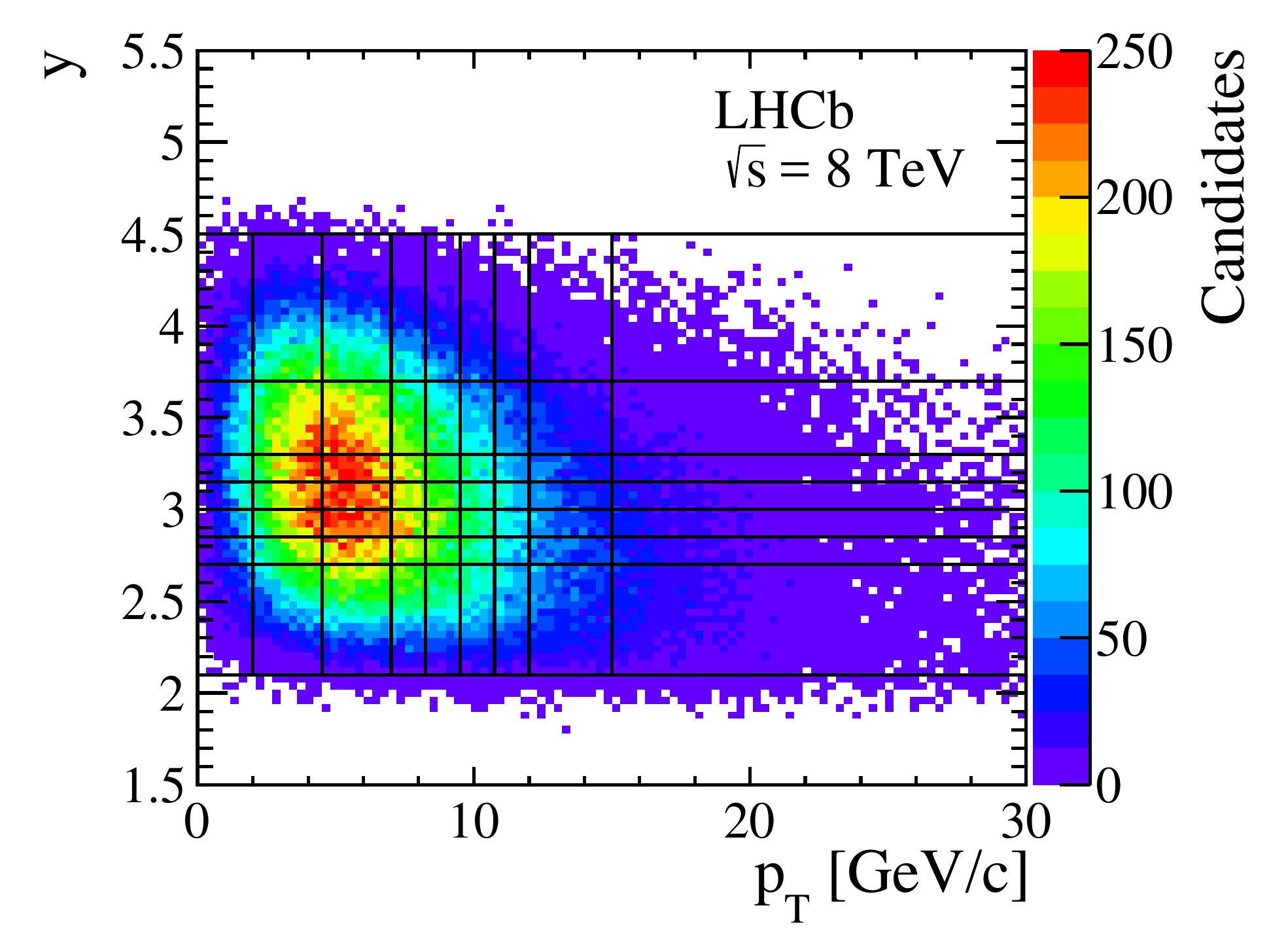}
    \includegraphics[width=0.48\linewidth]{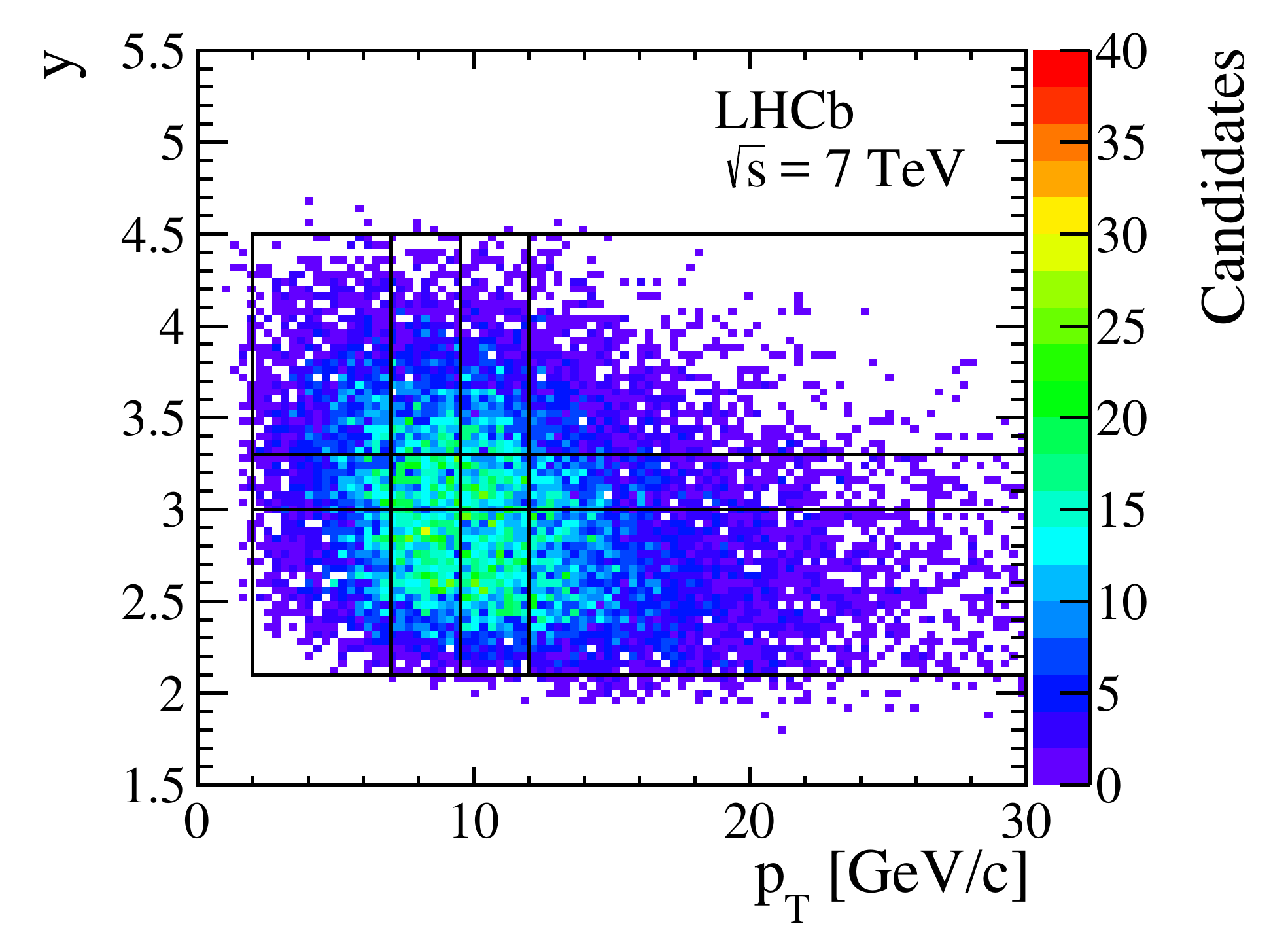}
    \includegraphics[width=0.48\linewidth]{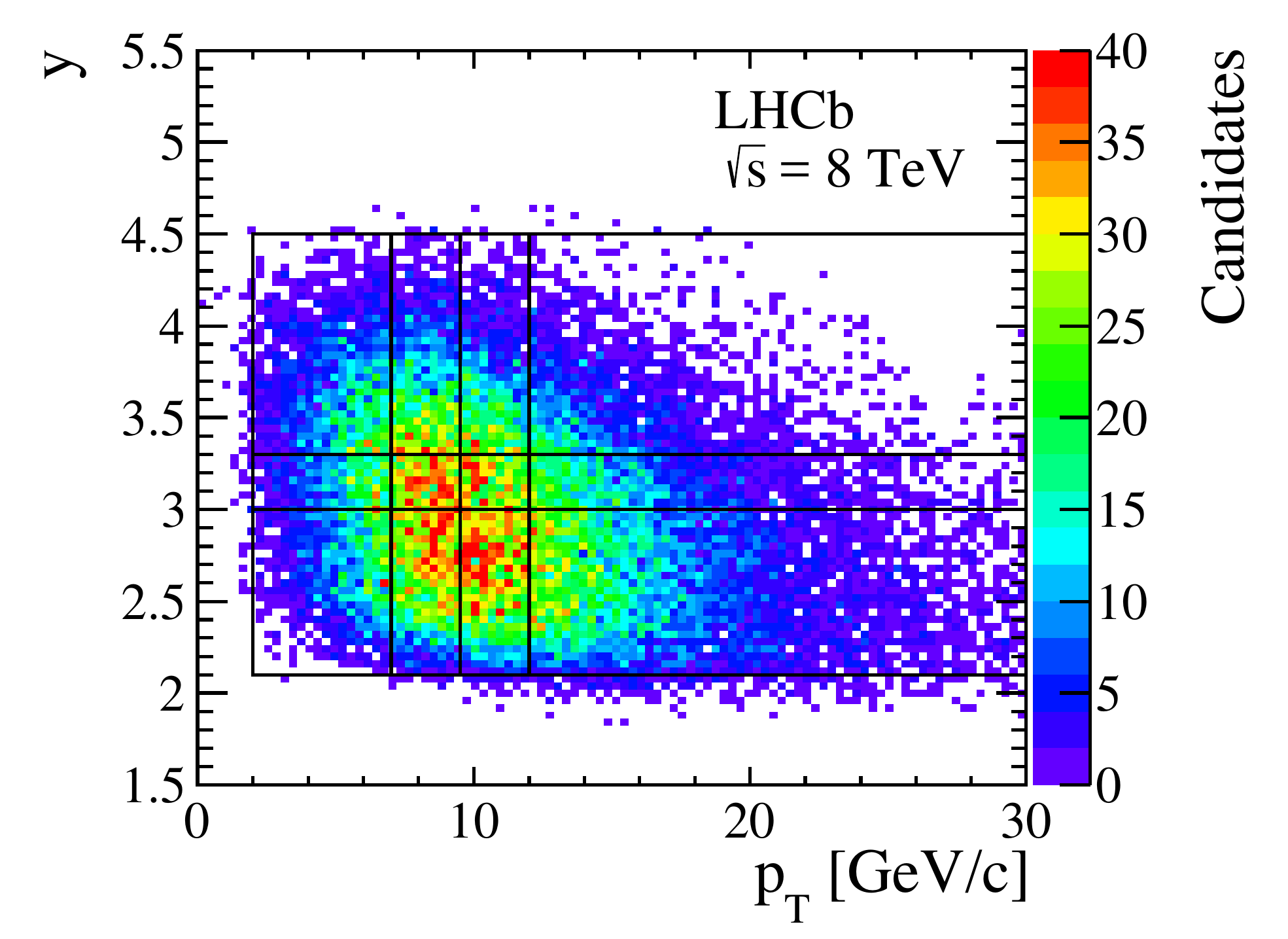}
 \end{center}
  \caption{Distributions of $\pt$ and $y$ for background-subtracted (top) \Bp, (middle) \Bz and (bottom) \Bs decays for data collected in proton-proton collisions at the centre-of-mass energies of (left) 7 and (right)  8\tev. The binning schemes are superimposed.}
  \label{fig:B_pt_eta}
\end{figure}

The asymmetry $A_\mathrm{P}\left(\Bp\right)$ is measured by means of a time-integrated analysis of $\Bp \to \jpsi \Kp$ decays, with $\jpsi \to \mup \mun$, starting from the raw asymmetry defined as
\begin{equation}
A_{\rm raw} \equiv \frac{N(\Bm \to \jpsi \Km) - N(\Bp \to \jpsi \Kp)}{N(\Bm \to \jpsi \Km) + N(\Bp \to \jpsi \Kp)}, \nonumber
\end{equation}
where $N$ denotes the observed yields. The raw asymmetry can be written, up to $\order(10^{-6})$ corrections, as 
\begin{eqnarray}
A_{\rm raw} (\Bp \to \jpsi \Kp) = A_{\rm P}(\Bp) + A_{\rm D} (\Kp) + A_{C\!P} (\Bp \to \jpsi \Kp),
\label{eq:prodasym}
\end{eqnarray}
where $A_{\rm D} (\Kp)$ is the \Kp detection asymmetry, measured by means of charm control samples as in Ref.~\cite{LHCb-PAPER-2014-013}, and $A_{C\!P} (\Bp \to \jpsi \Kp)$ is the \CP asymmetry in the decay, measured by \babar, \belle and \dzero~\cite{Abazov:2013sqa,Sakai:2010ch,Aubert:2004rz}. An improved measurement of the \CP asymmetry was also made recently by LHCb~\cite{LHCb-PAPER-2016-054}, using an independent data sample selected with different trigger requirements.
The $A_{\mathrm P}$ values obtained from Eq.~\ref{eq:test} and Eq.~\ref{eq:prodasym} are detector-independent quantities only if measured in kinematic regions where the reconstruction efficiencies are constant. 
To account for the dependence of the production asymmetries on the kinematics of the \Bp, \Bz and \Bs mesons, each data sample is divided into bins of (\pt, $y$), and the measurement is performed for each bin. Figure~\ref{fig:B_pt_eta} shows the distribution of (\pt, $y$) for $\Bp \to \jpsi \Kp$, $\Bz\to \jpsi \Kstarz$ and $\Bs\to\Dsm\pip$ decays, where the background components are subtracted using the \sPlot technique~\cite{Pivk:2004ty} and the definition of the various kinematic bins is overlaid. For the \Bp and \Bz decays a common set of bins is used, defined in Table~\ref{tab:resultsBpB02011} of the Appendix,
and in the case of the \Bs decay, the binning scheme is reported in Table~\ref{tab:resultsBsLb2011}.

In proton-proton collisions at the LHC, $\bquark$ and $\bquarkbar$ quarks are predominantly pair-produced via strong interaction processes. This leads to a relation between the $\Lb$ production asymmetry and the other $\bquark$-hadron production asymmetries, namely
\begin{eqnarray}
 A_{\rm P}(\Lb) = - \left[ \frac{f_u}{f_{\Lb}} A_{\rm P}(\Bp) + \frac{f_d}{f_{\Lb}} A_{\rm P}(\Bz) + \frac{f_s}{f_{\Lb}} A_{\rm P}(\Bs) + \frac{f_c}{f_{\Lb}} A_{\rm P}(\Bc) + \frac{f_{ \rm other}}{f_{\Lb}} A_{\rm P}(\rm other) \right] \nonumber
\label{eq:APLBrelation}
\end{eqnarray}
where $f_{u}$, $f_{d}$, $f_{s}$, $f_c$, $f_{\Lb}$ and $f_{\rm other}$ are the fragmentation fractions of a \bquark quark hadronizing into weakly-decaying \Bp, \Bz, \Bs, \Bc mesons, \Lb baryons and all the other $\bquark$-baryon species.
The ratios of the fragmentation fractions, $f_u/f_{\Lb}$, $f_d/f_{\Lb}$ and $f_s/f_{\Lb}$  are taken from LHCb measurements reported in Refs.~\cite{LHCb-PAPER-2012-037, LHCb-PAPER-2014-004}. Their dependence on \pt and $y$ is taken into account.
The terms $(f_c/f_{\Lb}) \cdot A_P(\Bc)$ and $(f_{\rm other}/f_{\Lb}) \cdot A_P(\rm other)$ are of the order of $3\cdot 10^{-5}$ and $2\cdot 10^{-3}$, respectively. This is estimated assuming that the value of $A_P(\Bc)$ and $A_P(\rm other)$ are of the same order as the \B-meson production asymmetries ($\simeq 10^{-2}$) and taking the values of $f_c/f_{\Lb}$ and $f_{\rm other}/f_{\Lb}$ from simulation. Neglecting these terms, the \Lb production asymmetry can be measured using the approximate relation
\begin{eqnarray}
 A_{\rm P}(\Lambda_b^0) \simeq - \left[ \frac{f_u}{f_{\Lb}} A_{\rm P}(B^+) +
  \frac{f_d}{f_{\Lb}} A_{\rm P}(B^0) + \frac{f_s}{f_{\Lb}} A_{\rm P}(B^0_s)
  \right].
\label{eq:APLBrelation_final}
\end{eqnarray}
Possible small deviations from this approximation, due in particular to contributions from other $b$ baryons, are taken into account in the evaluation of systematic uncertainties.

\subsection{Integrated production asymmetries}

In addition to the measurements in bins, integrated production asymmetries, where efficiency corrections have been applied, are also provided. The integration of the $A_{\rm P}$ values is performed in the ranges $0 < \pt < 30$\gevc and $2.1 < y < 4.5$ for the $\Bp$ and \Bz decays and in the ranges $2 < p_\mathrm{T} < 30 $\gevc and $2.1 < y < 4.5$ for the $\Bs$ and \Lb decays. The integrated value of $A_{\rm P}$ is given by
\begin{eqnarray}
A_{\rm{P}} = \sum_i \frac{N_{i}}{\varepsilon_{i}} A_{{\rm P},i}/ \sum_i \frac{N_{i}}{\varepsilon_{i}} \label{eq:APintegrated}
\end{eqnarray}
where the index $i$ runs over the bins,
$N_{i}$ is the number of observed signal events in the $i$-th bin and $\varepsilon_{i}$ is the efficiency defined as the number of selected events divided by the number of produced events in the $i$-th bin. The signal yield in each bin can be expressed as
\begin{eqnarray}
N_{i} = \mathcal{L} \,\, \sigma_{b \bar{b}} \,\, 2 \,\,f_{q} \,\, \mathcal{B} \,\, F_{i} \,\, \varepsilon_i \label{eq:yields}
\end{eqnarray}
where $\mathcal{L}$ is the integrated luminosity, $\sigma_{b \bar{b}}$ is the $b \bar{b}$ cross section, $f_q$ is the fragmentation fraction for quark flavour $q$, with $q\in\{\uquark,\dquark,\squark\}$, $F_{i}$ stands for the fraction of the \bquark hadrons produced in the $i$-th bin and $\mathcal{B}$ is the branching fraction of the $\bquark$-hadron decay being considered. By substituting $N_{i} / \varepsilon_{i}$ from Eq.~\ref{eq:yields} into Eq.~\ref{eq:APintegrated}, the integrated value of $A_{\rm{P}}$ becomes
\begin{eqnarray}
A_{\mathrm{P}} = \sum_i \omega_{i} A_{\mathrm{P},i} \ .\label{eq:AP} \nonumber
\end{eqnarray}
where $\omega_{i} = F_{i} / \sum_i F_{i}$.
The $\omega_i$ values are determined using simulated events, generated with proton-proton collisions at the centre-of-mass energies of 7 and 8\tev.

\section{Data set and event selections}
The selections of $\Bp \to \jpsi \Kp$ and $\Bz \to \jpsi \Kstarz$ decays are based on the reconstruction of $\jpsi \to \mun \mup$ decays combined with either a track identified as a kaon or with a \Kstarz decaying to $\Kp\pim$.
The $\jpsi$ candidates are formed from two oppositely charged tracks originating from a common vertex, identified as muons with $\pt > 500$\mevc.
The $\Kstarz$ candidates are formed from two oppositely charged tracks, one identified as a kaon and the other as a pion, originating from the same vertex. They are required to have $\pt > 1$\gevc and the \Kp\pim invariant mass in the range $826\text{--}966$\mevcc.
The invariant mass of \Bz and \Bp candidates, calculated constraining the two muon candidates to have the known \jpsi mass, is required to be in the range $5150\text{--}5450$\mevcc.
The proper decay time of the \B-meson candidate is calculated from a fit that constrains the candidate to originate from the PV with the smallest \chisqip with respect to the \B candidate.  Only \B-meson candidates with a decay time greater than 0.2 ps are retained. This lower bound on the decay time rejects a large fraction of the combinatorial background.

In the case of $\Bs \rightarrow \Dsm\pip$ decays, the \Dsm candidates are reconstructed using the $\Kp\Km\pim$ decay channel. Requirements are applied to the \Dsm decay products before combining them to form a common vertex, namely the scalar \pt sum of the tracks must exceed 1.8\gevc and the largest distance of closest approach between all possible pairs of tracks must be less than 0.5 mm. 
The \Dsm candidates are then required to be significantly detached from the PV and to have the invariant mass within the range  $1949\text{--}1989$\mevcc. Each \Dsm candidate is subsequently combined with a second charged pion, referred to as the accompanying pion in the following, to form the $\B$-meson decay vertex. The sum of the \pt values of the \Dsm and accompanying \pip must be larger than 5\gevc and the decay time of $\B$-meson candidates must be greater than 0.2 ps. Furthermore, the cosine of the angle between the $B$-meson candidate momentum vector and the vector connecting the PV and $B$-meson candidate vertex is required to be larger than 0.999.

Stringent particle identification criteria are required to be satisfied for the kaons and pions forming the $\Kstarz$ and \Dsm candidates, the kaon from the \Bp decay and the accompanying pion, in order
to reduce to a negligible level the background from other $\B$-meson decays with a misidentified kaon or pion, and from \Lb decays with a misidentified proton.

A final selection is applied using a multivariate analysis method
based on a Boosted Decision Tree~\cite{Breiman,*Roe}, where the variables used in the selection are: the \pt and the IP of the $B$ decay products, the flight distance and the IP of the $B$ candidate, and, in the case of \Bs, the flight distance of the \Dsm meson. The multivariate selection is trained using simulated events as a proxy for the signal, and \B-meson candidates from data selected in the upper mass sidebands to represent the background.

\section{Fit model}
\label{sec:fitmodel}

For each signal and background component, the invariant mass distribution of all \B candidates, and, in the case of $B^{0}_{(s)}$, the decay time, is modelled by defining  appropriate probability density functions (PDFs). Two categories of background are considered: the combinatorial background, due to the random association of tracks, and the partially reconstructed background, due to decays with a topology similar to that of the signal, but with one or more particles not reconstructed. The latter is only relevant for $\Bs \to \Dsm\pip$ decays.

\subsection{Invariant mass parameterization}

The signal component for \B mesons is modelled by convolving a sum of two Gaussian functions with a function
parameterizing the final-state QED radiation (FSR). The PDF of the invariant mass, $m$, is given by the convolution
\begin{eqnarray}
\label{eq:sigmodel}
g(m) \propto \int_{0}^{+\infty} (m^{\prime})^s G\left(m+m^{\prime};\mu\right)\mathrm{d}m^{\prime}
\end{eqnarray}
where $G$ is the sum of two Gaussian functions with different widths
and common mean $\mu$ that represents the $\B$-meson mass. 
The parameter $s$ governs the amount of FSR, and using simulation 
is found to be $s = -0.9966 \pm 0.0005$ for the \Bp decay,  $s = -0.9945 \pm 0.0003$ for the \Bz decay
and $s = -0.9832 \pm 0.0004$ for the \Bs decay. The invariant mass shape of the combinatorial background is well described by an exponential PDF.

Regarding the $\jpsi\Kp$ invariant mass spectrum, common parameters are used for both \Bp and \Bm mesons. In the case of the $\Dsm \pip$ spectrum, a background component due to partially reconstructed \Bs decays is also present in the low invariant mass region. The contributions with the highest branching fractions are from the $\Bs \rightarrow D_s^{*-}\pip$ decay, with $D_s^{*-}\to\Dsm\gamma$ or $D_s^{*-}\to \Dsm\piz$, where the $\gamma$ or $\pi^0$ is not reconstructed, and from the $\Bs \rightarrow D_s^{-}\rho^{+}$ decay, with $\rho^{+}\to\pip \piz$, where the $\piz$ is not reconstructed. The partially reconstructed components are parameterized by means of a kernel estimation technique~\cite{Cranmer:2000du} based on invariant mass distributions obtained from simulated events, where the same selection applied to data is used and differences in invariant mass resolution between data and simulation are taken into account. The yields are obtained from the fits. 

In the case of the $\Bs\to\Dsm\pip$ decay, an irreducible background component due to the $\Bz\to\Dsp\pim$ decay is also present. This component is accounted for in the fits using the same parameterization adopted for the signal, where the mean values of the two signal PDFs are separated by the difference in the known masses between \Bz and \Bs mesons~\cite{PDG2016} and the production asymmetry is fixed to the \Bz measured value. The yield of this component is fixed according to the known branching fraction~\cite{PDG2016}. 

\subsection{Decay time parameterization}

Starting from Eq.~\ref{eq:test} and summing over $\xi$, the decay rate to a flavour-specific final state of a neutral $\B$ meson is parameterized by the convolution 
\begin{eqnarray}
S\left(t,\psi\right) & \propto  &  \left[1-\psi\left( A_{\CP}+A_D\right)\right] \nonumber \label{eq:untaggedasymmetry}\\ 
&& \left\{ e^{-\Gamma_{\dquark(\squark)} t}  \left[\Lambda_{+}\cosh\left(\frac{\Delta\Gamma_{\dquark(\squark)} t}{2}\right)+\psi \Lambda_{-} \cos\left(\Delta m_{\dquark(\squark)} t\right) \right] \otimes R\left( t\right) \right\} \epsilon \left( t \right),\nonumber
\end{eqnarray}
where $R(t)$ is a function describing the decay-time resolution, as discussed in Sec.~\ref{decaytime}, and $\epsilon(t)$ is the reconstruction efficiency as a function of the decay time determined from simulation and parameterized for the \Bz decay by
\begin{equation}\label{eq:acceptance1}
  \epsilon\left(t\right) = \frac{1}{2}
  \left[1-\mathrm{erf}\left(p_{1} t^{p_{2}} \right)\right]
  \left(1+p_{3} t\right), \nonumber
\end{equation}
and for the $\Bs$ decay by
\begin{equation}\label{eq:acceptance2}
  \epsilon\left(t\right) =
  \frac{1}{2}\left[1-\frac{1}{2}\mathrm{erf}\left(\frac{n_{1}-t}{t}\right)-\frac{1}{2}\mathrm{erf}\left(\frac{n_{2}-t}{t}\right)\right]
    \left(1+n_{3}  t\right), \nonumber
\end{equation}
where erf is the error function, and $p_i$ and $n_i$ are parameters determined from simulation.
The terms $\Lambda_{+}$ and $\Lambda_{-}$ are defined as
\begin{equation}
\Lambda_{\pm}\equiv\left(1-A_\mathrm{P}\right)\left|\frac{q}{p}\right|^{1-\psi}\pm\left(1+A_\mathrm{P}\right)\left|\frac{q}{p}\right|^{-1-\psi}, \nonumber
\end{equation}
and the term $A_{\CP}A_{\rm D} $ is neglected, as $A_{\rm D}$ is $\order(10^{-2})$~\cite{LHCb-PAPER-2015-034} and $A_{\CP}$ is very small for the decays under study. For this reason, it is only possible for the fit to determine the sum of $A_{\rm D}$ and $A_{\CP}$, but not their individual values. 

The decay-time PDF of the combinatorial background is studied using events from a high invariant mass window where the signal is not present, namely in the range $5310\text{--}5340$\mevcc for \BdToJPsiKst and $5450\text{--}5900$\mevcc for $\Bs \to \Dsm \pip$ decays. The partially reconstructed component for the $\Bs \to \Dsm \pip$ decay is determined from simulated events.

\subsection{Decay time resolution}
\label{decaytime}
The decay-time resolutions of $\Bz$ and $\Bs$ mesons are estimated by studying the decay time of  fake \B candidates, formed from a \Dm decaying to \Kp\pim\pim and a pion track, both coming from the same PV. These \B candidates are called fake, as the probability to form a real decay with this technique is negligible. In order to avoid the introduction of biases in the decay-time measurements, the accompanying pion is selected with requirements on momentum and \pt, rather than on IP. The decay-time distribution of these fake \B candidates yields an estimate of the decay-time resolution of a real decay. This method is verified by means of simulated events, both for signal and fake \B decays. 
The resolution model, $R(t)$, consisting of a sum of three Gaussian functions with zero mean and three different widths, characterized by an average width of 49 fs, is used. The resolution is found to be overestimated by about 4 fs and to be dependent on the decay time. 
Taking these effects into account, an uncertainty of 8~fs on the average width is considered as a systematic uncertainty. It is estimated from simulation that the measurement of the decay time is biased by no more than 2 fs, and this effect is also accounted for as a systematic uncertainty.

\section{Determination of the production asymmetries}
\label{sec:fitresults}
The production asymmetries are determined by means of unbinned ($B^0_{(s)}$) and binned (\Bp) maximum likelihood fits, for each kinematic bin, to the invariant mass (\Bp) and invariant mass and decay time ($B^0_{(s)}$) distributions, using the models described in the previous section. The models are validated with a series of fits to the mass and lifetime distributions of events obtained from pseudoexperiments. No evidence of biases on central values nor on the uncertainty is found. 
Furthermore, a global fit to the total sample of selected candidates is performed for each of the three decay modes to validate the fitting model on data. In the case of the time-dependent analysis, the mass differences \dmd and \dms, the mixing parameters $|q/p|_{\Bz}$ and $|q/p|_{\Bs}$, the average decay widths \Gd and \Gs, and the width difference \DGs are fixed to the central values of the measurements reported in Table~\ref{tab:input}. The width difference \DGd is fixed to zero.

\begin{table}[!ht]
 \begin{center}
  \caption{Values of the various physical inputs used in the fits, as reported in Ref.~\cite{HFAG}.}   
 \label{tab:input}
   \begin{tabular}{l|c}
Parameter & Value \\ 
\hline
$\dmd\,[\!\invps]$ &  $0.5065\pm 0.0019$ \\
$\dms\,[\!\invps]$ &  $17.757 \pm 0.021\,\,\,$ \\
$\Gd\,[\!\invps]$ &  $0.6579 \pm 0.0017$ \\
$\Gs\,[\!\invps]$ & $0.6645\pm 0.0018$ \\
$\DGs\,[\!\invps]$ &  $0.083 \pm 0.006$ \\
$|q/p|_{ \Bz }$ & $1.0007\pm 0.0009$ \\
$|q/p|_{\Bs}$ & $1.0038 \pm 0.0021$ \\
 \end{tabular}
 \end{center}
\end{table}

\begin{figure}[!ht]
  \begin{center}
   \includegraphics[width=0.42\linewidth]{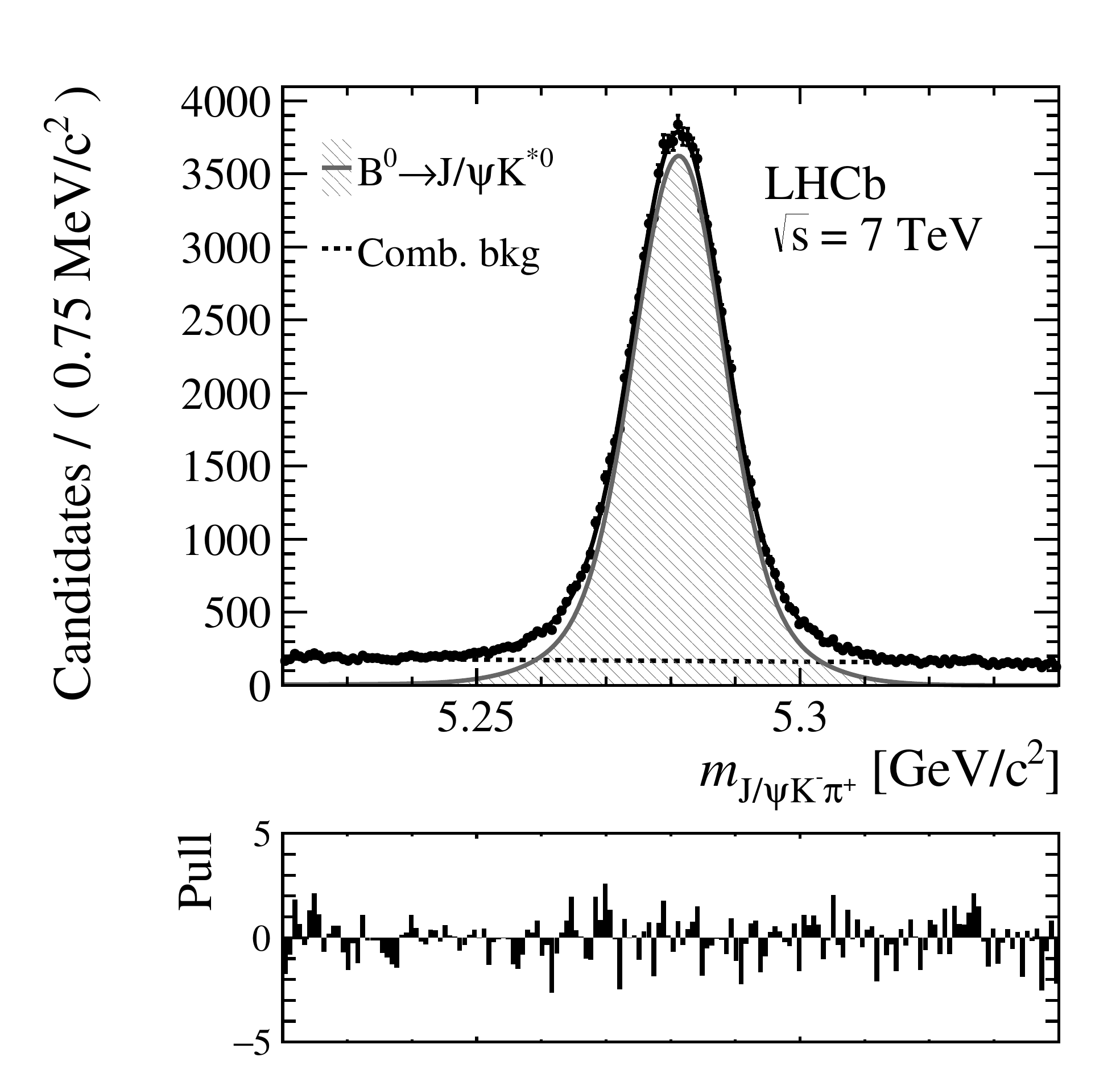}
   \includegraphics[width=0.42\linewidth]{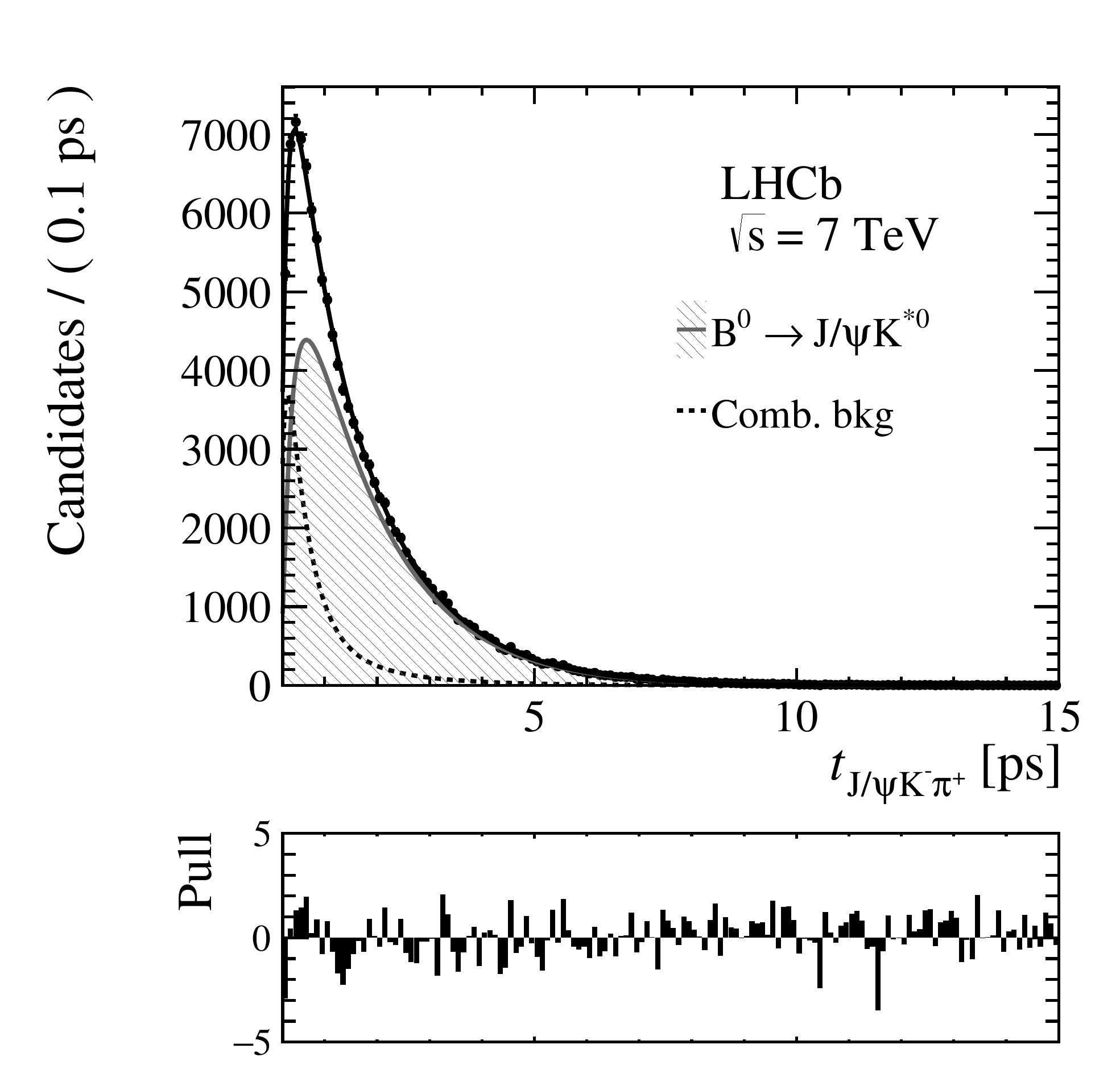}
   \includegraphics[width=0.42\linewidth]{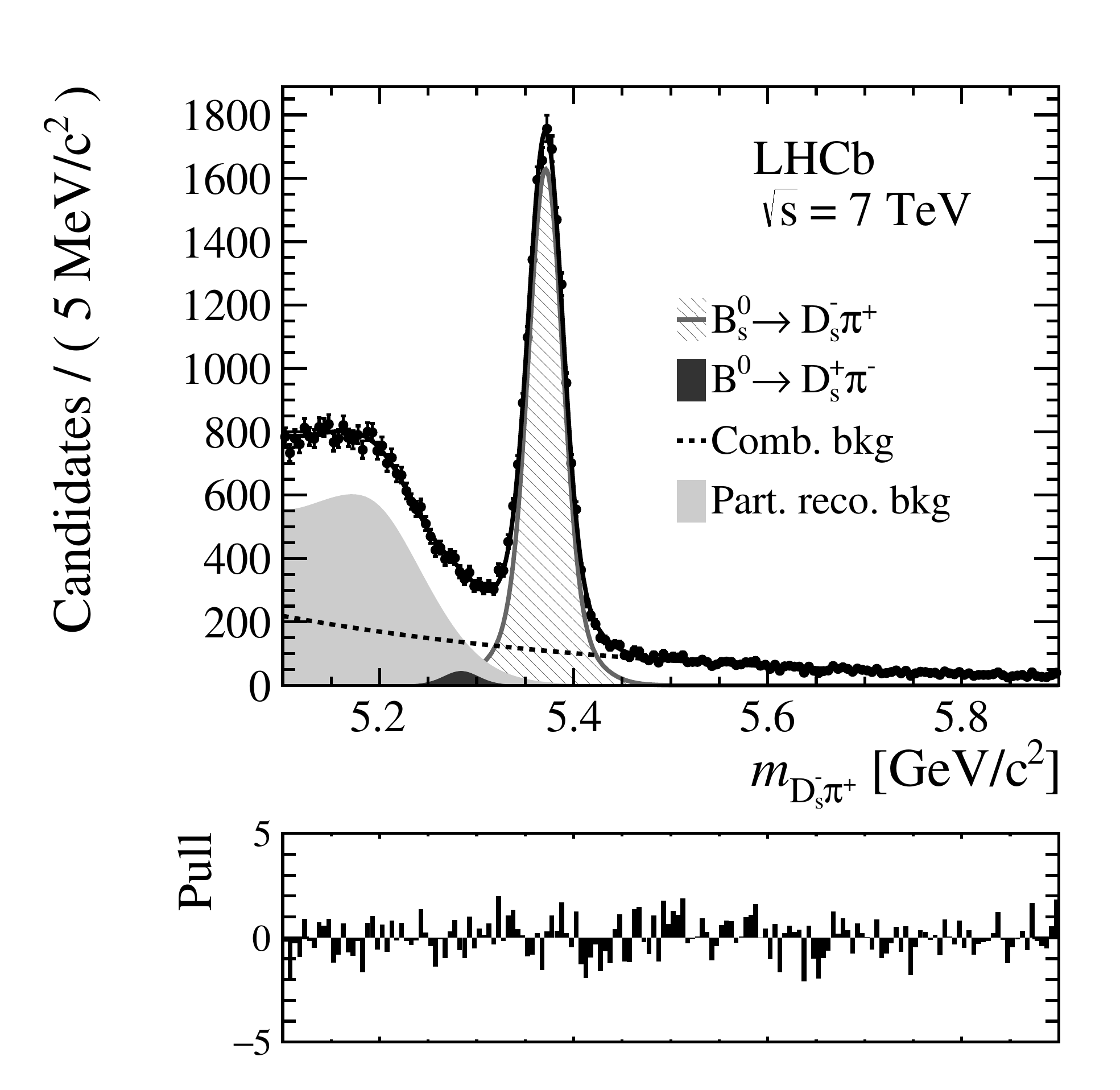}
   \includegraphics[width=0.42\linewidth]{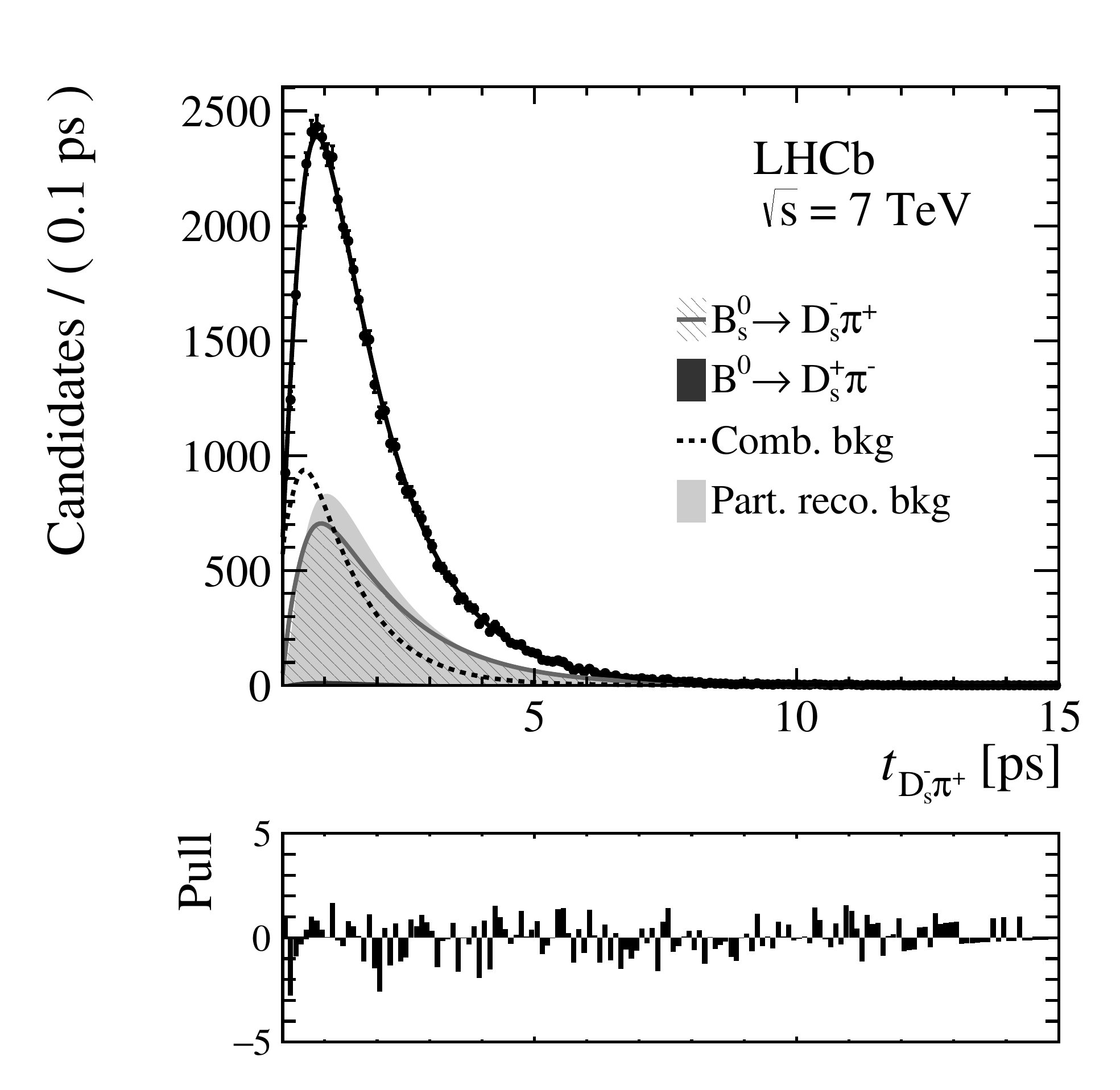}
   \includegraphics[width=0.42\textwidth]{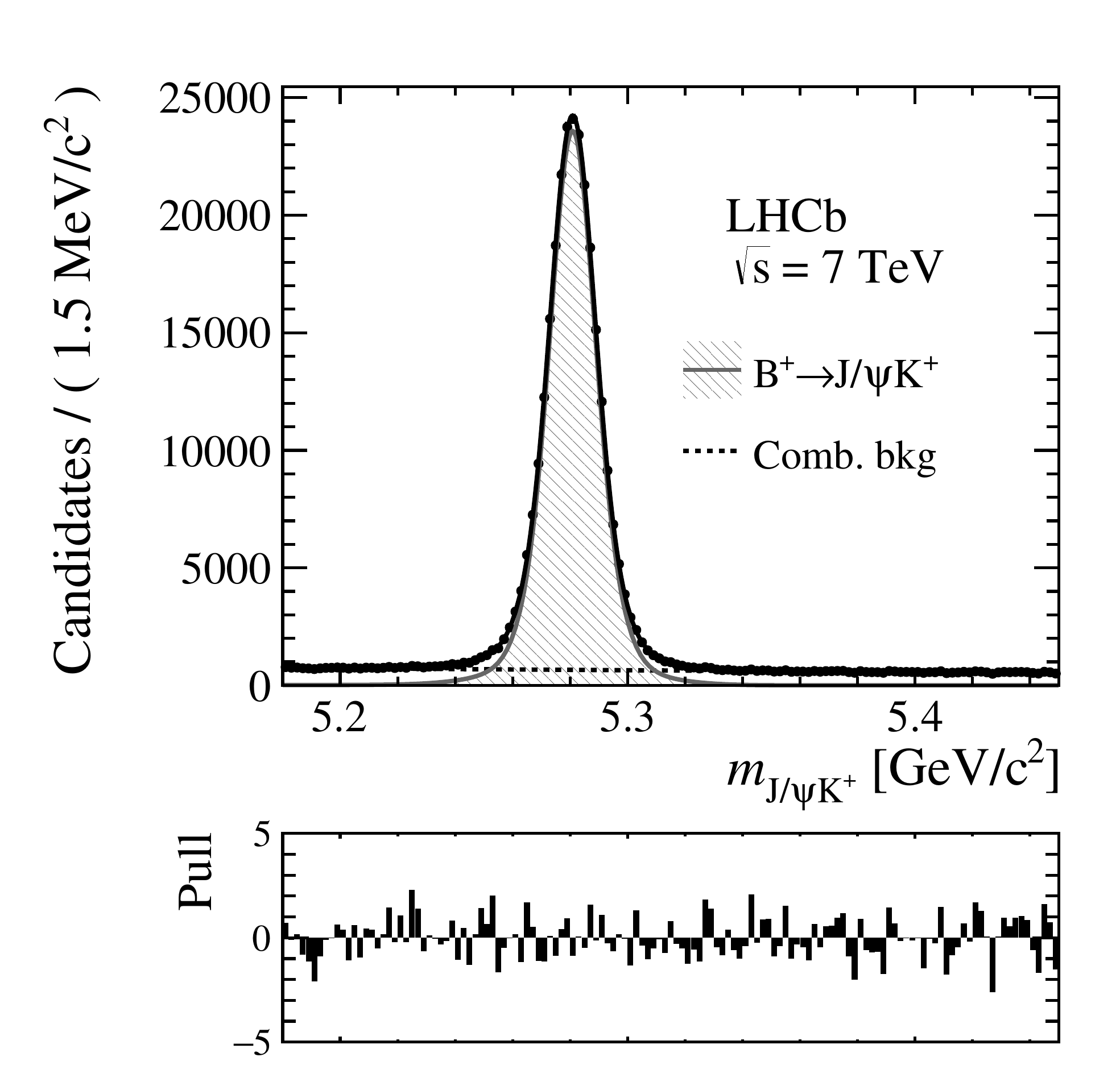}
 \end{center}
  \caption{Distributions of (left) invariant mass and (right) decay time for (top) $\Bz\to\jpsi\Kstarz$, (middle) $\Bs\to\Dsm\pip$ and (bottom) of invariant mass for $\Bp\to \jpsi\Kp$ decays, with the results of the fit overlaid. The data were collected in proton-proton collisions at the centre-of-mass energy of 7\tev. The contributions of the various background sources are also shown. Below each plot are the normalized residual distributions.}
  \label{fig:globalfits_2011}
\end{figure}

\begin{figure}[!ht]
  \begin{center}
  \includegraphics[width=0.42\linewidth]{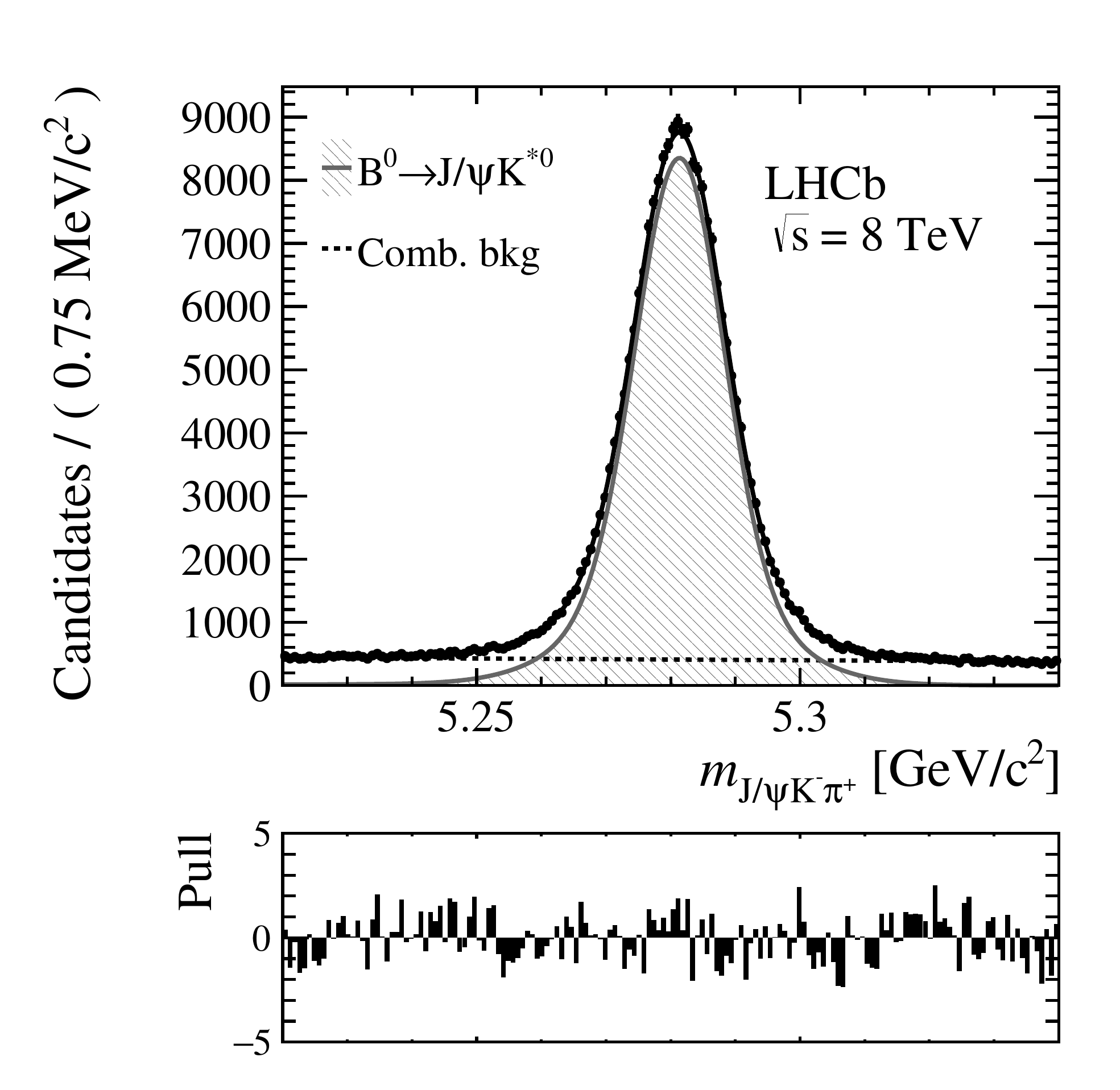}
   \includegraphics[width=0.42\linewidth]{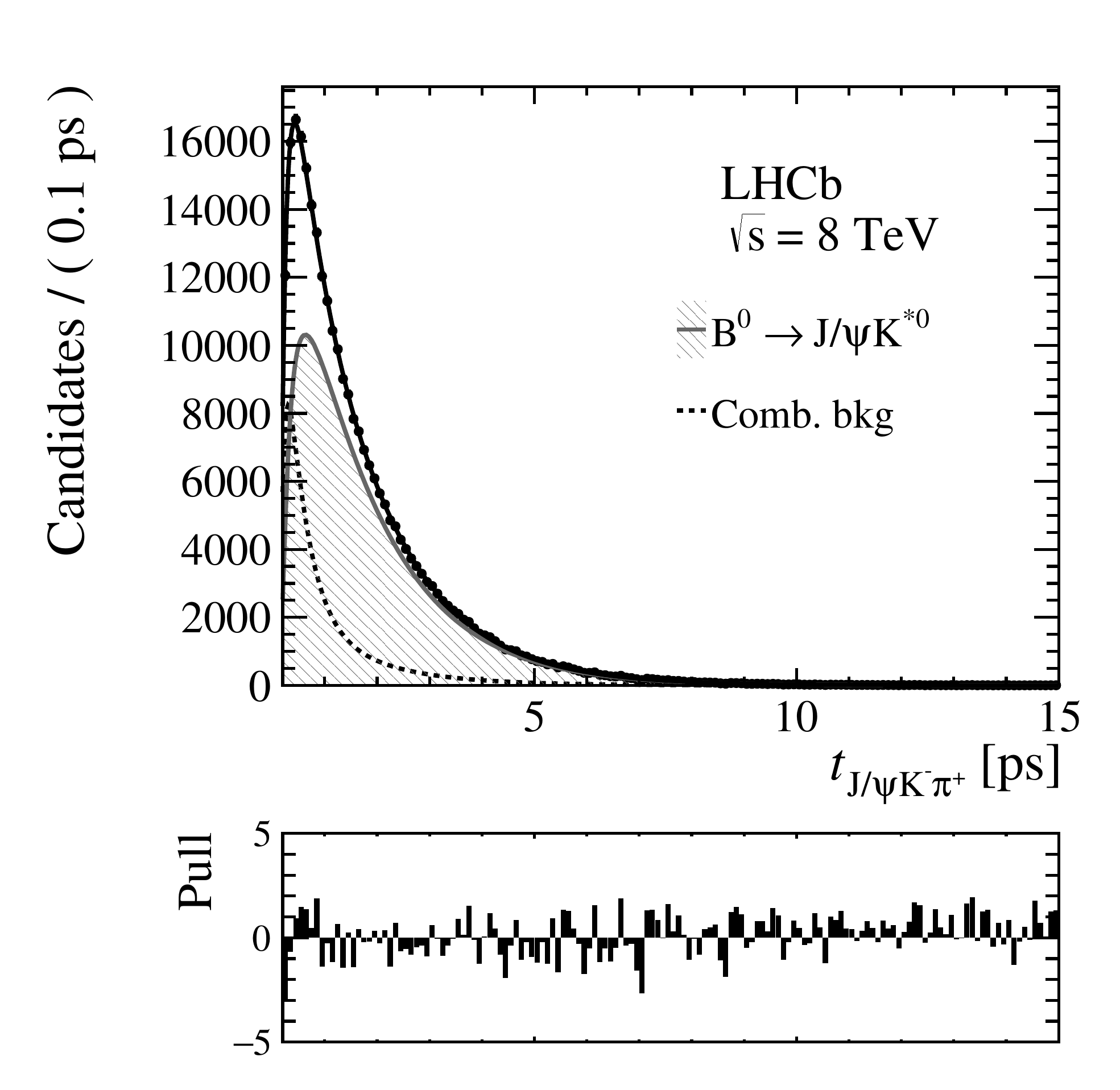}
   \includegraphics[width=0.42\linewidth]{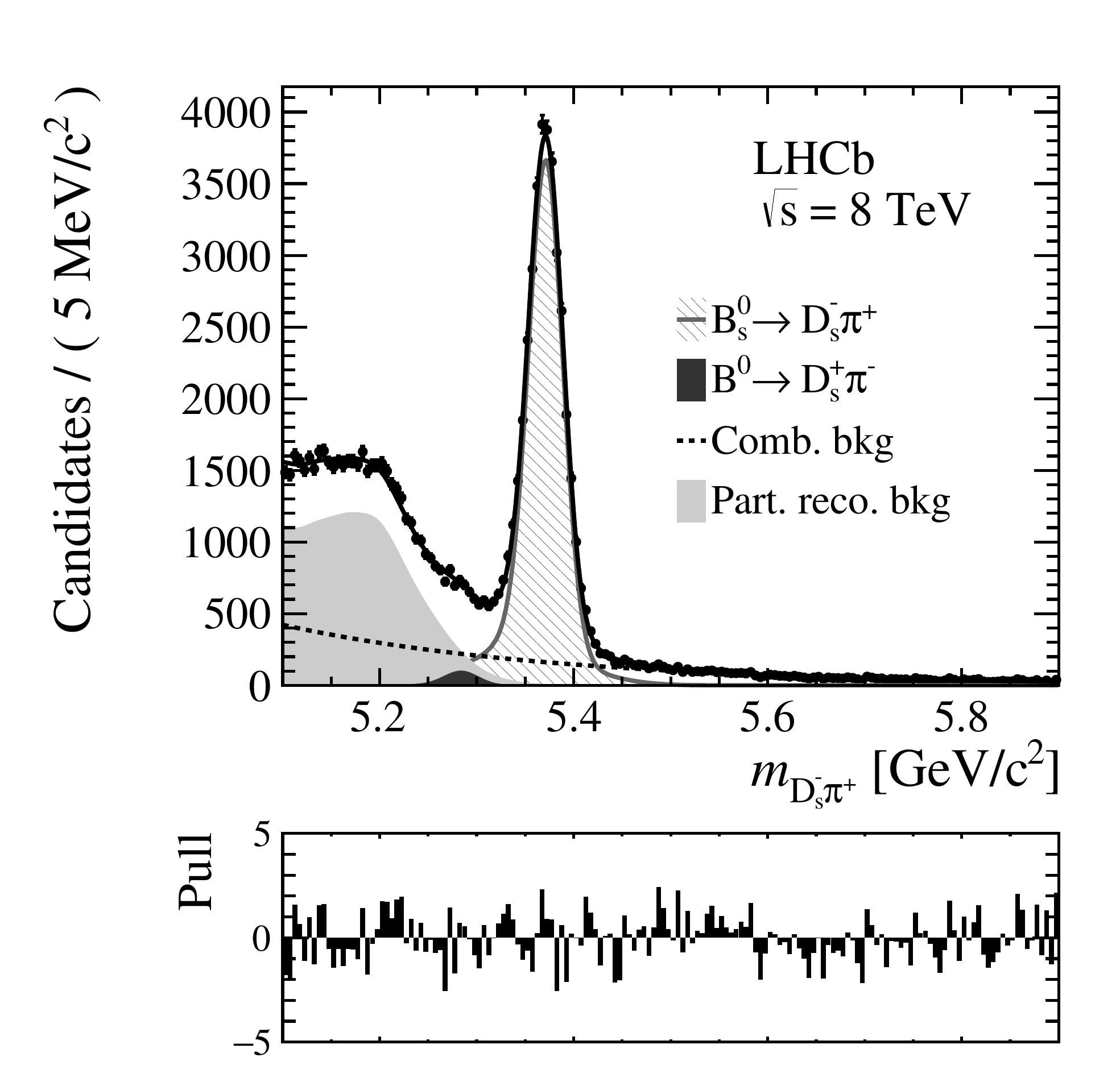}
   \includegraphics[width=0.42\linewidth]{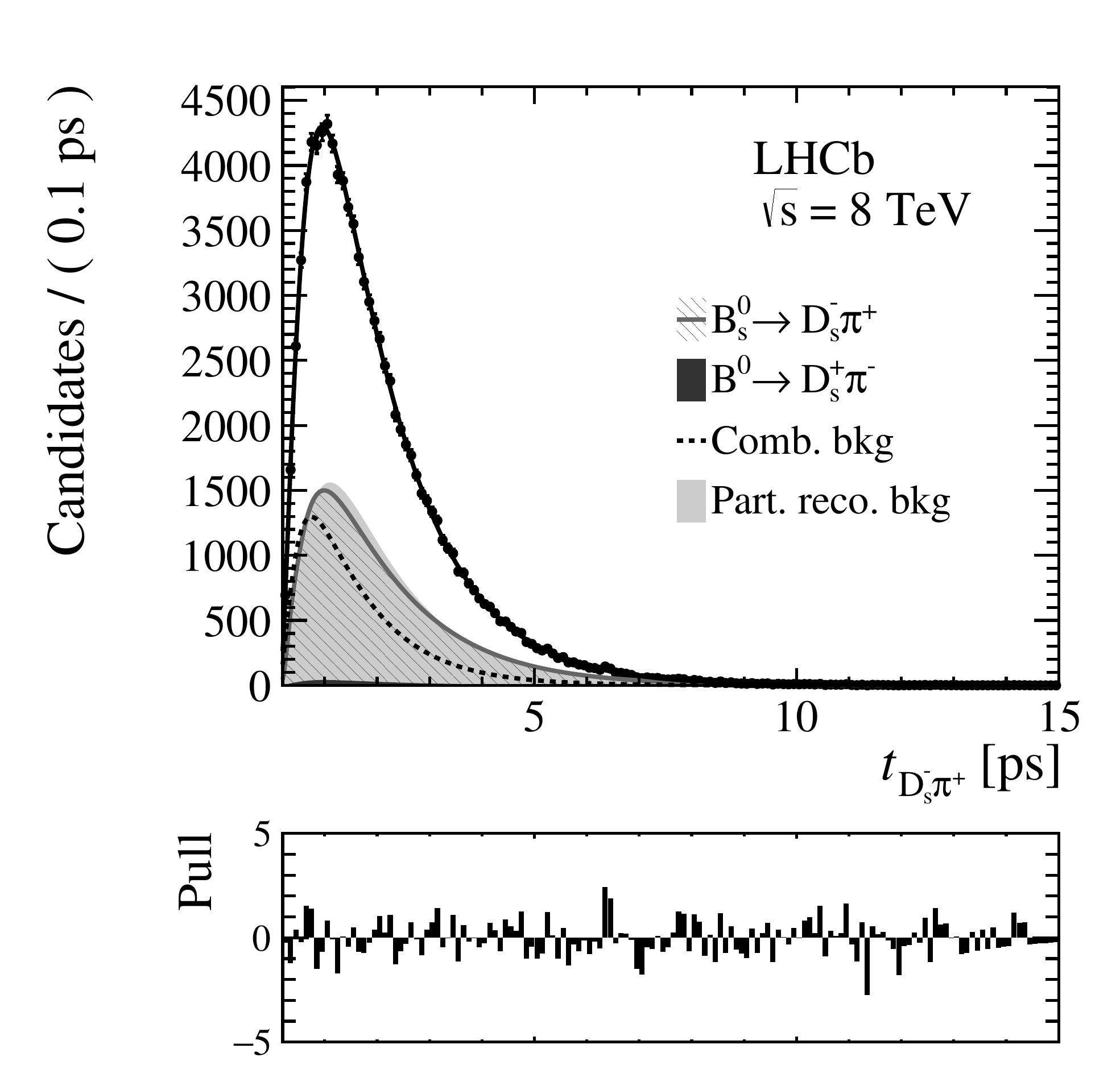}
   \includegraphics[width=0.42\textwidth]{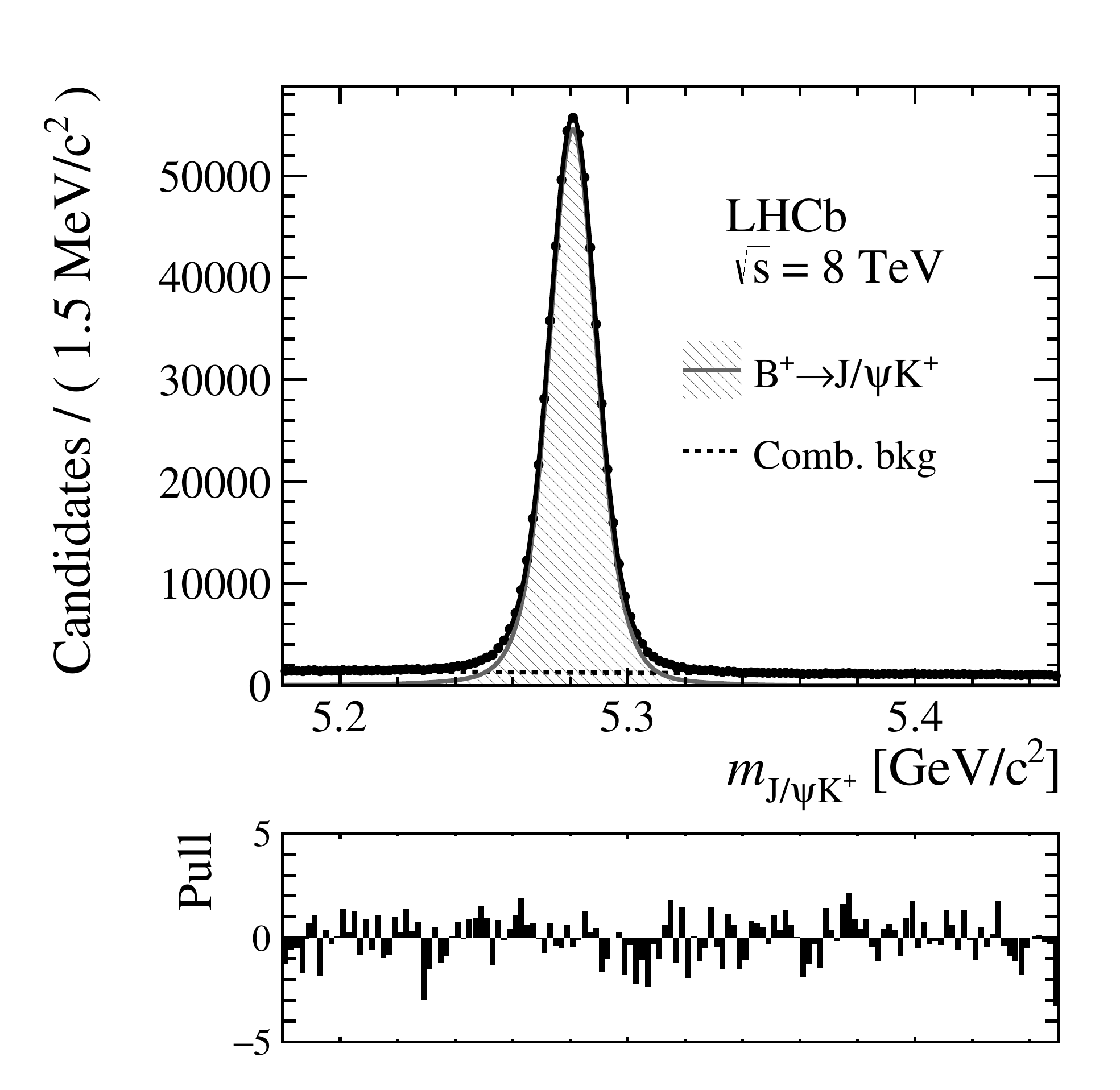}
 \end{center}
  \caption{Distributions of (left) invariant mass and (right) decay time for (top) $\Bz\to \jpsi\Kstarz$, (middle) $\Bs\to\Dsm\pip$ and (bottom) of invariant mass for $\Bp\to \jpsi\Kp$ decays, with the results of the fit overlaid. The data were collected in proton-proton collisions at the centre-of-mass energy of 8\tev. The contributions of the various background sources are also shown. Below each plot are the normalized residual distributions.}
  \label{fig:globalfits_2012}
\end{figure}

As already mentioned, for small values of $A_{\CP}$ and $A_{\rm D}$, the decay rate is to first order only sensitive to the sum of these two quantities. For this reason, $A_{\CP}$ is fixed to zero and $A_{\rm D}$ is left as a free parameter in the fits and hence measured from data, oppositely to the \Bp case, where an external input is necessary for $A_{\rm D}$. It is verified that the choice of different $A_{\CP}$ values, up to the few percent level, leads to negligible variations of $A_{\rm P}$.
Figures~\ref{fig:globalfits_2011} and~\ref{fig:globalfits_2012} show the $\jpsi\Kp$, $\jpsi\Kp\pim$ and $\Dsm\pip$ invariant mass distributions and, in the case of the neutral \B meson, the time distributions with the results of the global fits overlaid, for data recorded at centre-of-mass energies of 7 and 8\tev.

\begin{figure}[t]
\begin{center}
    \includegraphics[width=0.47\linewidth]{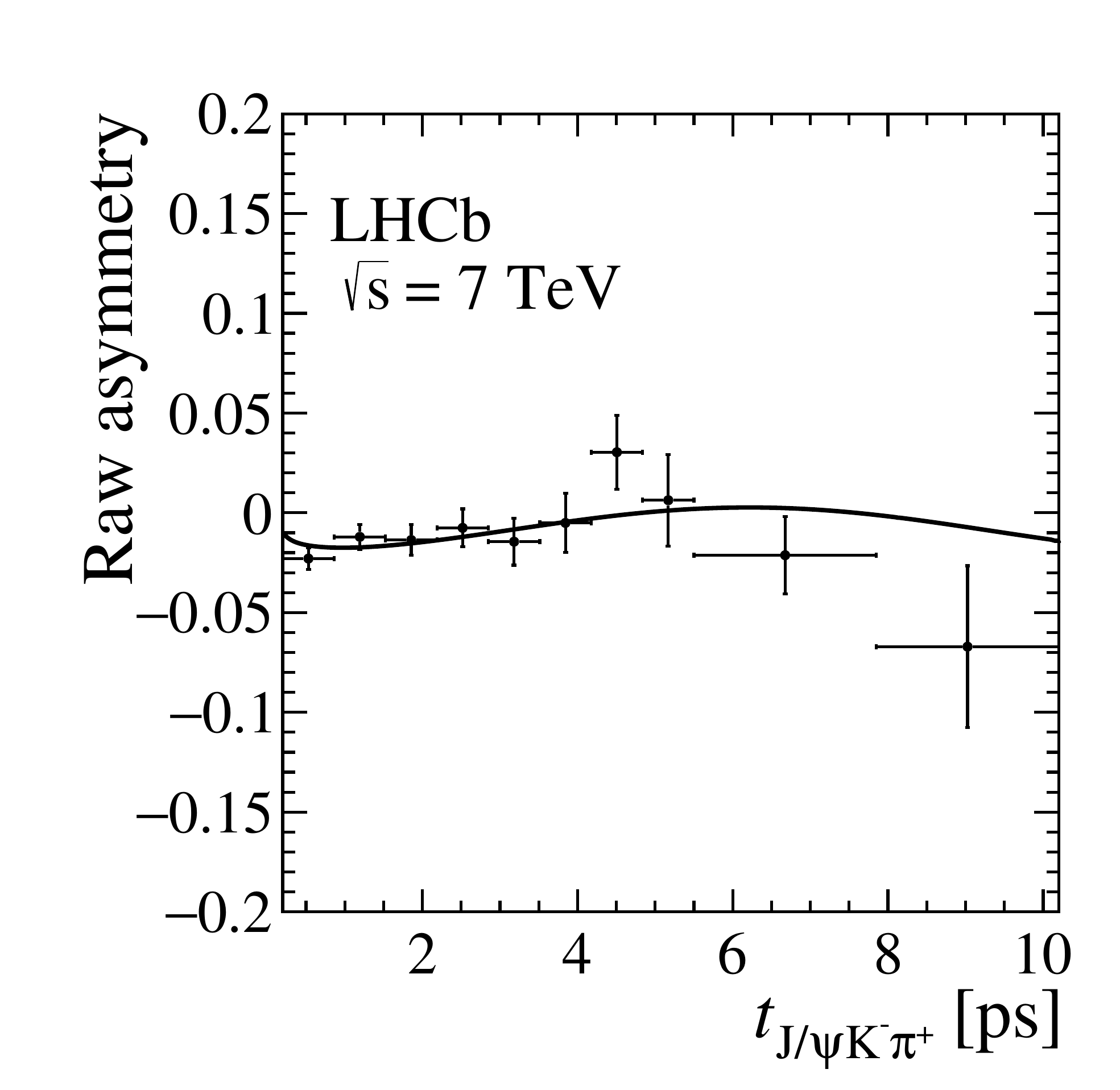}
    \includegraphics[width=0.47\linewidth]{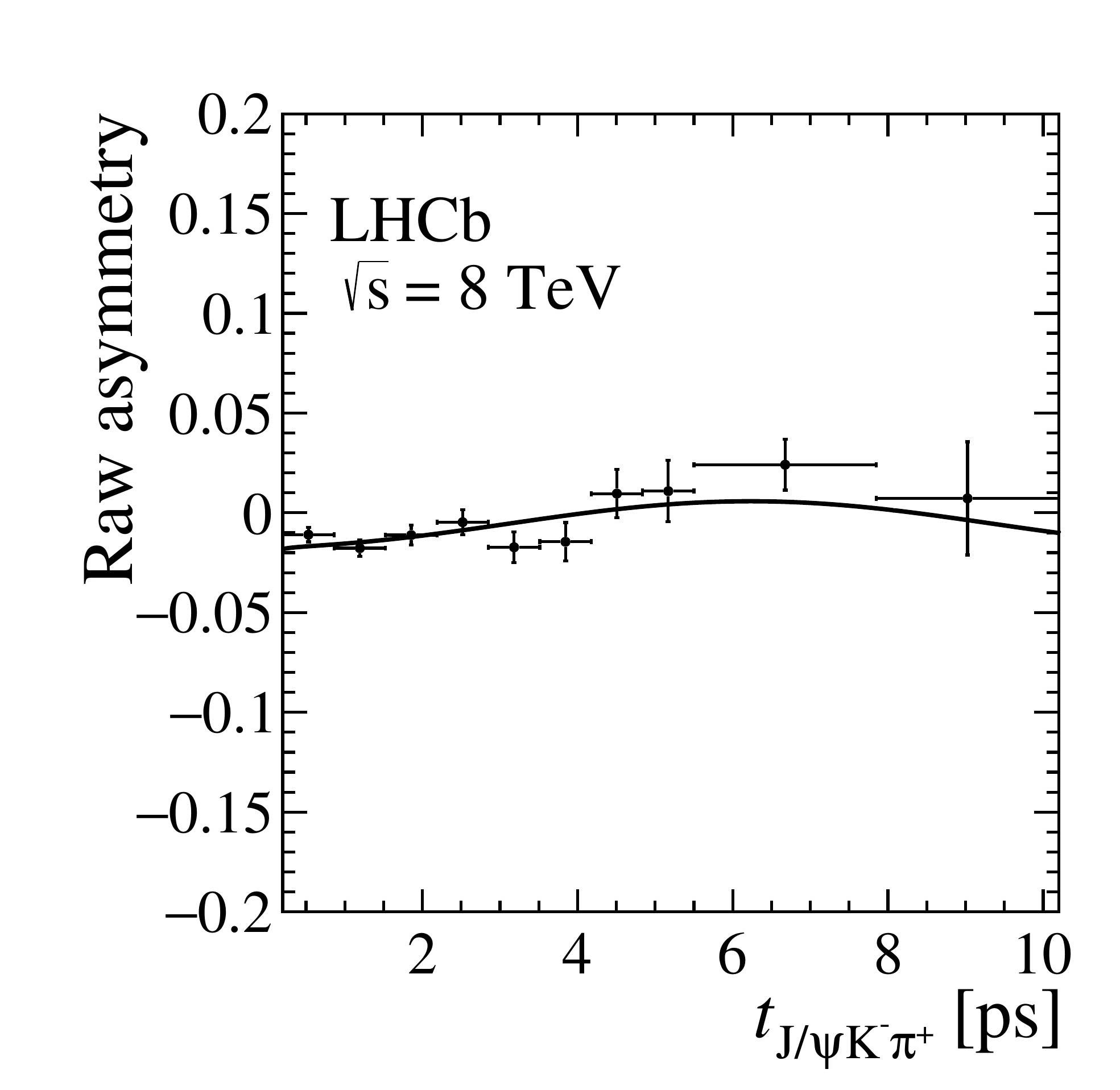}
    \includegraphics[width=0.47\linewidth]{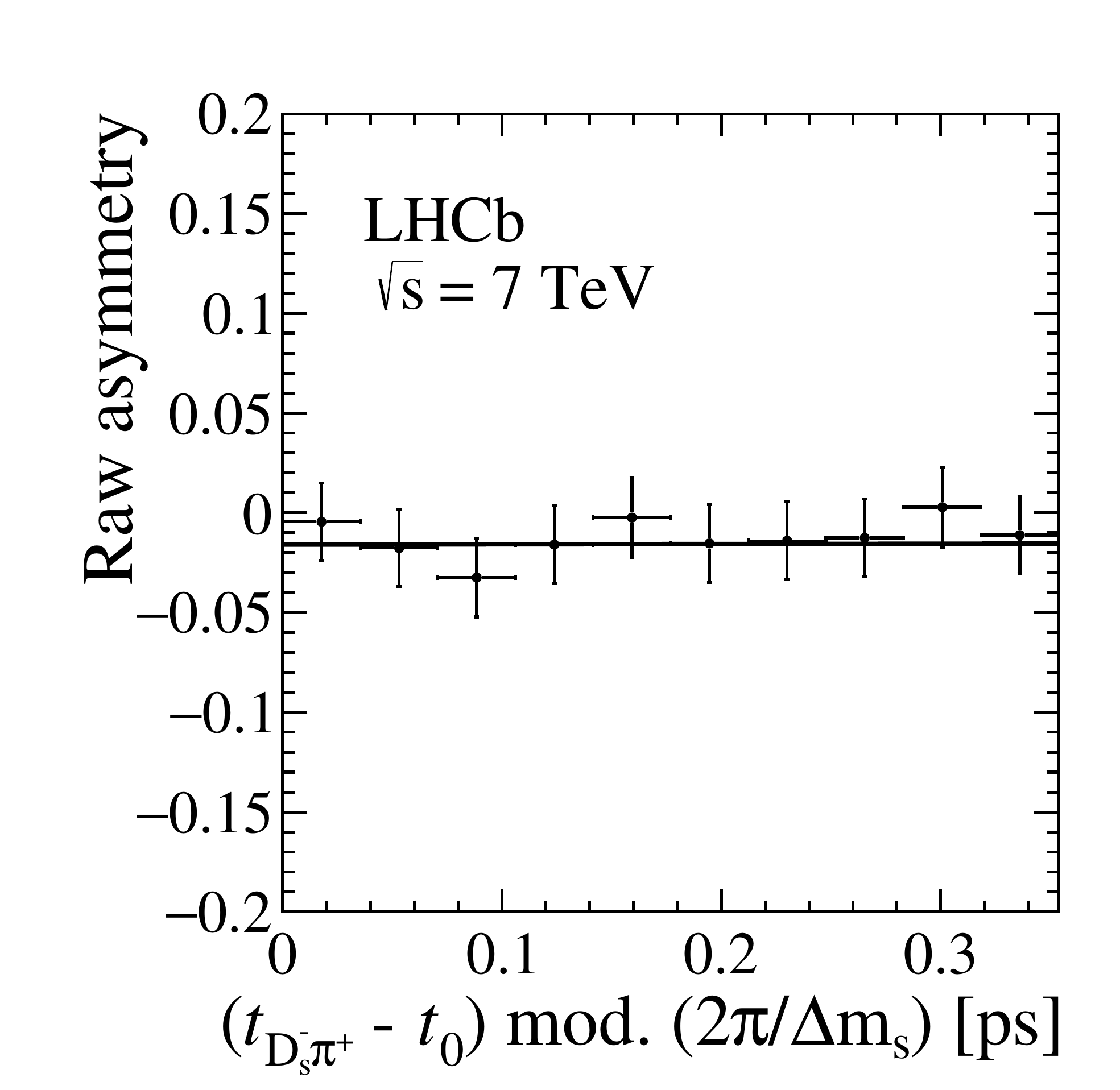}
    \includegraphics[width=0.47\linewidth]{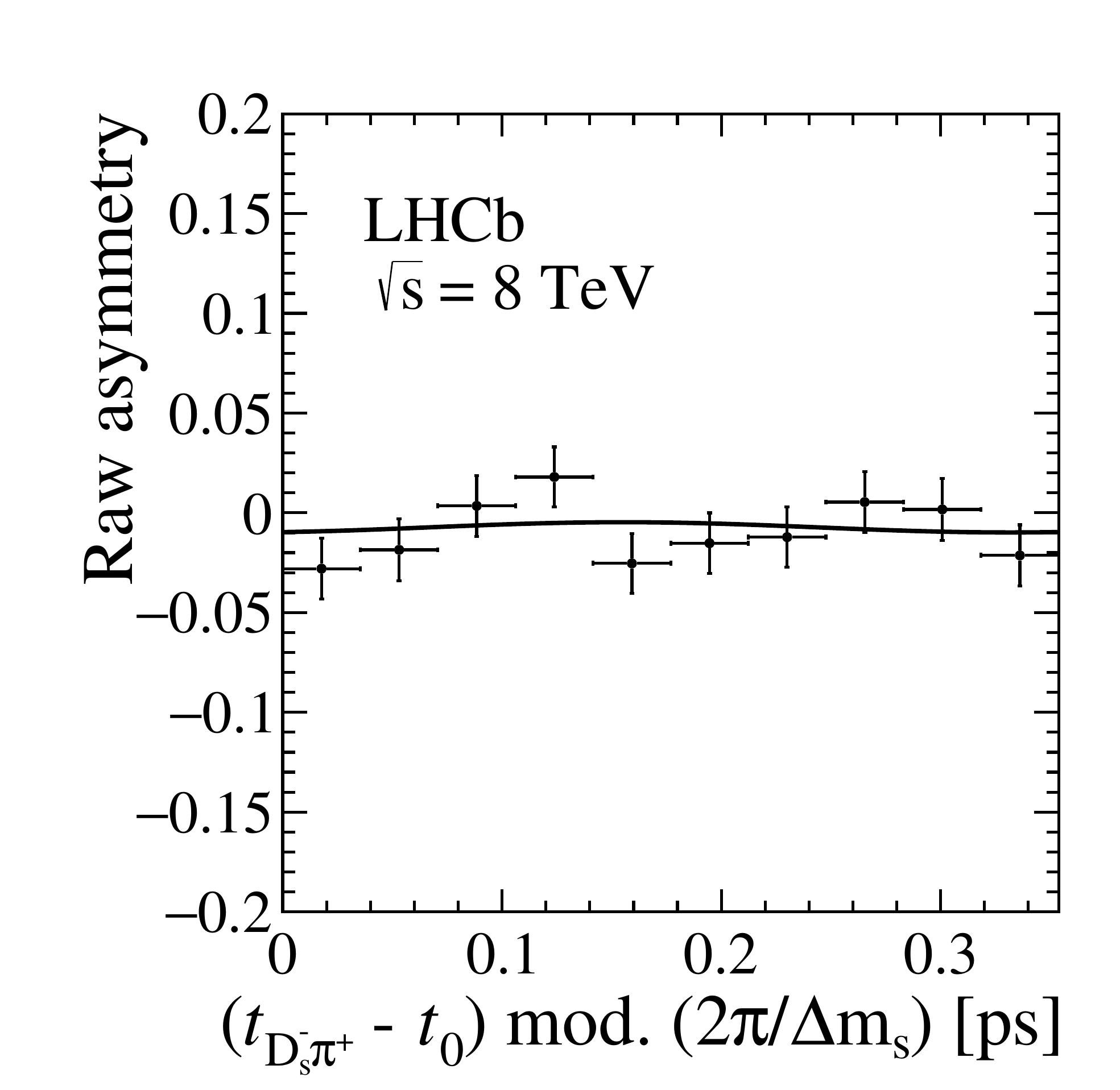}
  \end{center}
  \caption{Time-dependent raw asymmetries for candidates in the (top) $\Bz\to \jpsi\Kstarz$ and (bottom) $\Bs\to\Dsm\pip$ signal mass regions with the results of the global fits overlaid. Left and right plots correspond to data recorded in proton-proton collisions at centre-of-mass energies of  7 and  8\tev, respectively. For the \Bs decay, the asymmetries are obtained by folding the decay-time distributions into one oscillation period, and the offset $t_{0}= 0.2$~ps corresponds to the selection requirement on the decay time.}
  \label{fig:asymmetries}
\end{figure}

Figure~\ref{fig:asymmetries} shows the raw asymmetries for neutral \B-meson decays, defined as the ratio between the difference and the sum of the overall decay-time distributions, as a function of the decay time for candidates in the signal invariant mass regions, defined as the ranges $5250\text{--}5310$\mevcc for \Bz decay and $5290\text{--}5450$\mevcc for \Bs decay. The results of the global fits are overlaid.

The signal yields, $A_{\rm P}$ values and detection asymmetries obtained from the global fits are reported in Table~\ref{tab:signalyieldsB0}, for the neutral \B-meson decays, while the signal yield and $A_{\rm raw}$ for the \Bp decay are reported in Table~\ref{tab:signalyieldsBp}. The $A_{\rm P}$ values obtained from the time-dependent global fits, reported here for illustrative purposes, are detector-independent quantities only if efficiency corrections as a function of $\pt$ and $y$ are applied.
An accurate knowledge of the decay-time resolution is important for the $\Bs\to\Dsm\pip$ decay, due to the fast oscillation of the \Bs meson. For this reason the decay-time resolution is determined using the method previously described, applied to candidates in each (\pt, $y$) bin.

\begin{table}[!ht]
\begin{center}
\caption{Values of signal yields, $A_{\rm P}$ and $A_{\rm D}$ obtained from global fits for the two neutral \B-meson decays under study.}
 \label{tab:signalyieldsB0}
\resizebox{\columnwidth}{!}{
\begin{tabular}{c|c|c|c|c}
Parameter & \multicolumn{2}{c}{\sqs = 7\tev} & \multicolumn{2}{|c}{\sqs = 8\tev}\\
\hline
       & $\Bz \to \jpsi \Kstarz$ & $\Bs \to \Dsm \pip$ & $\Bz \to \jpsi \Kstarz$ & $\Bs \to \Dsm \pip$\\
$N_{\rm sig}$ &   $95\,122 \pm 369\,\,$ &  $16\,932 \pm 174\,\,$ &   $221\,973 \pm 569\,\,\,\,\,$ &  $36\,726 \pm 250\,\,$\\
$A_{\rm P}$   & $-0.0113 \pm 0.0063$ &  $-0.0001 \pm 0.0166$ &  $-0.0109 \pm 0.0042$ &  $\,\,\,\,\,0.0081 \pm 0.0111$\\
$A_{\rm D}$   &  $-0.0098 \pm 0.0046$ & $-0.0143 \pm 0.0086$ &   $-0.0056 \pm 0.0030$ &  $-0.0103\pm 0.0058$  \\
\end{tabular}
}
\end{center}
\end{table}

\begin{table}[!ht]
\begin{center}
\caption{Values of signal yields and raw asymmetries obtained from global fits in the case of the $\Bp\to\jpsi\Kp$ decay.}
 \label{tab:signalyieldsBp}
\begin{tabular}{c|c|c}
Parameter        &  \sqs = 7\tev & \sqs = 8\tev \\
\hline
$N_{\rm sig}$ &   $265\,574 \pm 576\,\,\,\,\,$ & $619\,800 \pm 908\,\,\,\,\,$ \\
$A_{\rm raw}$   & $-0.017 \pm 0.002$ & $-0.014 \pm 0.001$\ \\
\end{tabular}
\end{center}
\end{table}

According to Eq.~\ref{eq:prodasym}, the measurement of $A_{\rm P}(\Bp)$ requires knowledge of the \CP asymmetry $A_{\CP}(\Bp\to\jpsi\Kp)$ and $A_{\rm D}(\Kp)$. The value recently measured by LHCb~\cite{LHCb-PAPER-2016-054} with an independent data set is used for the former and corresponds to $A_{\CP}(\Bp\to\jpsi\Kp) = \left(0.09 \pm 0.27\, \stat \pm 0.07\, \syst\right) \times 10^{-2}$.
The measurement of the kaon detection asymmetry is obtained from $\D$-meson decays produced directly in proton-proton collisions, using the same technique reported in Ref.~\cite{LHCb-PAPER-2014-013}. It consists of measuring raw asymmetries from the two decay modes, $\Dp \to \Km\pip\pip$ and $\Dp\to\KS\pip$ with $\KS \to\pip\pim$, to obtain the \Kp\pim detection asymmetry, $A_{\rm D}(\Kp\pim)$, in each ($\pt$, $y$) bin of the \Bp mesons. 
Additionally, $A_{\rm D}(\Kp)$ is obtained by subtracting from $A_{\rm D}(\Kp\pim)$ the pion detection asymmetry, $A_{\rm D}(\pim)$,
measured by means of a sample of partially and fully reconstructed $D^{*+} \to \Dz(\Km \pip \pim \pip)\pip $ decays, as described in Ref.~\cite{LHCb-PAPER-2012-009}. It is estimated that the pion detection asymmetries across the various \Bp meson bins of ($\pt$, $y$) are in the range $0\text{--}0.2$ \%. Finally, the detection asymmetry of the $\Kz \to \pip\pim$ final state, $A_{\mathrm{D}}(\Kz)$, measured by LHCb to be $A_{\mathrm{D}}(\Kz) = (0.054 \pm 0.014) \%$~\cite{LHCb-PAPER-2014-013}, is also subtracted.

\section{Systematic uncertainties}

Several sources of systematic uncertainty are considered. They are evaluated for each kinematic bin and for each decay mode.
For the invariant mass model, the effects of the uncertainty on the shapes of all components (signals, combinatorial and partially reconstructed backgrounds) are investigated. For the decay-time model, systematic effects related to the decay-time resolution and reconstruction efficiency are studied. The effects of the uncertainties on the external inputs used in the fits, reported in Table~\ref{tab:input}, are evaluated by repeating the fits with each parameter varied by $\pm 1$ standard deviation ($\sigma$). Alternative decay-time parameterizations of the background components are also considered. To estimate the contribution of each single source, the fit is repeated for each (\pt, $y$) bin after having modified the baseline fit model. The shifts from the relevant baseline values are taken as the systematic uncertainties. A detailed description follows. To estimate a systematic uncertainty related to the parameterization of final-state radiation effects on the signal mass distributions, the parameter $s$ of Eq.~\ref{eq:sigmodel} is varied by $\pm 1 \sigma$ of the corresponding value obtained from fits to simulated decays. A systematic uncertainty related to the invariant mass resolution model is estimated by repeating the fit using a simplified model with a single Gaussian function. The systematic uncertainty related to the parameterization of the mass distribution for the combinatorial background is investigated by replacing the exponential function with a linear function. Concerning the partially reconstructed background, a systematic uncertainty is assessed by repeating the fits while excluding the low invariant mass region, applying the requirement $m > 5330$\mevcc to $\Bs\to\Dsm\pip$ candidates. To estimate the uncertainty related to the parameterization of signal decay-time reconstruction efficiency, different functions are considered. Effects of inaccuracies in the knowledge of the decay-time resolution are estimated by rescaling the widths of the baseline model to obtain an average resolution width differing by $\pm 8$ fs. The impact of the small bias in the reconstructed decay time 
is assessed by introducing a corresponding bias of $\pm 2$ fs in the decay-time resolution model.
The determination of the systematic uncertainties related to the $|q/p|$ input values requires special treatment, as $A_{\rm P}$ is correlated with $|q/p|$. For this reason, any variation of $|q/p|$ produces the same shift of $A_{\rm P}$ in each of the kinematic bins. Such a correlation is taken into account when  integrating over \pt and $y$. The values of the systematic uncertainties related to the knowledge of $|q/p|$ are 0.0009 in the case of $A_{\rm P}(\Bz)$ and 0.0021 in the case of $A_{\rm P}(\Bs)$. 
For the \Bp decay, the uncertainties on $A_{\CP}(\Bp\to\jpsi\Kp)$ and $A_{\rm D}(\Kp)$ are considered as systematic uncertainties. They introduce correlations among the bins that are considered when the integrated results are calculated.

The \Lb production asymmetry is calculated, in each kinematic bin, assuming that the number of produced hadrons of any species in the $i$-th bin containing
a $\bquark$ quark, $N_{i,\,\bquark}$, is equal to the number of produced hadrons containing a $\bquarkbar$ quark, $N_{i,\,\bquarkbar}$, \ie relying on Eq.~\ref{eq:APLBrelation_final}. This assumption is strictly valid in the full phase space, but not necessarily in a specific bin. In the event that $N_{i,\,\bquark} \ne N_{i,\,\bquarkbar}$, $A_{\rm P}(\Lb)$  is biased by the quantity
\begin{equation}
\delta z_i = \frac{N_{i,\,\bquark} - N_{i,\,\bquarkbar}}{N_{i,\,\bquark} + N_{i,\,\bquarkbar}} \cdot \frac{1}{f_{\Lb}}. \nonumber
\end{equation}
Values for $\delta z_i$ are studied using simulated events. Systematic uncertainties on $A_{\rm P}(\Lb)$, in each kinematic bin, are assigned as half of the maximum variation from zero of the quantities $\delta z_i \pm \sigma(\delta z_i)$, where $\sigma(\delta z_i)$ is the related uncertainty. 

The term $(f_c/ f_{\Lb}) \cdot A_{\rm P}(\Bc)$, estimated to be $3 \cdot 10^{-5}$, can be safely neglected, while a systematic uncertainty related to neglecting the term $(f_{\rm other}/f_{\Lb}) \cdot A_{\rm P}(\rm other)$ has to be assessed. Amongst all other $\bquark$ baryons, the production rate of $\Xi_b$ baryons is estimated from the simulation (which well reproduces the \Bp, \Bz, \Bs and \Lb fragmentation fractions) to be dominant, corresponding to about 1\% of all $b$-hadron species produced in the primary collisions. On this basis, the neglected term can be evaluated as
\begin{equation}
\frac{f_{\mathrm{other}}}{f_{\Lb}} A_{\mathrm{P}}(\mathrm{other}) \simeq \frac{f_{\Xi_b}}{f_{\Lb}}A_{\rm P}(\Xi_b). \nonumber
\end{equation}
The value of $A_{\rm P}(\Xi_b)$ is found to be double that of $A_{\rm P}(\Lb)$ in the simulation. A systematic uncertainty on $A_{\rm P}(\Lb)$ is obtained by assuming $A_{\rm P}(\Xi_b) =  2\,A_{\rm P}(\Lb)$.

\begin{table}[H]
 \begin{center}
  \caption{Absolute values of systematic uncertainties for integrated production asymmetries. 
The total systematic uncertainties are obtained by summing the individual contributions in quadrature.}
 \label{tab:systemIntegral}
   \begin{tabular}{l|c|c|c|c|}
\multicolumn{1}{c}{}& \multicolumn{4}{c}{Uncertainty [\sqs = 7\ensuremath{\mathrm{\,Te\kern -0.1em V}}]}  \\
\cline{2-5}
Source & $A_{\rm P}(\Bp)$ & $A_{\rm P}(\Bz)$ & $A_{\rm P}(\Bs)$ & $A_{\rm P}(\Lb)$\\
\hline
Signal mass shape                                                                     & 0.0016   & 0.0005 & 0.0036 &  0.0024 \\
Decay-time bias                                                                       & 0.0000   & 0.0000      & 0.0008 &  0.0004  \\
\dmd, \dms                                                                            & 0.0000        & 0.0001 & 0.0014 &  0.0007  \\
Decay-time resolution                                                                 & 0.0000        & 0.0000      & 0.0026 &  0.0014  \\
Final-state radiation                                                                 & 0.0000        & 0.0001 & 0.0000      &  0.0001      \\
Decay-time reconstruction efficiency                                                  & 0.0000        & 0.0001 & 0.0000      &  0.0001   \\
Combinatorial background mass shape                                                   & 0.0003   & 0.0000      & 0.0004 &  0.0003      \\
Partially reconstructed background mass shape                                         & 0.0000        & 0.0000      & 0.0029 &  0.0015       \\
\DGs                                                                                  & 0.0000        & 0.0000      & 0.0000      &  0.0000         \\
$A_{\rm D}(\Kp)$                                                                      & 0.0018   & 0.0000      & 0.0000      &  0.0013           \\
$|q/p|_{\Bz}$, $|q/p|_{\Bs}$                                                                     & 0.0000        & 0.0009 & 0.0021 &  0.0013        \\
Uncertainties from fragmentation fractions                                            & 0.0000        & 0.0000      & 0.0000      &  0.0058 \\
Difference between $\omega_i$ or $\omega_i^{\rm data}$                                & 0.0003   & 0.0003 & 0.0003 &  0.0003   \\
Neglecting term with $A_{\rm P}(\Xi_{b})$ in Eq.~\ref{eq:APLBrelation_final}          &  0.0000       &  0.0000     & 0.0000      &  0.0071 \\
Validity of $N_{\bquark} = N_{\bquarkbar}$ in each bin                                &  0.0000       &  0.0000     & 0.0000      &  0.0032 \\
$A_{\CP}(\Bp\to\jpsi\Kp)$                                                             & 0.0028   &  0.0000     & 0.0000      &  0.0028  \\
$A_{\rm D}(\Kzb)$                                                                     & 0.0001   &  0.0000     & 0.0000      &  0.0002 \\
\hline
Total systematic uncertainty                                                          & 0.0037   & 0.0011 & 0.0059& 0.0108  \\
\cline{2-5}
\multicolumn{5}{c}{}  \\
\multicolumn{1}{c}{} & \multicolumn{4}{c}{Uncertainty [\sqs = 8\ensuremath{\mathrm{\,Te\kern -0.1em V}}]}  \\
\cline{2-5}
Source & $A_{\rm P}(\Bp)$ & $A_{\rm P}(\Bz)$ & $A_{\rm P}(\Bs)$ & $A_{\rm P}(\Lb)$\\
\hline
Signal mass shape                                                                     & 0.0006   & 0.0004 & 0.0035 & 0.0021 \\
Decay-time bias                                                                       & 0.0000        & 0.0000      & 0.0008 & 0.0004\\
\dmd, \dms                                                                             & 0.0000      & 0.0001 & 0.0015 & 0.0008 \\
Decay-time resolution                                                                 & 0.0000        & 0.0000      & 0.0028 &  0.0016 \\
Final-state radiation                                                                 & 0.0000        & 0.0001 & 0.0001 & 0.0001 \\
Decay-time reconstruction efficiency                                                  & 0.0000        & 0.0001 & 0.0001 & 0.0001  \\
Combinatorial background mass shape                                                   & 0.0002   & 0.0000      & 0.0004 & 0.0003 \\
Partially reconstructed background mass shape                                         & 0.0000        & 0.0000      & 0.0027 & 0.0015 \\
\DGs                                                                                  & 0.0000        & 0.0000      & 0.0001 & 0.0001 \\
$A_{\rm D}(\Kp)$                                                                      & 0.0014   & 0.0000      &  0.0000     & 0.0011\\
$|q/p|_{\Bz}$, $|q/p|_{\Bs}$                                                                     & 0.0000        & 0.0009 & 0.0021 &0.0014           \\
Uncertainties from fragmentation fractions                                            & 0.0000        & 0.0000      & 0.0000      & 0.0025 \\
Difference between $\omega_i$ or $\omega_i^{\rm data}$                                & 0.0002  & 0.0003  & 0.0003 & 0.0003   \\
Neglecting term with $A_{\rm P}(\Xi_{b})$ in Eq.~\ref{eq:APLBrelation_final}          &  0.0000      &  0.0000      & 0.0000      & 0.0046\\
Validity of $N_{\bquark} = N_{\bquarkbar}$ in each bin                                &  0.0000      &  0.0000      & 0.0000      & 0.0033 \\
$A_{\CP}(\Bp\to\jpsi\Kp)$                                                             & 0.0028  &  0.0000      & 0.0000      & 0.0027  \\
$A_{\rm D}(\Kzb)$                                                                     & 0.0001  &  0.0000      & 0.0000      & 0.0002 \\
\hline
Total systematic uncertainty                                                          &  0.0032 & 0.0010 & 0.0059  & 0.0076 \\
\cline{2-5}
\end{tabular}
 \end{center}
\end{table}

The dominant systematic uncertainties for the \Bp and \Bz cases are related to the measured value of $A_{\CP}(\Bu\to\jpsi\Kp)$ and to $|q/p|_{\Bz}$, respectively. The systematic uncertainty associated with the signal mass shape is the main source for the $\Bs$ case, while it is the one related to neglecting the term $f_{\rm other}/f_{\Lb} \cdot A_{\rm P}(\rm other)$ in Eq.~\ref{eq:APLBrelation} in the case of the $\Lb$ decay.
All the systematic uncertainties are summed in quadrature for each kinematic bin. Their values are reported, together with the final measurements, in Tables~\ref{tab:resultsBpB02011}--\ref{tab:resultsBsLb2012} in the Appendix.

When the integrated results are calculated, all the systematic uncertainties estimated for each bin are propagated according to Eq.~\ref{eq:APintegrated} and correlations among the bins are taken into account. An additional systematic uncertainty is considered by studying how the integrated values vary in the case that 
the values of $\omega_i$ are measured using a data driven approach. In this case $\omega_{i}^{\mathrm{data}}$ is measured as
\begin{equation}
\omega_{i}^{\mathrm{data}} = \frac{N_i}{\varepsilon_{i}^{\mathrm{total}}} / \sum_i \frac{N_i}{\varepsilon_{i}^{
\mathrm{total}}} \nonumber
\end{equation}
where $\varepsilon_i^{\mathrm{total}}$ is the total reconstruction efficiency, obtained as a combination of the selection efficiency, determined from simulation, and PID and trigger efficiencies, measured from data. Differences in the central values between $A_{\mathrm{P}}$ calculated using either $\omega_i$ or $\omega_i^{\mathrm{data}}$ are found to be small for all the decay modes. Table~\ref{tab:systemIntegral} summarizes systematic uncertainties associated with the integrated measurements.  

\section{Results and conclusions}

Using a data sample corresponding to an integrated luminosity of 3.0\invfb, the \Bp, \Bz and \Bs hadron production asymmetries have been determined independently for each (\pt,$y$) bin and then combined using Eq.~\ref{eq:APLBrelation_final} to derive the \Lb production asymmetry. Tables~\ref{tab:resultsBpB02011}--\ref{tab:resultsBsLb2012}, in the Appendix, report the final results. 

The \Bp, \Bz, \Bs and \Lb hadron production asymmetries are also determined integrating over \pt or $y$, in the range $0 < \pt < 30 $\gevc and $2.1 < y < 4.5$ for \Bp and \Bz decays, and in the range $2 < \pt < 30 $\gevc and $2.1 < y < 4.5$ for \Bs and \Lb decay. The corresponding numerical values are reported in Tables~\ref{tab:AP_BpB0_2011_pt}--\ref{tab:AP_BsLb_2012_eta}, in the Appendix, and in Figs.~\ref{fig:AP_Bp}--\ref{fig:AP_Lb}, where the results of the fits with a constant and a first-order polynomial function are also shown. Table~\ref{tab:polfit} reports the values of the fit parameters.
No evidence for any dependence is observed.
Finally, integrating over both \pt and $y$, the \bquark-hadron production asymmetries are found to be
\begin{eqnarray}
A_{\rm P} (\Bp)_{\sqs = 7\tev}&=&-0.0023 \pm 0.0024\, \stat \pm 0.0037\, \syst,  \nonumber\\
A_{\rm P} (\Bp)_{\sqs = 8\tev}&=&-0.0074 \pm 0.0015\,\stat \pm 0.0032\, \syst,  \nonumber \\
A_\mathrm{P}(\Bz)_{\sqs = 7\tev}&=& \phantom{-}0.0044 \pm 0.0088 \,\stat \pm 0.0011 \,\syst, \nonumber \\
A_\mathrm{P}(\Bz)_{\sqs = 8\tev}&=&-0.0140 \pm 0.0055 \,\stat \pm 0.0010 \,\syst, \nonumber \\
A_\mathrm{P}(\Bs)_{\sqs = 7\tev}&=&-0.0065 \pm 0.0288 \,\stat \pm 0.0059 \,\syst, \nonumber \\
A_\mathrm{P}(\Bs)_{\sqs = 8\tev}&=&\phantom{-}0.0198 \pm 0.0190 \,\stat \pm 0.0059 \,\syst,\nonumber \\
A_\mathrm{P}(\Lb)_{\sqs = 7\tev}&=&-0.0011 \pm 0.0253\, \stat \pm 0.0108\, \syst,  \nonumber \\
A_\mathrm{P}(\Lb)_{\sqs = 8\tev}&=&\phantom{-}0.0344 \pm  0.0161\, \stat \pm  0.0076\, \syst. \nonumber
\end{eqnarray}
All the results are consistent with zero 
within $2.5$ standard deviations.
The results of this analysis supersede the previous LHCb results of Ref.~\cite{LHCb-PAPER-2014-042}.
These measurements, once integrated using appropriate weights for any reconstructed \Bp, \Bz, \Bz, \Lb decay in LHCb, can be used to determine effective production asymmetries,
as inputs for \CP violation measurements with the LHCb data.

\begin{figure}[H]
  \begin{center}
    \includegraphics[width=0.48\textwidth]{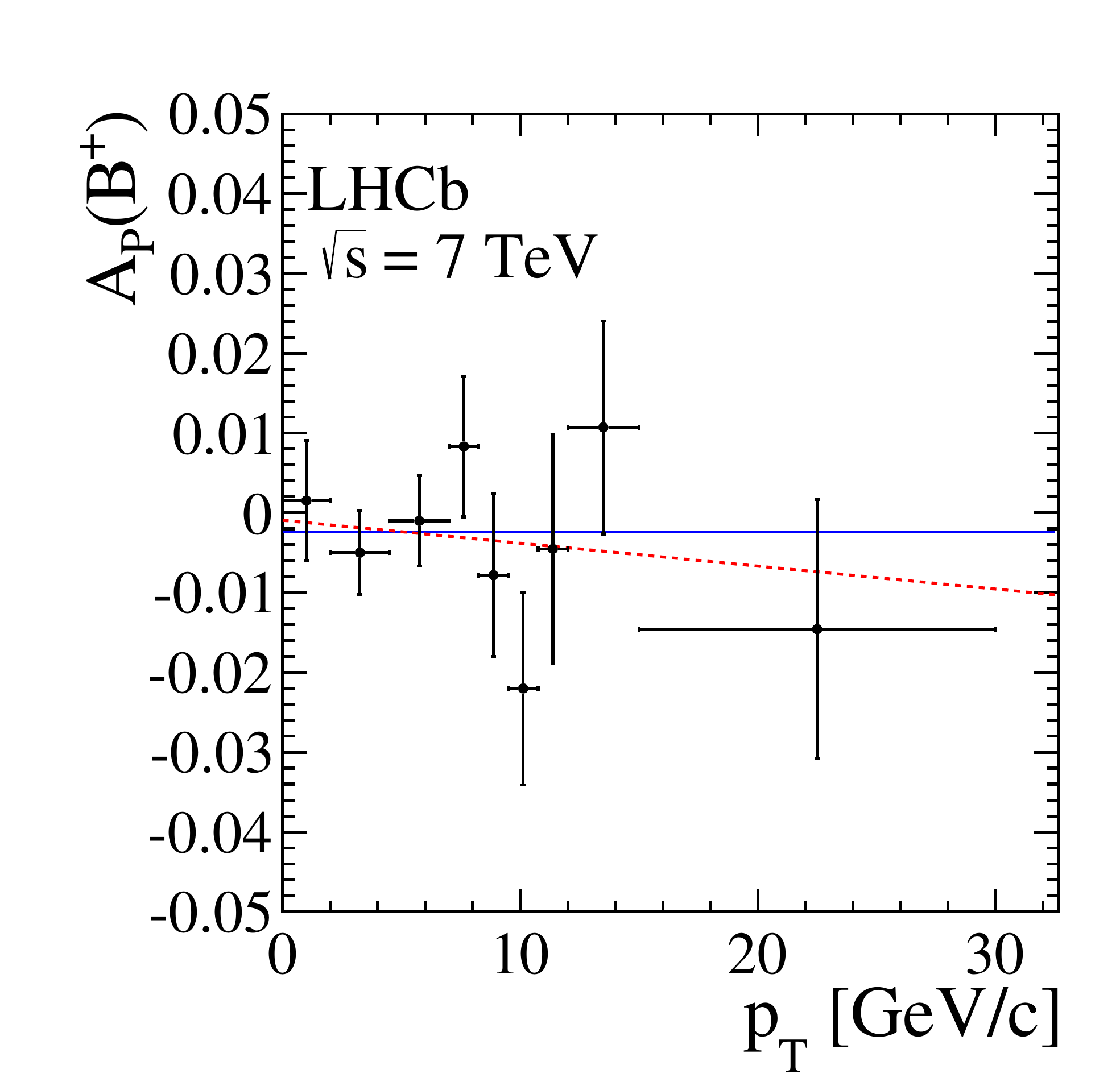}
    \includegraphics[width=0.48\textwidth]{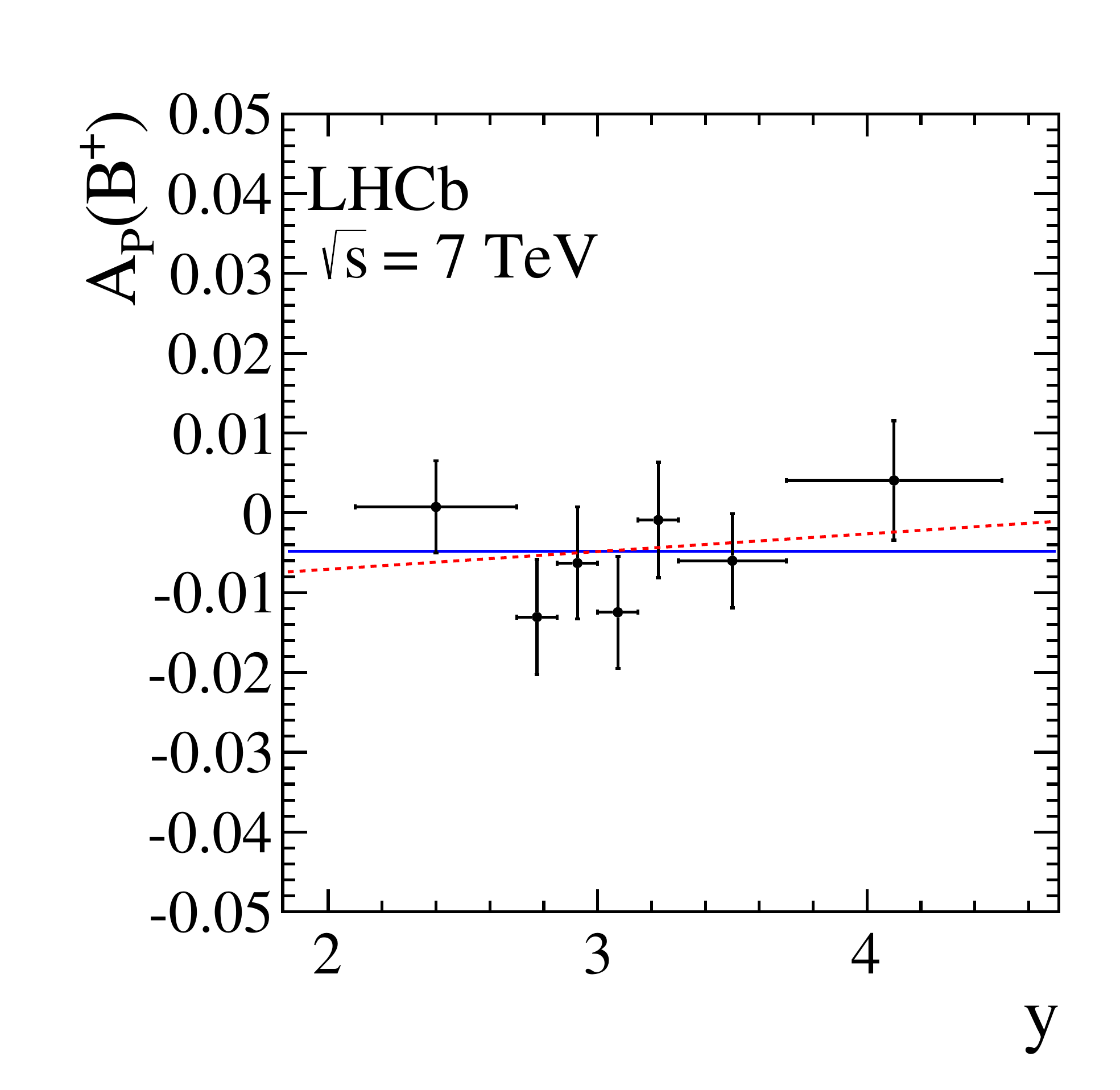}
    \includegraphics[width=0.48\textwidth]{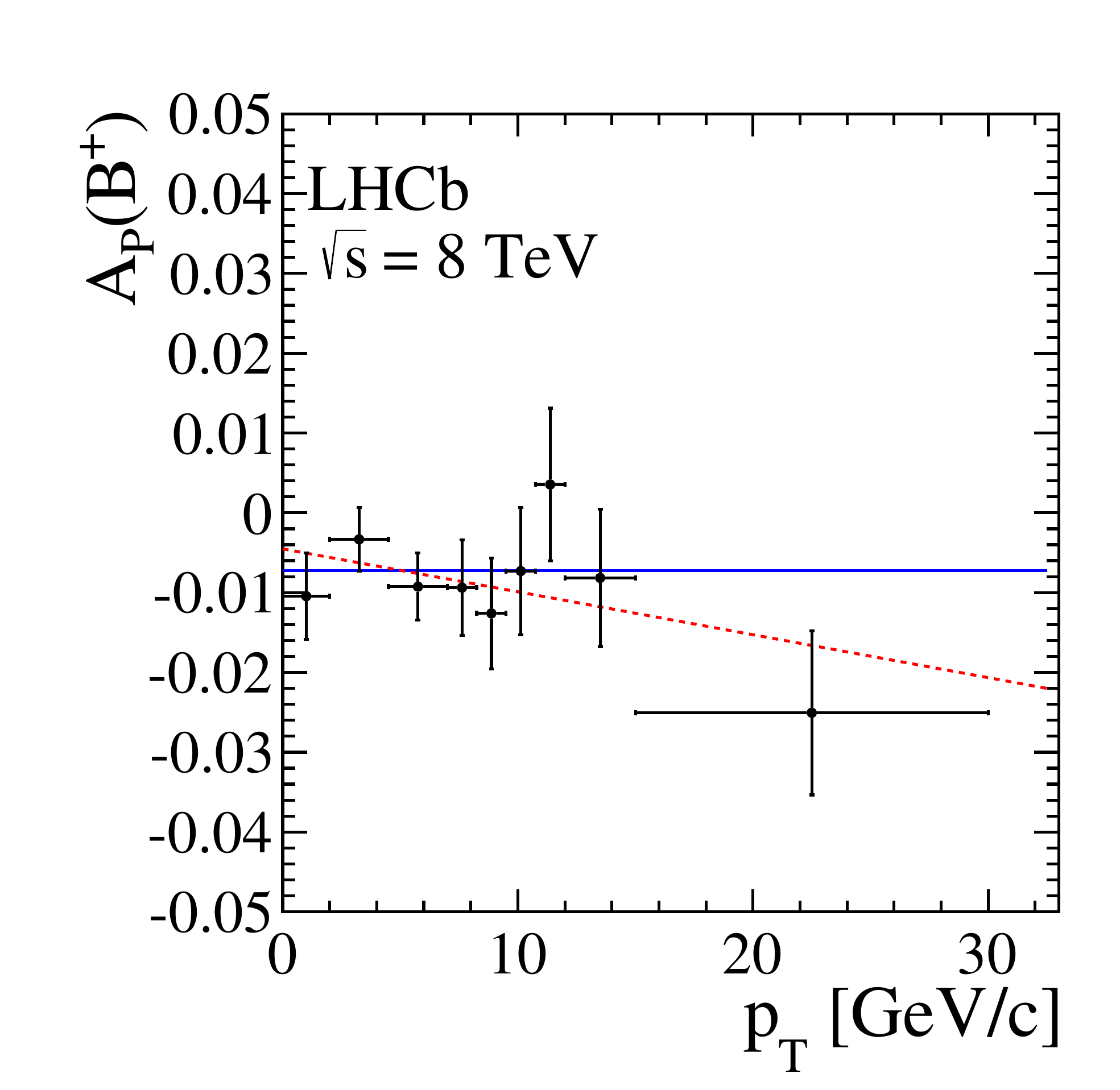}
    \includegraphics[width=0.48\textwidth]{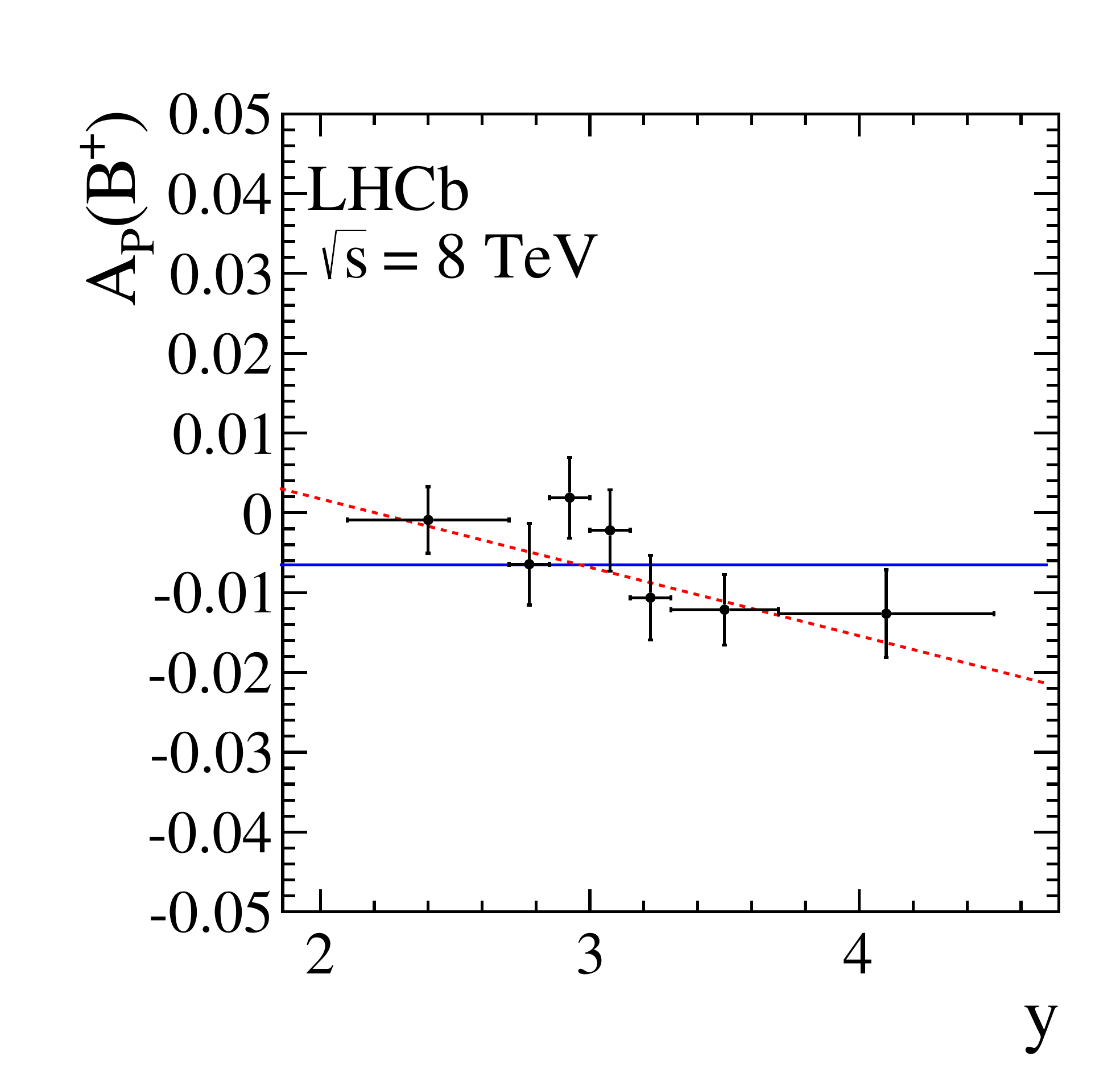}
 \end{center}
  \caption{Dependence of $A_\mathrm{P}(B^+)$, for data collected in proton-proton collisions with centre-of-mass of energies of (top) 7 and (bottom) 8 \tev, on (left) \pt and (right) $y$. The results of fits using a straight line with zero (solid line) or floating slope parameter (dashed line) are also shown. The fits take into account the correlations amongst the bins.}
  \label{fig:AP_Bp}
\end{figure}

 \begin{figure}[th!]
  \begin{center}
    \includegraphics[width=0.48\textwidth]{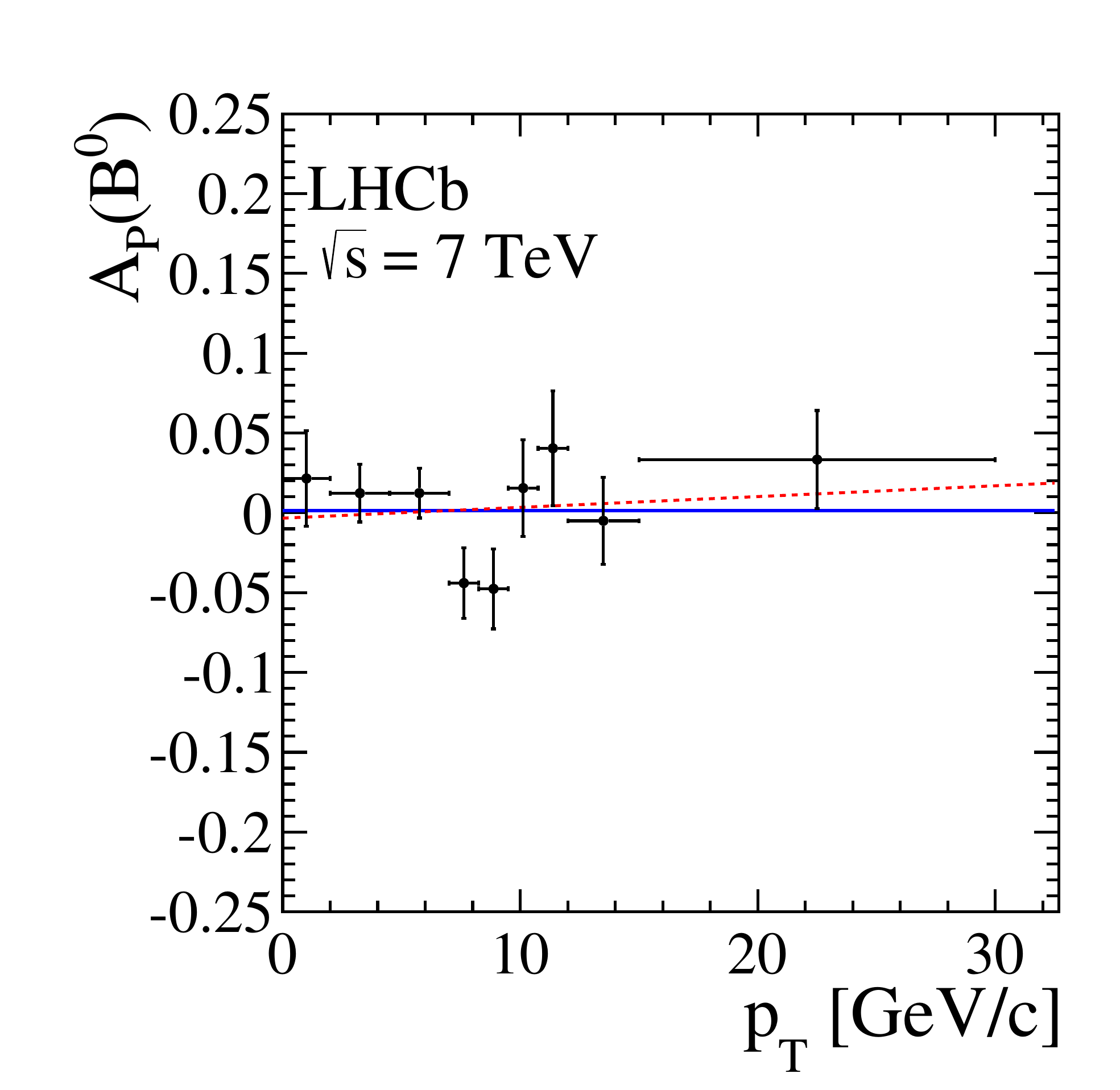}
    \includegraphics[width=0.48\textwidth]{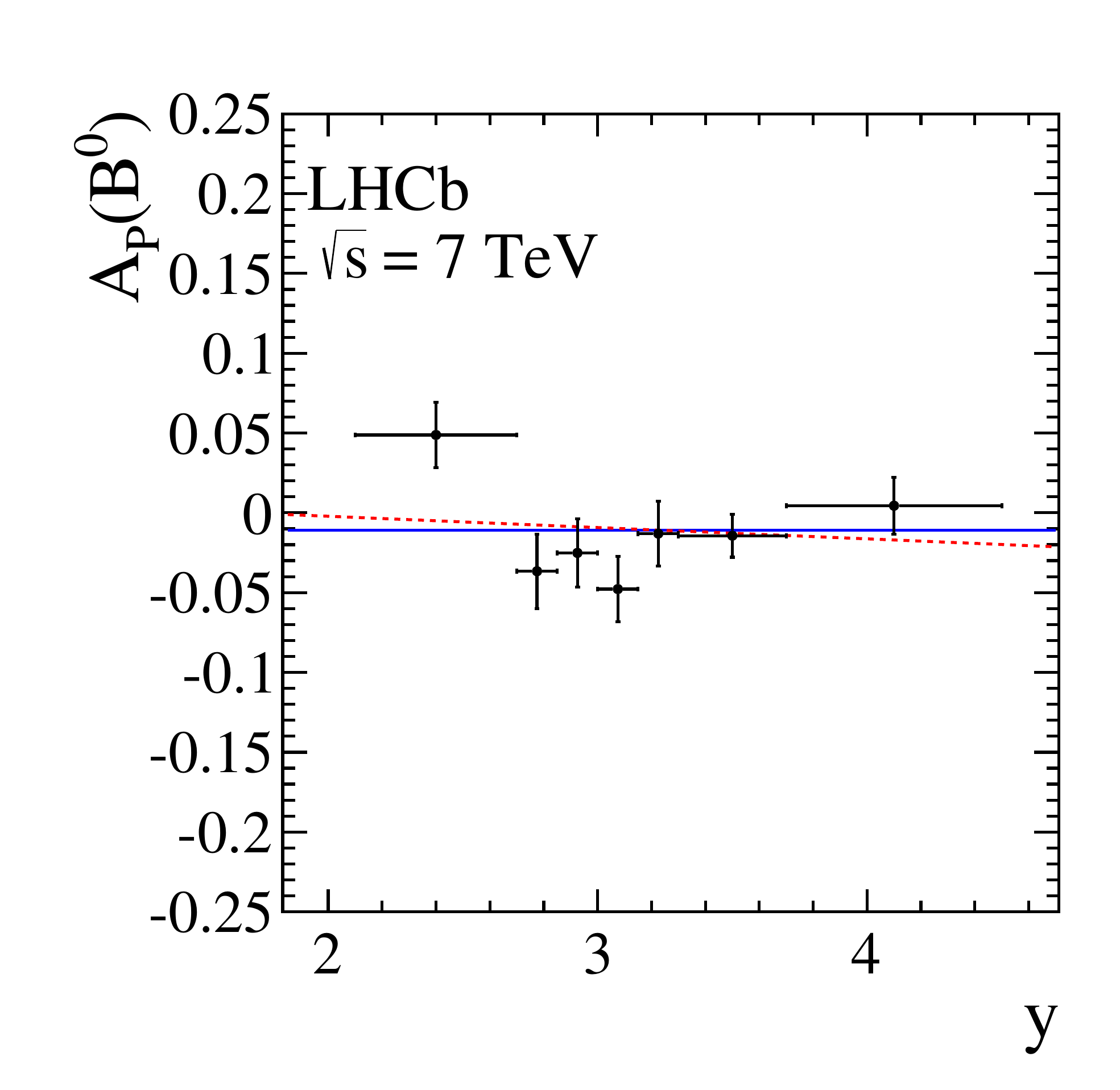}
    \includegraphics[width=0.48\textwidth]{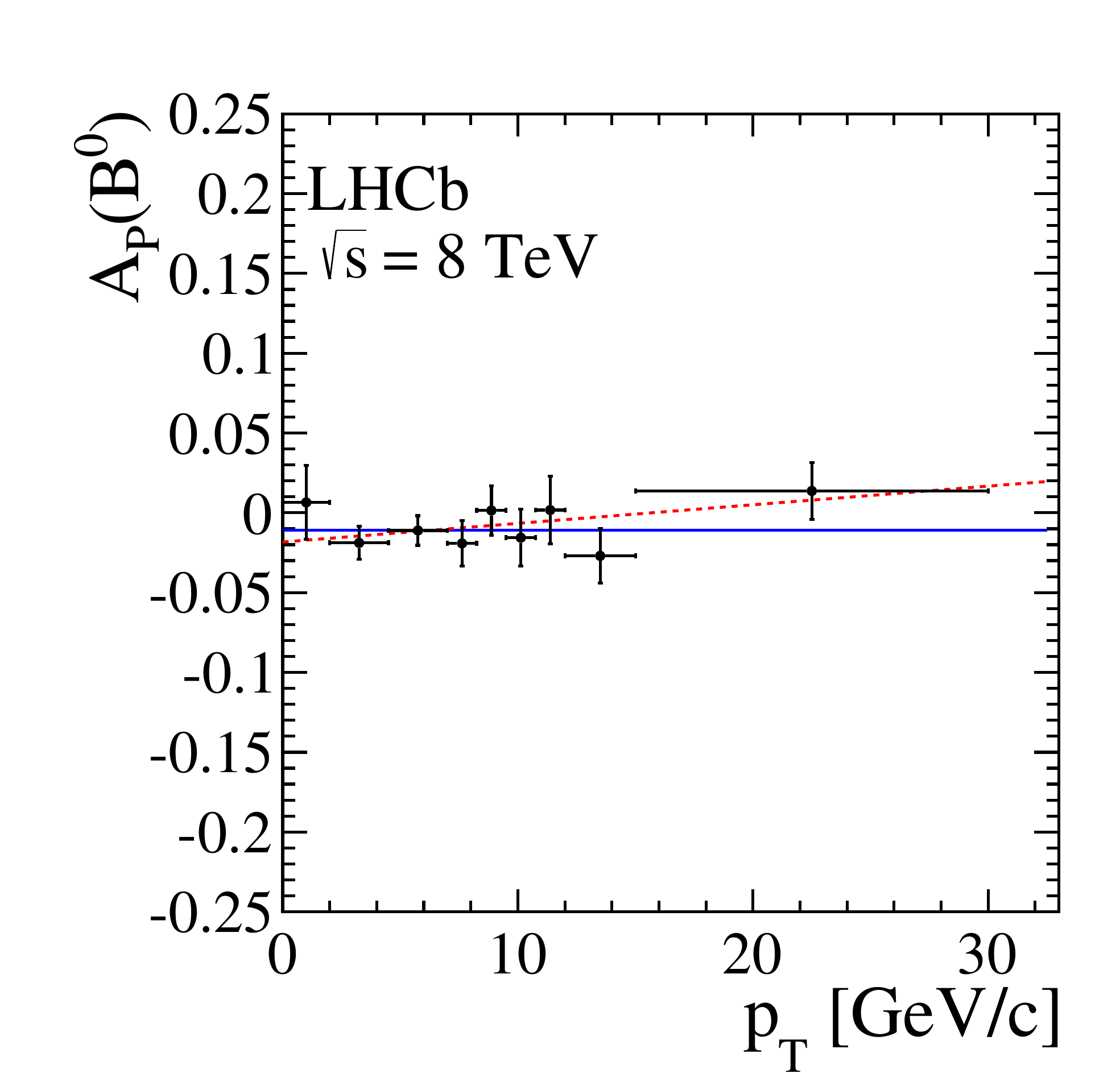}
    \includegraphics[width=0.48\textwidth]{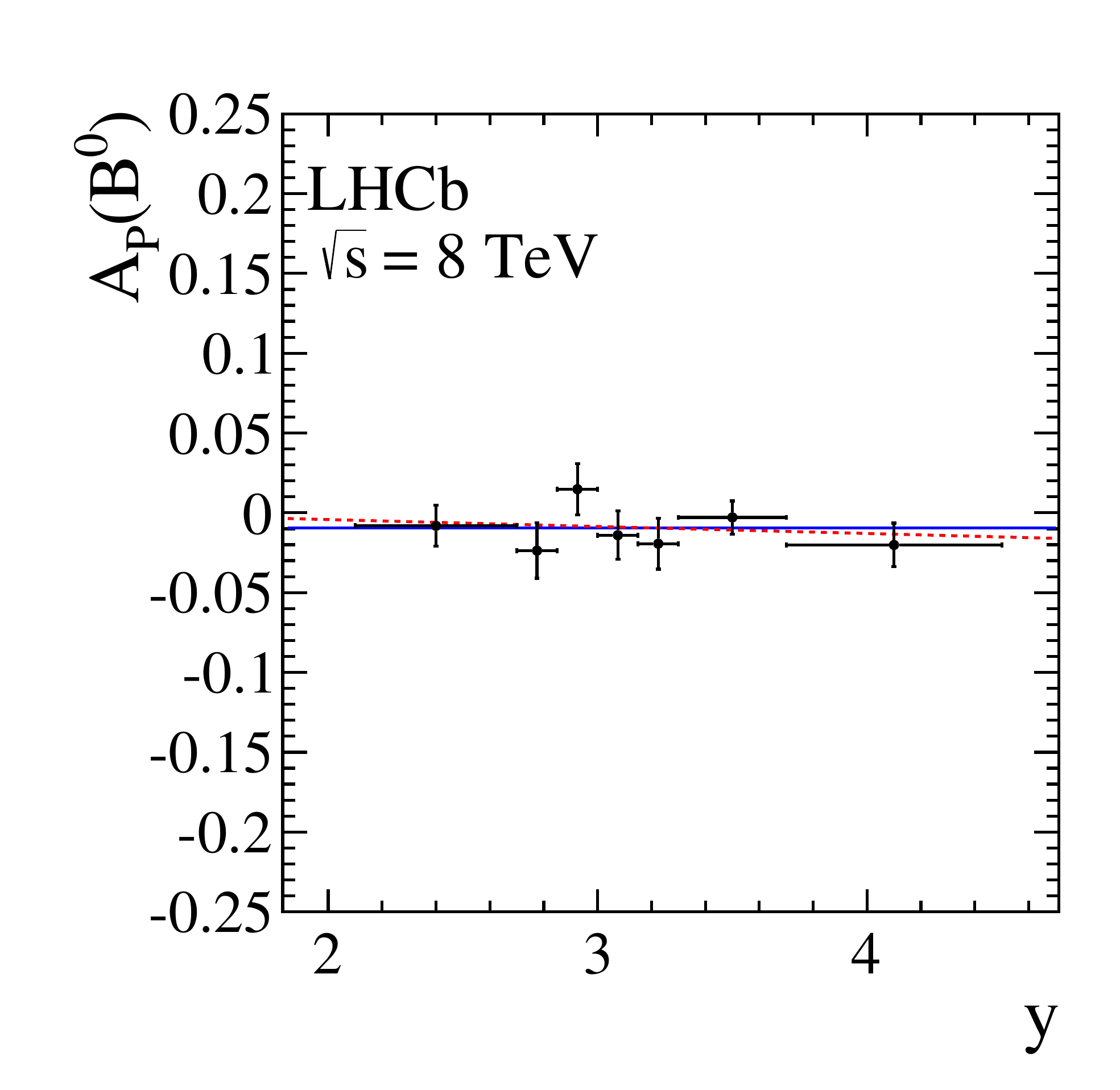}
 \end{center}
  \caption{Dependence of $A_\mathrm{P}(B^0)$, for data collected in proton-proton collisions with centre-of-mass of energies of (top) 7 and (bottom) 8 \tev, on (left) \pt and (right) $y$. The results of fits using a straight line with zero (solid line) or floating slope parameter (dashed line) are also shown. The fits take into account the correlations amongst the bins.}
  \label{fig:AP_B0}
\end{figure}

 \begin{figure}[th!]
  \begin{center}
    \includegraphics[width=0.48\textwidth]{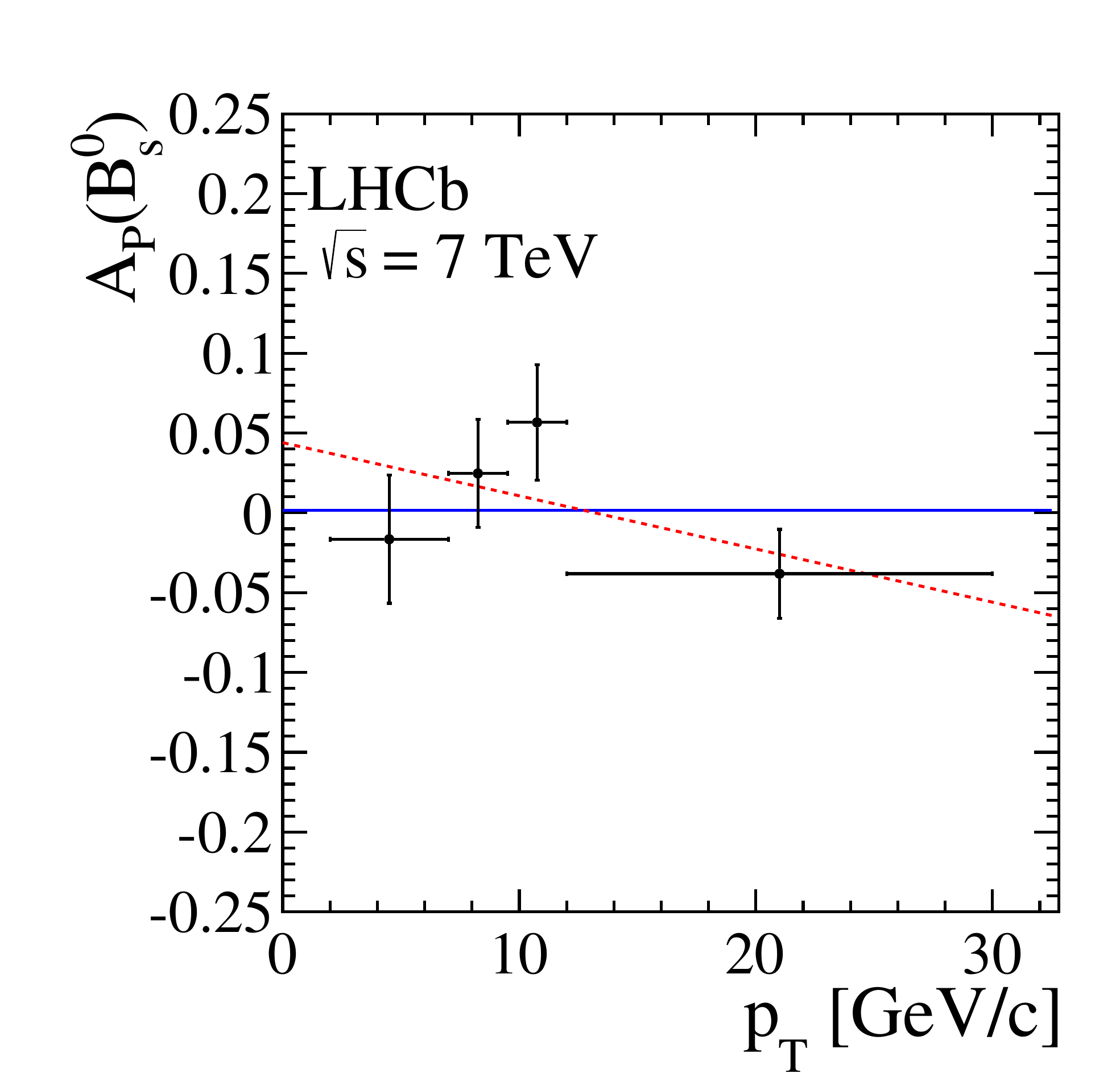}
    \includegraphics[width=0.48\textwidth]{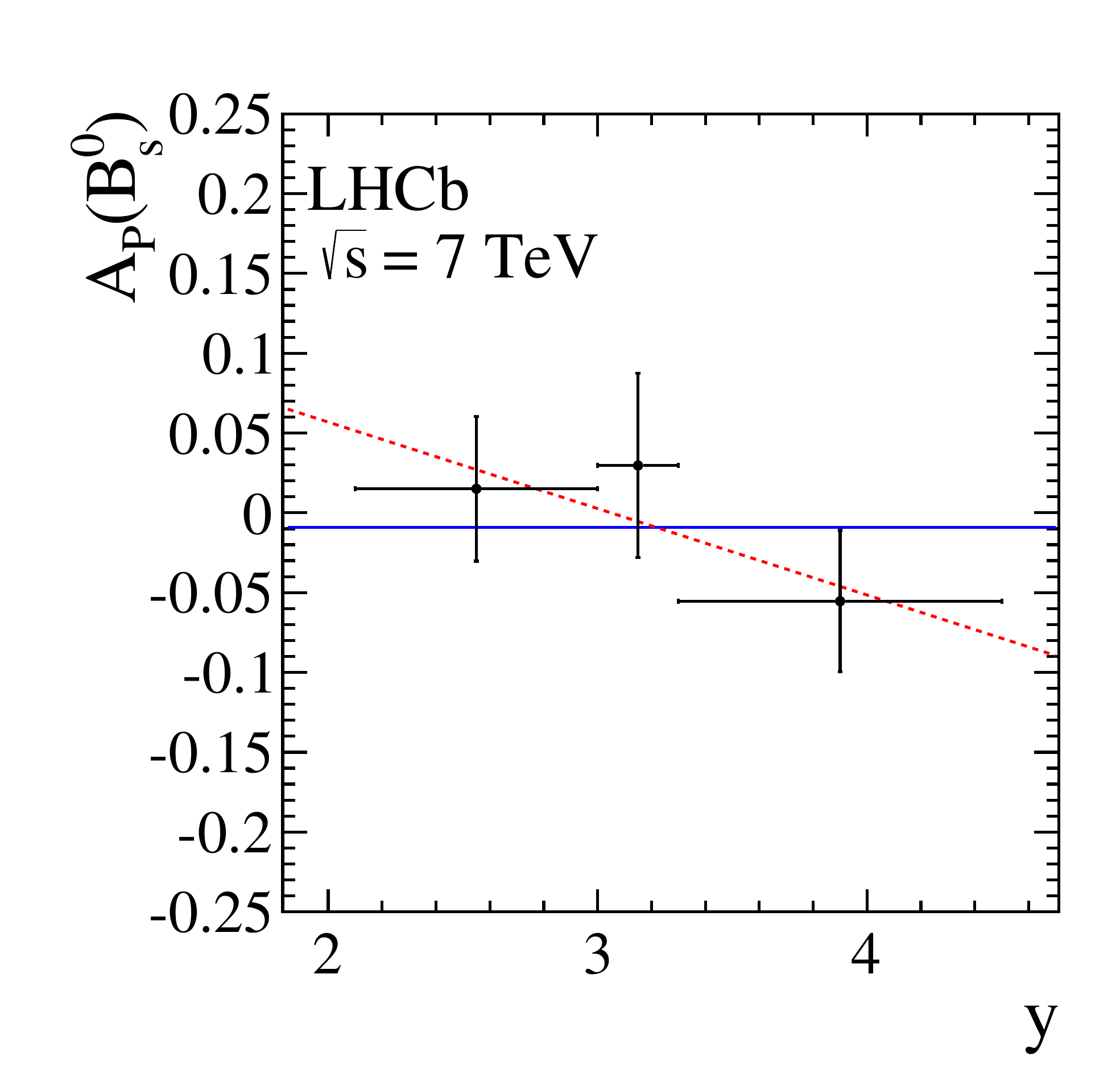}
    \includegraphics[width=0.48\textwidth]{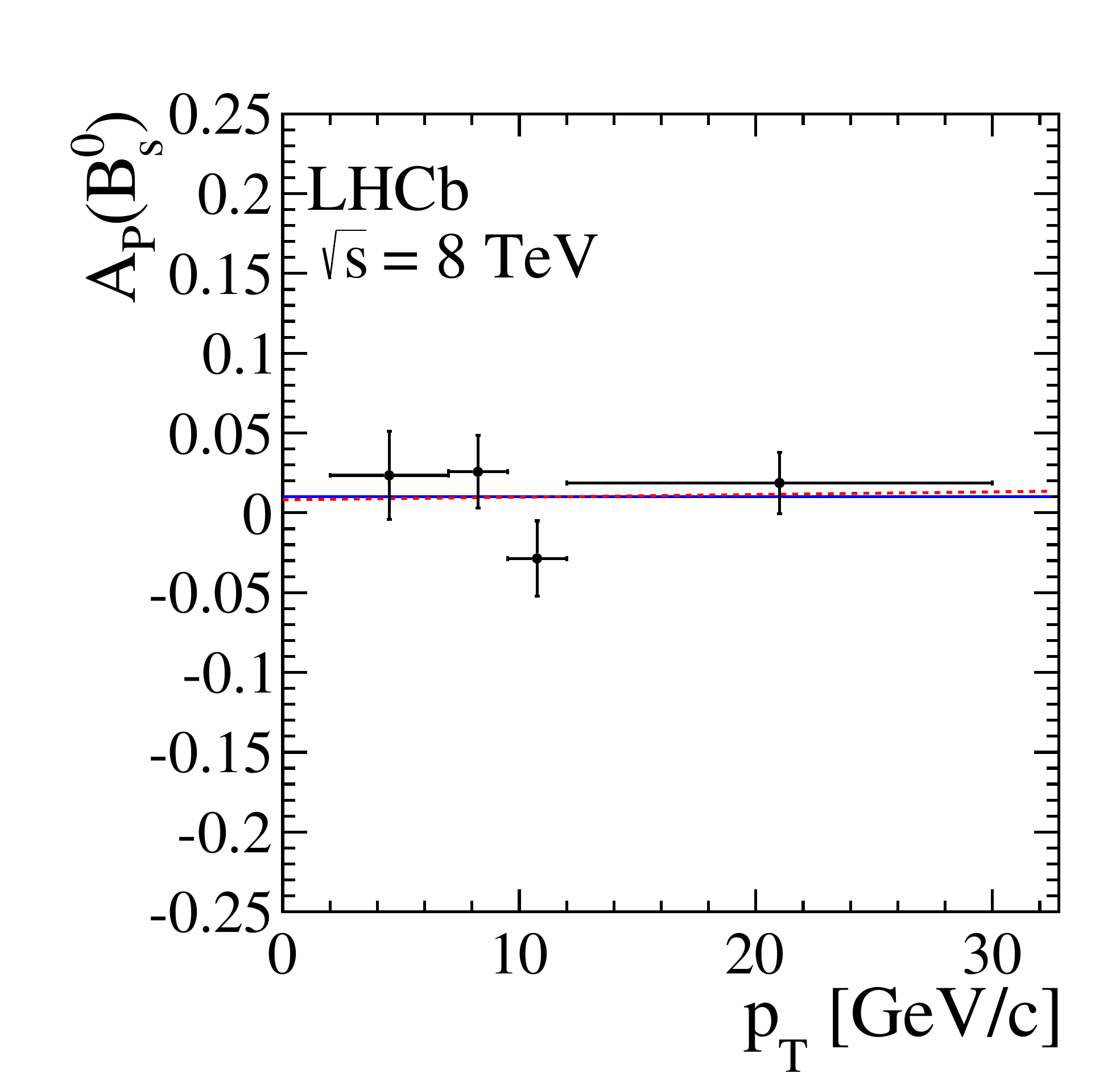}
    \includegraphics[width=0.48\textwidth]{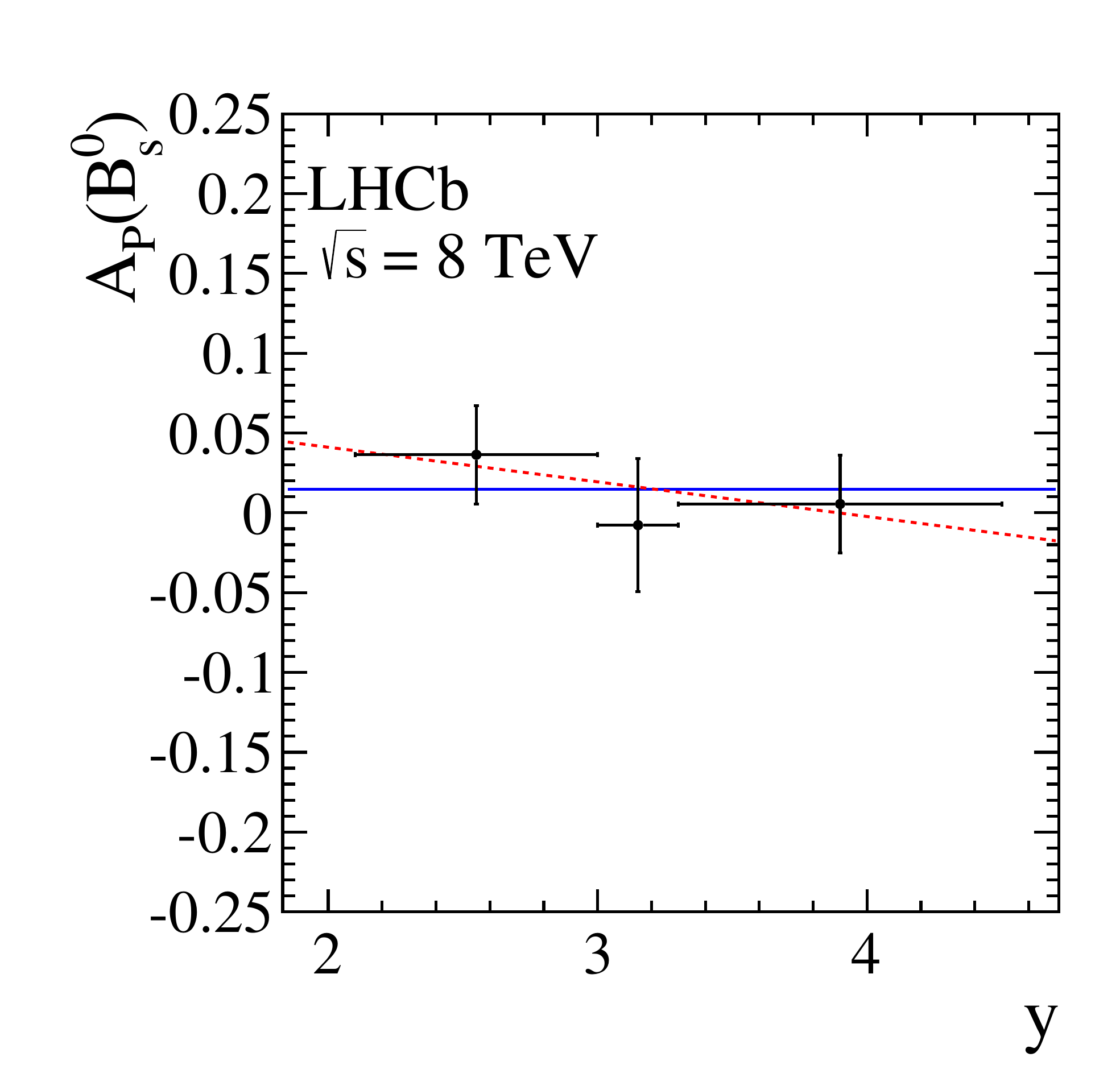}
 \end{center}
  \caption{Dependence of $A_\mathrm{P}(\Bs)$, for data collected in proton-proton collisions with centre-of-mass of energies of (top) 7 and (bottom) 8 \tev, on (left) \pt and (right) $y$. The results of fits with a straight line with zero (solid line) or floating slope parameter (dashed line) are also shown. The fits take into account the correlations amongst the bins.}
  \label{fig:AP_Bs}
\end{figure}

 \begin{figure}[th!]
  \begin{center}
    \includegraphics[width=0.48\textwidth]{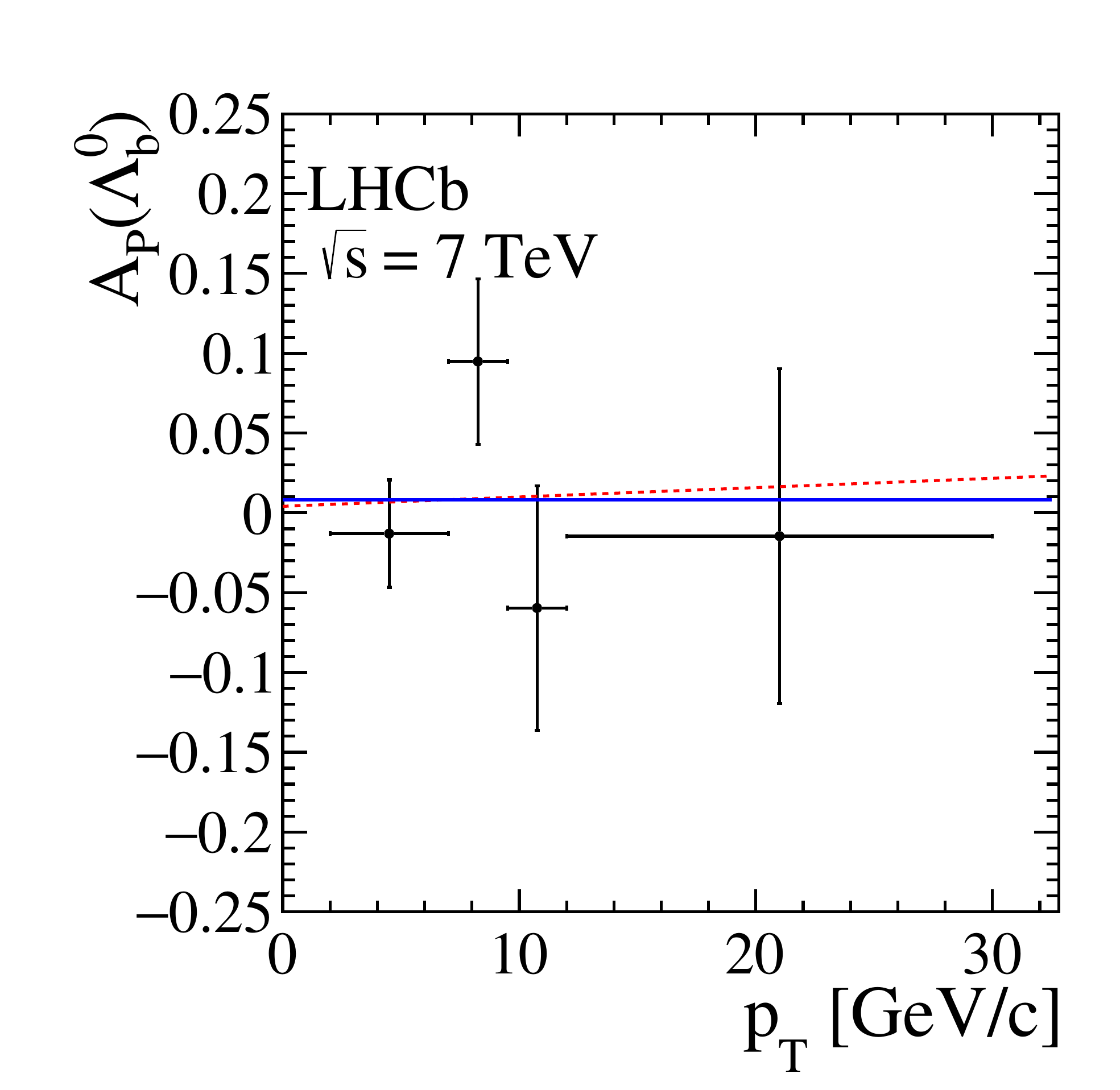}
    \includegraphics[width=0.48\textwidth]{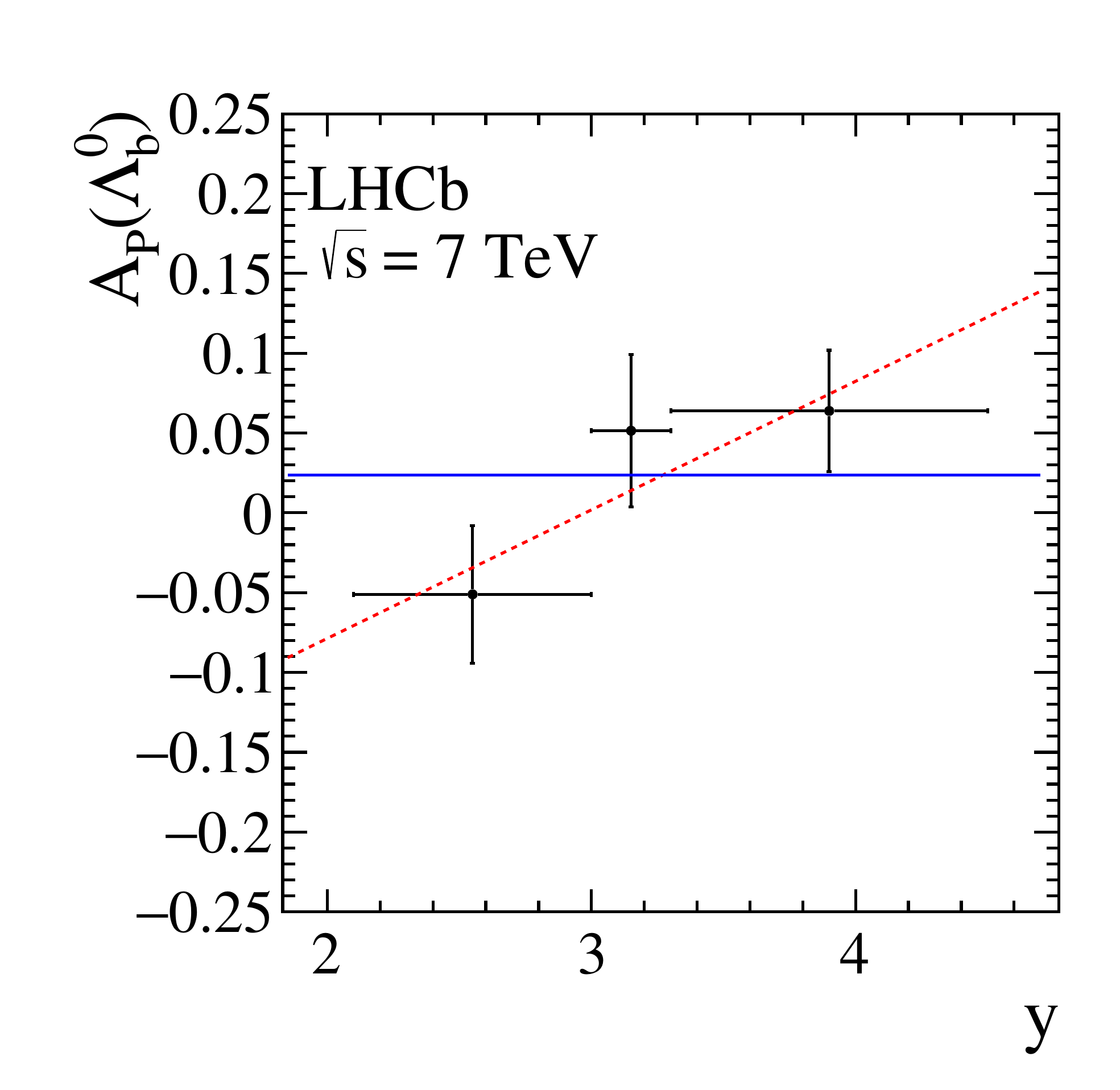}
    \includegraphics[width=0.48\textwidth]{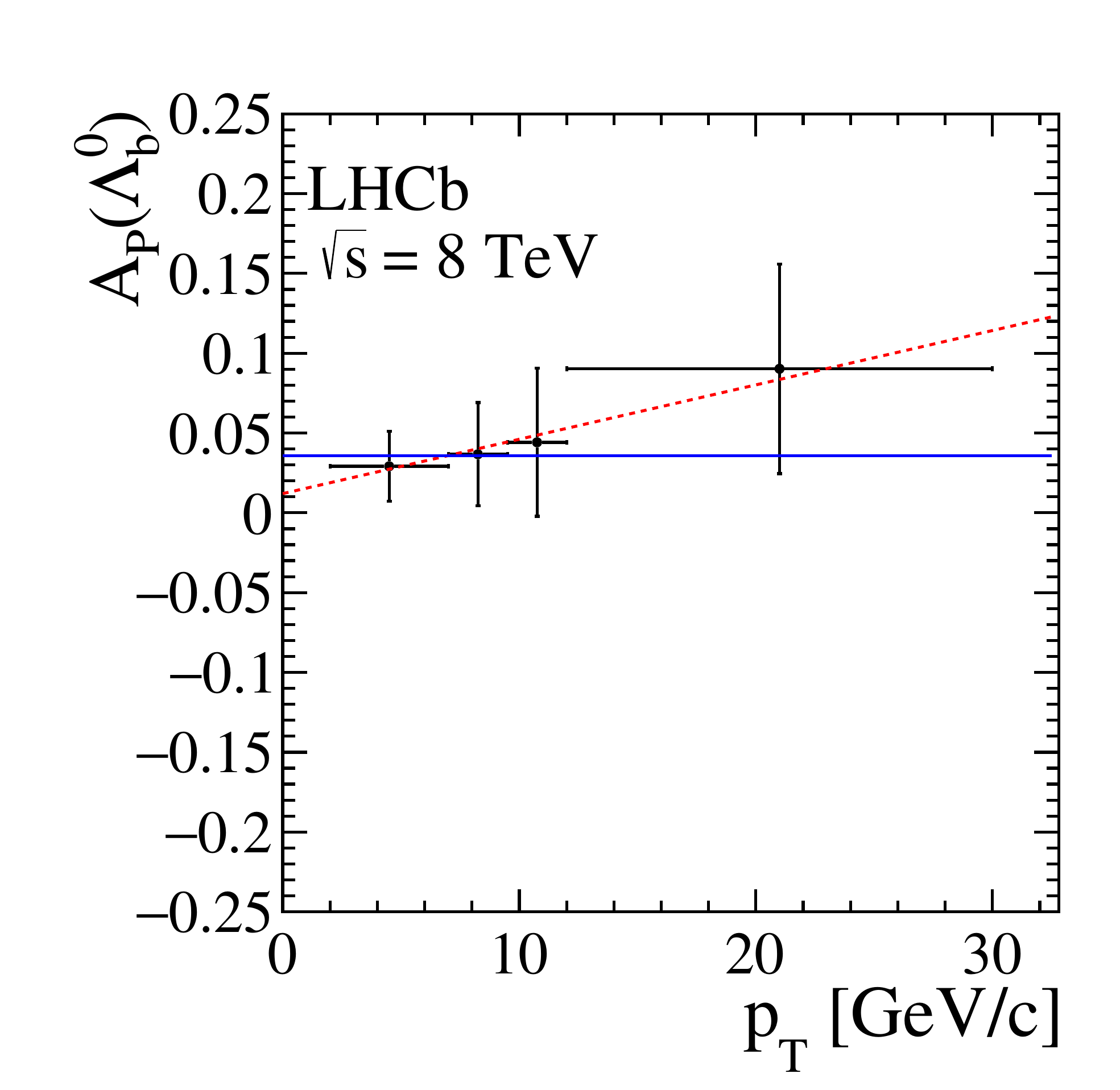}
    \includegraphics[width=0.48\textwidth]{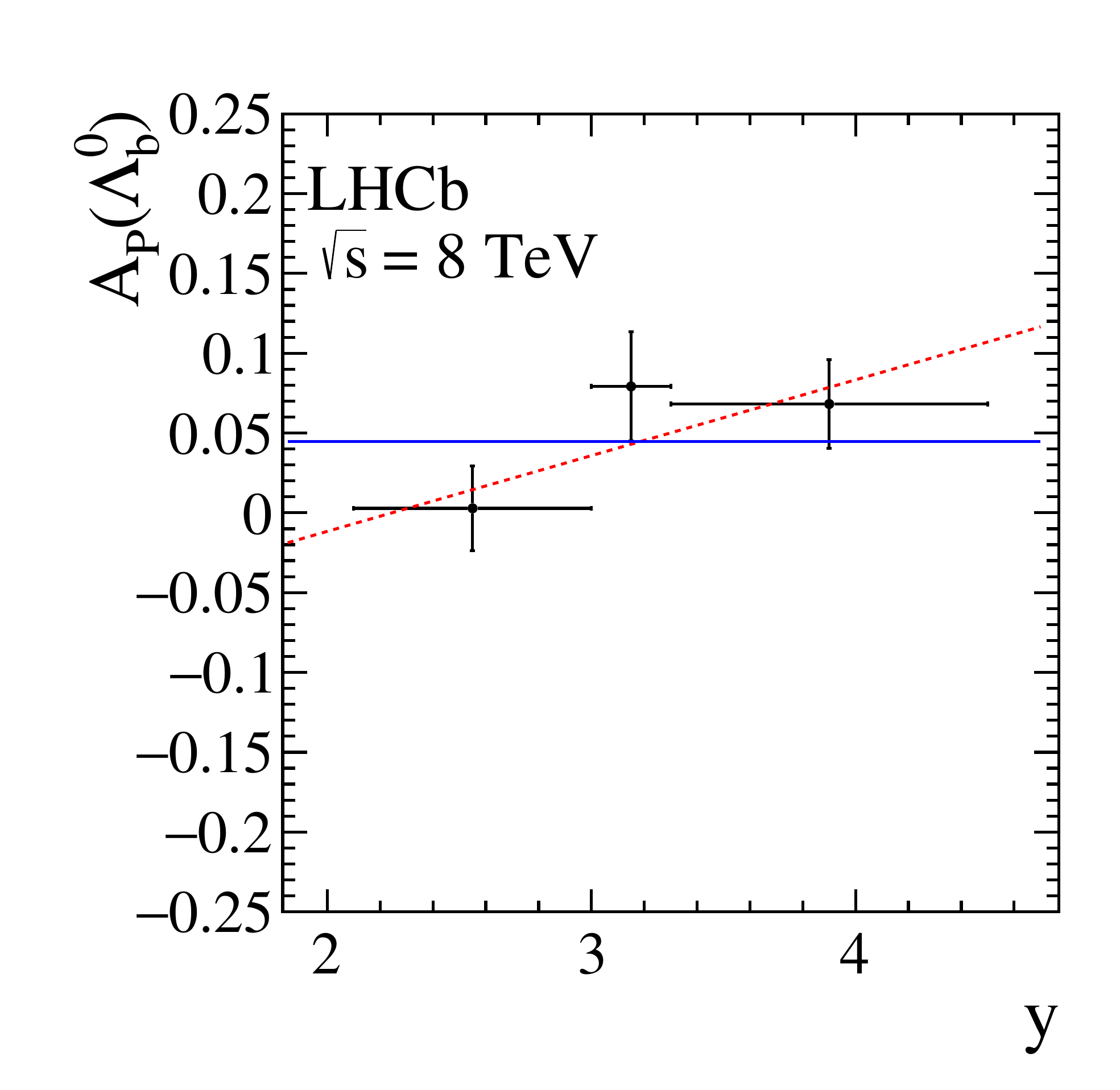}
 \end{center}
  \caption{Dependence of $A_\mathrm{P}(\Lb)$, for data collected in proton-proton collisions with centre-of-mass of energies of (top) 7 and (bottom) 8 \tev, on (left) \pt and (right) $y$. The results of fits with a straight line with zero (solid line) or floating slope parameter (dashed line) are also shown. The fits take into account the correlations amongst the bins.}
  \label{fig:AP_Lb}
\end{figure}

\begin{table}[!ht]
 \begin{center}
  \caption{Values for $m$ and $q$ and their correlation coefficient ($\rho$) obtained from fits to the values reported in Tables~\ref{tab:AP_BpB0_2011_pt}--\ref{tab:AP_BsLb_2012_eta} with a first order polynomial function (FOPF), $A_{\mathrm{P}}(b\,\mathrm{hadron}) = a\,x + b$ with $x = \pt,y$. The label SL indicates the fit to the values with a straight line.}
 \label{tab:polfit}
   \begin{tabular}{c|c|c|c|c}
\pt & \multicolumn{4}{c}{\sqs = 7\tev} \\
\hline
 & $\Bp$ & $\Bz$ & $\Bs$ & $\Lb$  \\
\hline
$a\,[c/\mathrm{GeV} \cdot 10^{-4}] $ & $-3 \pm 6$ & $\phantom{-}7 \pm 14$  & $-33 \pm 26$            & $ 10 \pm 60$  \\ 
$b\,[10^{-3}]$                        & $-1 \pm 5$ & $-3 \pm 12$            & $ \phantom{-}44 \pm 37$ & $\,\,\,7 \pm 50$  \\
$\rho(m,q)$                          & $-0.59$    & $-0.78 $               & $-0.89$                 & $ -0.85$\\ 
\hline
fit \chisqndf (SL)   & $\,\,\,$1.05 & $\,\,\,$1.67 & $\,\,\,$1.67 & $\,\,\,$2.03  \\
fit \chisqndf (FOPF) & $\,\,\,$0.95 &$\,\,\,$1.50  & $\,\,\,$1.68 & $\,\,\,$1.34 \\
\multicolumn{5}{c}{}  \\
\pt & \multicolumn{4}{c}{\sqs = 8\tev} \\
\hline
 & $\Bp$ & $\Bz$ & $\Bs$ & $\Lb$  \\
\hline
$a\,[c/\mathrm{GeV} \cdot 10^{-4}] $ & $-5 \pm 4$ & $\phantom{-}12 \pm 9$  & $2 \pm 18$ & $ 30 \pm 30$ \\
$b\,[10^{-3}]$                        & $-5 \pm 4$ & $-18 \pm 8$            & $8 \pm 25$ & $ 12 \pm 32$ \\
$\rho(m,q)$                          & $-0.49$    & $-0.78 $               & $-0.89$                 & $-0.84 $\\ 
\hline
fit \chisqndf (SL)   & $\,\,\,$0.99 & $\,\,\,$0.54 & $\,\,\,$1.80 & $\,\,\,$0.07  \\
fit \chisqndf (FOPF) & $\,\,\,$1.12 &$\,\,\,$0.67  & $\,\,\,$1.19 & $\,\,\,$0.28 \\
\multicolumn{5}{c}{}  \\
$y$ & \multicolumn{4}{c}{\sqs = 7\tev} \\
\hline
 & $\Bp$ & $\Bz$ & $\Bs$ & $\Lb$  \\
\hline
$a\,[10^{-4}] $ & $\phantom{-}22  \pm 45$ & $-71 \pm 141$  & $-542 \pm 469$           & $ \phantom{-}810 \pm 420$ \\
$b\,[10^{-3}]$                        & $-12 \pm 14$            & $-12 \pm 46 $  & $\phantom{-}165\pm 153$ & $-240 \pm 130$  \\
$\rho(m,q)$                          & $-0.96$    & $-0.99 $               & $-0.99$                 & $ -0.98$\\ 
\hline
fit \chisqndf (SL)   & $\,\,\,$1.36 & $\,\,\,$2.79 & $\,\,\,$0.48 & $\,\,\,$0.86  \\
fit \chisqndf (FOPF) & $\,\,\,$1.17 &$\,\,\,$2.36  & $\,\,\,$0.91 & $\,\,\,$2.29 \\
\multicolumn{5}{c}{}  \\
$y$ & \multicolumn{4}{c}{\sqs = 8\tev} \\
\hline
 & $\Bp$ & $\Bz$ & $\Bs$ & $\Lb$  \\
\hline
$a\,[10^{-4}] $ & $-86 \pm 29$            & $-44 \pm 100$              & $-217 \pm 321$           & $\,\,\,\,\,\,470 \pm 280$ \\
$b\,[10^{-3}]$                        & $ \phantom{,}19 \pm  9$ & $-4 \pm  32$  & $\phantom{-2}85 \pm 105$ & $-111 \pm  90$  \\
$\rho(m,q)$                          & $-0.93$    & $-0.99 $               & $-0.98$                 & $-0.98$\\ 
\hline
fit \chisqndf (SL)   & $\,\,\,$1.10 & $\,\,\,$0.86 & $\,\,\,$0.42 & $\,\,\,$1.35  \\
fit \chisqndf (FOPF) & $\,\,\,$2.43 &$\,\,\,$0.75  & $\,\,\,$0.44 & $\,\,\,$2.23 \\
\end{tabular}
\end{center}
\end{table}

\clearpage

\section*{Acknowledgements}
\noindent We express our gratitude to our colleagues in the CERN
accelerator departments for the excellent performance of the LHC. We
thank the technical and administrative staff at the LHCb
institutes. We acknowledge support from CERN and from the national
agencies: CAPES, CNPq, FAPERJ and FINEP (Brazil); MOST and NSFC (China);
CNRS/IN2P3 (France); BMBF, DFG and MPG (Germany); INFN (Italy); 
NWO (The Netherlands); MNiSW and NCN (Poland); MEN/IFA (Romania); 
MinES and FASO (Russia); MinECo (Spain); SNSF and SER (Switzerland); 
NASU (Ukraine); STFC (United Kingdom); NSF (USA).
We acknowledge the computing resources that are provided by CERN, IN2P3 (France), KIT and DESY (Germany), INFN (Italy), SURF (The Netherlands), PIC (Spain), GridPP (United Kingdom), RRCKI and Yandex LLC (Russia), CSCS (Switzerland), IFIN-HH (Romania), CBPF (Brazil), PL-GRID (Poland) and OSC (USA). We are indebted to the communities behind the multiple open 
source software packages on which we depend.
Individual groups or members have received support from AvH Foundation (Germany),
EPLANET, Marie Sk\l{}odowska-Curie Actions and ERC (European Union), 
Conseil G\'{e}n\'{e}ral de Haute-Savoie, Labex ENIGMASS and OCEVU, 
R\'{e}gion Auvergne (France), RFBR and Yandex LLC (Russia), GVA, XuntaGal and GENCAT (Spain), Herchel Smith Fund, The Royal Society, Royal Commission for the Exhibition of 1851 and the Leverhulme Trust (United Kingdom).

\newpage
\appendix
\section*{Appendix} 
\label{App:AppendixA}

\begin{center}
\begin{longtable}{c|c|c|c}
  \caption{Values of $A_{\rm P}(\Bp)$ and $A_{\rm P}(\Bz)$ in each kinematic bin for data collected in proton-proton collisions at centre-of-mass energy of  7\tev. The first uncertainties are statistical and the second systematic. }  \label{tab:resultsBpB02011}\\
\pt [\gevc] & $y$ & $A_{\rm P}(\Bp)_{\sqs = 7\, \tev}$ & $A_{\rm P}(\Bz)_{\sqs = 7\, \tev}$ \\
\hline
$(0.00,   2.00)$   &  $(2.10,  2.70)$  &  $  \phantom{-}0.0085  \pm  0.0156  \pm  0.0036  $  &  $  \phantom{-}0.0722  \pm  0.0770  \pm  0.0010  $  \\
$(0.00,   2.00)$   &  $(2.70,  2.85)$  &  $  -0.0014            \pm  0.0191  \pm  0.0036  $  &  $  -0.1108            \pm  0.0815  \pm  0.0020  $  \\
$(0.00,   2.00)$   &  $(2.85,  3.00)$  &  $  \phantom{-}0.0016  \pm  0.0177  \pm  0.0036  $  &  $  -0.0300            \pm  0.0733  \pm  0.0024  $  \\
$(0.00,   2.00)$   &  $(3.00,  3.15)$  &  $  -0.0052            \pm  0.0171  \pm  0.0036  $  &  $  -0.0849            \pm  0.0624  \pm  0.0038  $  \\
$(0.00,   2.00)$   &  $(3.15,  3.30)$  &  $  -0.0006            \pm  0.0171  \pm  0.0037  $  &  $  -0.0662            \pm  0.0638  \pm  0.0035  $  \\
$(0.00,   2.00)$   &  $(3.30,  3.70)$  &  $  \phantom{-}0.0107  \pm  0.0110  \pm  0.0040  $  &  $  \phantom{-}0.0116  \pm  0.0397  \pm  0.0011  $  \\
$(0.00,   2.00)$   &  $(3.70,  4.50)$  &  $  -0.0104            \pm  0.0141  \pm  0.0046  $  &  $  \phantom{-}0.0702  \pm  0.0462  \pm  0.0013  $  \\
$(2.00,   4.50)$   &  $(2.10,  2.70)$  &  $  \phantom{-}0.0007  \pm  0.0088  \pm  0.0036  $  &  $  \phantom{-}0.0691  \pm  0.0392  \pm  0.0017  $  \\
$(2.00,   4.50)$   &  $(2.70,  2.85)$  &  $  -0.0171            \pm  0.0112  \pm  0.0036  $  &  $  \phantom{-}0.0136  \pm  0.0409  \pm  0.0013  $  \\
$(2.00,   4.50)$   &  $(2.85,  3.00)$  &  $  -0.0120            \pm  0.0105  \pm  0.0036  $  &  $  -0.0284            \pm  0.0375  \pm  0.0010  $  \\
$(2.00,   4.50)$   &  $(3.00,  3.15)$  &  $  -0.0269            \pm  0.0101  \pm  0.0037  $  &  $  -0.0273            \pm  0.0360  \pm  0.0009  $  \\
$(2.00,   4.50)$   &  $(3.15,  3.30)$  &  $  \phantom{-}0.0043  \pm  0.0102  \pm  0.0038  $  &  $  \phantom{-}0.0137  \pm  0.0351  \pm  0.0015  $  \\
$(2.00,   4.50)$   &  $(3.30,  3.70)$  &  $  -0.0167            \pm  0.0071  \pm  0.0041  $  &  $  -0.0273            \pm  0.0230  \pm  0.0028  $  \\
$(2.00,   4.50)$   &  $(3.70,  4.50)$  &  $  \phantom{-}0.0053  \pm  0.0098  \pm  0.0045  $  &  $  -0.0269            \pm  0.0279  \pm  0.0013  $  \\
$(4.50,   7.00)$   &  $(2.10,  2.70)$  &  $  \phantom{-}0.0023  \pm  0.0087  \pm  0.0035  $  &  $  \phantom{-}0.0597  \pm  0.0329  \pm  0.0039  $  \\
$(4.50,   7.00)$   &  $(2.70,  2.85)$  &  $  -0.0002            \pm  0.0120  \pm  0.0037  $  &  $  -0.0177            \pm  0.0404  \pm  0.0010  $  \\
$(4.50,   7.00)$   &  $(2.85,  3.00)$  &  $  \phantom{-}0.0034  \pm  0.0116  \pm  0.0038  $  &  $  -0.0103            \pm  0.0362  \pm  0.0019  $  \\
$(4.50,   7.00)$   &  $(3.00,  3.15)$  &  $  \phantom{-}0.0092  \pm  0.0115  \pm  0.0039  $  &  $  -0.0696            \pm  0.0372  \pm  0.0016  $  \\
$(4.50,   7.00)$   &  $(3.15,  3.30)$  &  $  -0.0092            \pm  0.0120  \pm  0.0042  $  &  $  -0.0444            \pm  0.0359  \pm  0.0015  $  \\
$(4.50,   7.00)$   &  $(3.30,  3.70)$  &  $  -0.0168            \pm  0.0088  \pm  0.0044  $  &  $  -0.0214            \pm  0.0234  \pm  0.0010  $  \\
$(4.50,   7.00)$   &  $(3.70,  4.50)$  &  $  \phantom{-}0.0010  \pm  0.0129  \pm  0.0044  $  &  $  \phantom{-}0.0192  \pm  0.0316  \pm  0.0013  $  \\
$(7.00,   8.25)$   &  $(2.10,  2.70)$  &  $  \phantom{-}0.0031  \pm  0.0140  \pm  0.0036  $  &  $  -0.0239            \pm  0.0441  \pm  0.0025  $  \\
$(7.00,   8.25)$   &  $(2.70,  2.85)$  &  $  -0.0591            \pm  0.0208  \pm  0.0039  $  &  $  -0.2197            \pm  0.0602  \pm  0.0017  $  \\
$(7.00,   8.25)$   &  $(2.85,  3.00)$  &  $  -0.0089            \pm  0.0203  \pm  0.0040  $  &  $  -0.0619            \pm  0.0595  \pm  0.0031  $  \\
$(7.00,   8.25)$   &  $(3.00,  3.15)$  &  $  \phantom{-}0.0016  \pm  0.0213  \pm  0.0043  $  &  $  -0.0151            \pm  0.0590  \pm  0.0047  $  \\
$(7.00,   8.25)$   &  $(3.15,  3.30)$  &  $  -0.0205            \pm  0.0222  \pm  0.0044  $  &  $  -0.0037            \pm  0.0566  \pm  0.0039  $  \\
$(7.00,   8.25)$   &  $(3.30,  3.70)$  &  $  \phantom{-}0.0303  \pm  0.0172  \pm  0.0046  $  &  $  -0.0305            \pm  0.0406  \pm  0.0018  $  \\
$(7.00,   8.25)$   &  $(3.70,  4.50)$  &  $  \phantom{-}0.0603  \pm  0.0259  \pm  0.0047  $  &  $  -0.0348            \pm  0.0516  \pm  0.0018  $  \\
$(8.25,   9.50)$   &  $(2.10,  2.70)$  &  $  -0.0134            \pm  0.0157  \pm  0.0037  $  &  $  -0.0442            \pm  0.0477  \pm  0.0087  $  \\
$(8.25,   9.50)$   &  $(2.70,  2.85)$  &  $  -0.0099            \pm  0.0246  \pm  0.0039  $  &  $  -0.0506            \pm  0.0652  \pm  0.0028  $  \\
$(8.25,   9.50)$   &  $(2.85,  3.00)$  &  $  -0.0112            \pm  0.0246  \pm  0.0042  $  &  $  -0.0611            \pm  0.0674  \pm  0.0043  $  \\
$(8.25,   9.50)$   &  $(3.00,  3.15)$  &  $  -0.0613            \pm  0.0251  \pm  0.0044  $  &  $  -0.0015            \pm  0.0695  \pm  0.0024  $  \\
$(8.25,   9.50)$   &  $(3.15,  3.30)$  &  $  \phantom{-}0.0552  \pm  0.0279  \pm  0.0045  $  &  $  \phantom{-}0.0219  \pm  0.0731  \pm  0.0014  $  \\
$(8.25,   9.50)$   &  $(3.30,  3.70)$  &  $  -0.0038            \pm  0.0216  \pm  0.0046  $  &  $  -0.0621            \pm  0.0478  \pm  0.0032  $  \\
$(8.25,   9.50)$   &  $(3.70,  4.50)$  &  $  \phantom{-}0.0047  \pm  0.0342  \pm  0.0047  $  &  $  -0.0856            \pm  0.0637  \pm  0.0025  $  \\
$(9.50,   10.75)$  &  $(2.10,  2.70)$  &  $  -0.0249            \pm  0.0182  \pm  0.0037  $  &  $  \phantom{-}0.0408  \pm  0.0525  \pm  0.0089  $  \\
$(9.50,   10.75)$  &  $(2.70,  2.85)$  &  $  -0.0113            \pm  0.0292  \pm  0.0041  $  &  $  \phantom{-}0.0228  \pm  0.0759  \pm  0.0036  $  \\
$(9.50,   10.75)$  &  $(2.85,  3.00)$  &  $  -0.0241            \pm  0.0290  \pm  0.0045  $  &  $  \phantom{-}0.0102  \pm  0.0904  \pm  0.0017  $  \\
$(9.50,   10.75)$  &  $(3.00,  3.15)$  &  $  \phantom{-}0.0267  \pm  0.0318  \pm  0.0045  $  &  $  -0.0586            \pm  0.0847  \pm  0.0023  $  \\
\pt [\gevc] & $y$ & $A_{\rm P}(\Bp)_{\sqs = 7\, \tev}$ & $A_{\rm P}(\Bz)_{\sqs = 7\, \tev}$ \\
\hline
$(9.50,   10.75)$  &  $(3.15,  3.30)$  &  $  \phantom{-}0.0118  \pm  0.0352  \pm  0.0048  $  &  $  -0.0577            \pm  0.0775  \pm  0.0012  $  \\
$(9.50,   10.75)$  &  $(3.30,  3.70)$  &  $  -0.0164            \pm  0.0281  \pm  0.0048  $  &  $  \phantom{-}0.0624  \pm  0.0577  \pm  0.0016  $  \\
$(9.50,   10.75)$  &  $(3.70,  4.50)$  &  $  -0.0605            \pm  0.0411  \pm  0.0049  $  &  $  -0.0328            \pm  0.0946  \pm  0.0021  $  \\
$(10.75,  12.00)$  &  $(2.10,  2.70)$  &  $  -0.0200            \pm  0.0206  \pm  0.0038  $  &  $  \phantom{-}0.0154  \pm  0.0636  \pm  0.0023  $  \\
$(10.75,  12.00)$  &  $(2.70,  2.85)$  &  $  -0.0068            \pm  0.0344  \pm  0.0044  $  &  $  -0.0104            \pm  0.1017  \pm  0.0044  $  \\
$(10.75,  12.00)$  &  $(2.85,  3.00)$  &  $  -0.0017            \pm  0.0362  \pm  0.0045  $  &  $  \phantom{-}0.0179  \pm  0.0849  \pm  0.0040  $  \\
$(10.75,  12.00)$  &  $(3.00,  3.15)$  &  $  -0.0181            \pm  0.0411  \pm  0.0047  $  &  $  \phantom{-}0.1481  \pm  0.0890  \pm  0.0024  $  \\
$(10.75,  12.00)$  &  $(3.15,  3.30)$  &  $  -0.0239            \pm  0.0441  \pm  0.0047  $  &  $  \phantom{-}0.0478  \pm  0.0835  \pm  0.0025  $  \\
$(10.75,  12.00)$  &  $(3.30,  3.70)$  &  $  \phantom{-}0.0058  \pm  0.0362  \pm  0.0048  $  &  $  \phantom{-}0.0377  \pm  0.0731  \pm  0.0037  $  \\
$(10.75,  12.00)$  &  $(3.70,  4.50)$  &  $  \phantom{-}0.0485  \pm  0.0547  \pm  0.0051  $  &  $  \phantom{-}0.1058  \pm  0.1181  \pm  0.0018  $  \\
$(12.00,  15.00)$  &  $(2.10,  2.70)$  &  $  \phantom{-}0.0059  \pm  0.0174  \pm  0.0039  $  &  $  -0.0071            \pm  0.0446  \pm  0.0039  $  \\
$(12.00,  15.00)$  &  $(2.70,  2.85)$  &  $  \phantom{-}0.0210  \pm  0.0321  \pm  0.0046  $  &  $  \phantom{-}0.0264  \pm  0.0924  \pm  0.0042  $  \\
$(12.00,  15.00)$  &  $(2.85,  3.00)$  &  $  \phantom{-}0.0092  \pm  0.0334  \pm  0.0062  $  &  $  \phantom{-}0.0230  \pm  0.0775  \pm  0.0046  $  \\
$(12.00,  15.00)$  &  $(3.00,  3.15)$  &  $  -0.0267            \pm  0.0386  \pm  0.0050  $  &  $  -0.1190            \pm  0.0791  \pm  0.0040  $  \\
$(12.00,  15.00)$  &  $(3.15,  3.30)$  &  $  -0.0516            \pm  0.0420  \pm  0.0046  $  &  $  \phantom{-}0.1330  \pm  0.0909  \pm  0.0029  $  \\
$(12.00,  15.00)$  &  $(3.30,  3.70)$  &  $  \phantom{-}0.0071  \pm  0.0349  \pm  0.0052  $  &  $  \phantom{-}0.0469  \pm  0.0588  \pm  0.0021  $  \\
$(12.00,  15.00)$  &  $(3.70,  4.50)$  &  $  \phantom{-}0.0748  \pm  0.0542  \pm  0.0049  $  &  $  -0.1026            \pm  0.0854  \pm  0.0031  $  \\
$(15.00,  30.00)$  &  $(2.10,  2.70)$  &  $  \phantom{-}0.0116  \pm  0.0188  \pm  0.0040  $  &  $  \phantom{-}0.0703  \pm  0.0456  \pm  0.0014  $  \\
$(15.00,  30.00)$  &  $(2.70,  2.85)$  &  $  -0.0763            \pm  0.0401  \pm  0.0046  $  &  $  -0.0009            \pm  0.0748  \pm  0.0034  $  \\
$(15.00,  30.00)$  &  $(2.85,  3.00)$  &  $  -0.0541            \pm  0.0458  \pm  0.0047  $  &  $  -0.0550            \pm  0.0755  \pm  0.0049  $  \\
$(15.00,  30.00)$  &  $(3.00,  3.15)$  &  $  -0.0449            \pm  0.0512  \pm  0.0046  $  &  $  -0.1637            \pm  0.0925  \pm  0.0026  $  \\
$(15.00,  30.00)$  &  $(3.15,  3.30)$  &  $  \phantom{-}0.0011  \pm  0.0599  \pm  0.0073  $  &  $  \phantom{-}0.0456  \pm  0.1119  \pm  0.0018  $  \\
$(15.00,  30.00)$  &  $(3.30,  3.70)$  &  $  \phantom{-}0.0089  \pm  0.0502  \pm  0.0048  $  &  $  -0.0193            \pm  0.0777  \pm  0.0027  $  \\
$(15.00,  30.00)$  &  $(3.70,  4.50)$  &  $  -0.0662            \pm  0.0827  \pm  0.0186  $  &  $  \phantom{-}0.1690  \pm  0.1332  \pm  0.0030  $  \\
\end{longtable}
\end{center}

\clearpage

 \begin{center}
 \begin{longtable}{c|c|c|c}
  \caption{Values of $A_{\rm P}(\Bp)$ and $A_{\rm P}(\Bz)$ in each kinematic bin for data collected in proton-proton collisions at centre-of-mass energy of  8\tev. The first uncertainties are statistical and the second systematic.}     \label{tab:resultsBpB02012} \\
\pt [\gevc] & $y$ & $A_{\rm P}(\Bp)_{\sqs = 8\, \tev}$ & $A_{\rm P}(\Bz)_{\sqs = 8\, \tev}$ \\
\hline
$(0.00,   2.00)$   &  $(2.10,  2.70)$  &  $  -0.0178            \pm  0.0097  \pm  0.0031  $  &  $  \phantom{-}0.0068  \pm  0.0537  \pm  0.0009  $  \\
$(0.00,   2.00)$   &  $(2.70,  2.85)$  &  $  -0.0027            \pm  0.0126  \pm  0.0031  $  &  $  -0.0735            \pm  0.0719  \pm  0.0017  $  \\
$(0.00,   2.00)$   &  $(2.85,  3.00)$  &  $  \phantom{-}0.0093  \pm  0.0120  \pm  0.0031  $  &  $  \phantom{-}0.0503  \pm  0.0628  \pm  0.0011  $  \\
$(0.00,   2.00)$   &  $(3.00,  3.15)$  &  $  \phantom{-}0.0005  \pm  0.0119  \pm  0.0031  $  &  $  \phantom{-}0.0086  \pm  0.0549  \pm  0.0034  $  \\
$(0.00,   2.00)$   &  $(3.15,  3.30)$  &  $  -0.0230            \pm  0.0119  \pm  0.0033  $  &  $  \phantom{-}0.0817  \pm  0.0617  \pm  0.0016  $  \\
$(0.00,   2.00)$   &  $(3.30,  3.70)$  &  $  -0.0120            \pm  0.0080  \pm  0.0033  $  &  $  \phantom{-}0.0668  \pm  0.0367  \pm  0.0009  $  \\
$(0.00,   2.00)$   &  $(3.70,  4.50)$  &  $  -0.0077            \pm  0.0103  \pm  0.0037  $  &  $  -0.0419            \pm  0.0453  \pm  0.0010  $  \\
$(2.00,   4.50)$   &  $(2.10,  2.70)$  &  $  \phantom{-}0.0050  \pm  0.0054  \pm  0.0031  $  &  $  -0.0192            \pm  0.0234  \pm  0.0013  $  \\
$(2.00,   4.50)$   &  $(2.70,  2.85)$  &  $  -0.0076            \pm  0.0073  \pm  0.0031  $  &  $  -0.0070            \pm  0.0291  \pm  0.0009  $  \\
$(2.00,   4.50)$   &  $(2.85,  3.00)$  &  $  \phantom{-}0.0009  \pm  0.0070  \pm  0.0031  $  &  $  -0.0088            \pm  0.0278  \pm  0.0009  $  \\
$(2.00,   4.50)$   &  $(3.00,  3.15)$  &  $  -0.0046            \pm  0.0069  \pm  0.0032  $  &  $  -0.0213            \pm  0.0271  \pm  0.0009  $  \\
$(2.00,   4.50)$   &  $(3.15,  3.30)$  &  $  -0.0018            \pm  0.0070  \pm  0.0032  $  &  $  -0.0635            \pm  0.0260  \pm  0.0012  $  \\
$(2.00,   4.50)$   &  $(3.30,  3.70)$  &  $  -0.0081            \pm  0.0049  \pm  0.0034  $  &  $  -0.0169            \pm  0.0174  \pm  0.0009  $  \\
$(2.00,   4.50)$   &  $(3.70,  4.50)$  &  $  -0.0133            \pm  0.0067  \pm  0.0036  $  &  $  -0.0131            \pm  0.0203  \pm  0.0009  $  \\
$(4.50,   7.00)$   &  $(2.10,  2.70)$  &  $  -0.0045            \pm  0.0054  \pm  0.0031  $  &  $  -0.0074            \pm  0.0192  \pm  0.0028  $  \\
$(4.50,   7.00)$   &  $(2.70,  2.85)$  &  $  -0.0002            \pm  0.0077  \pm  0.0031  $  &  $  -0.0440            \pm  0.0264  \pm  0.0027  $  \\
$(4.50,   7.00)$   &  $(2.85,  3.00)$  &  $  -0.0019            \pm  0.0075  \pm  0.0032  $  &  $  \phantom{-}0.0315  \pm  0.0235  \pm  0.0028  $  \\
$(4.50,   7.00)$   &  $(3.00,  3.15)$  &  $  -0.0107            \pm  0.0076  \pm  0.0033  $  &  $  -0.0203            \pm  0.0233  \pm  0.0020  $  \\
$(4.50,   7.00)$   &  $(3.15,  3.30)$  &  $  -0.0175            \pm  0.0078  \pm  0.0034  $  &  $  -0.0248            \pm  0.0234  \pm  0.0010  $  \\
$(4.50,   7.00)$   &  $(3.30,  3.70)$  &  $  -0.0241            \pm  0.0059  \pm  0.0035  $  &  $  -0.0254            \pm  0.0159  \pm  0.0010  $  \\
$(4.50,   7.00)$   &  $(3.70,  4.50)$  &  $  -0.0101            \pm  0.0087  \pm  0.0036  $  &  $  -0.0015            \pm  0.0213  \pm  0.0010  $  \\
$(7.00,   8.25)$   &  $(2.10,  2.70)$  &  $  -0.0052            \pm  0.0086  \pm  0.0031  $  &  $  \phantom{-}0.0080  \pm  0.0276  \pm  0.0028  $  \\
$(7.00,   8.25)$   &  $(2.70,  2.85)$  &  $  -0.0177            \pm  0.0131  \pm  0.0033  $  &  $  -0.0383            \pm  0.0390  \pm  0.0014  $  \\
$(7.00,   8.25)$   &  $(2.85,  3.00)$  &  $  -0.0083            \pm  0.0132  \pm  0.0033  $  &  $  -0.0543            \pm  0.0382  \pm  0.0025  $  \\
$(7.00,   8.25)$   &  $(3.00,  3.15)$  &  $  \phantom{-}0.0065  \pm  0.0134  \pm  0.0035  $  &  $  -0.0575            \pm  0.0377  \pm  0.0012  $  \\
$(7.00,   8.25)$   &  $(3.15,  3.30)$  &  $  -0.0055            \pm  0.0144  \pm  0.0040  $  &  $  -0.0120            \pm  0.0379  \pm  0.0013  $  \\
$(7.00,   8.25)$   &  $(3.30,  3.70)$  &  $  -0.0003            \pm  0.0111  \pm  0.0036  $  &  $  -0.0089            \pm  0.0268  \pm  0.0044  $  \\
$(7.00,   8.25)$   &  $(3.70,  4.50)$  &  $  -0.0300            \pm  0.0168  \pm  0.0036  $  &  $  -0.0486            \pm  0.0364  \pm  0.0022  $  \\
$(8.25,   9.50)$   &  $(2.10,  2.70)$  &  $  -0.0038            \pm  0.0097  \pm  0.0031  $  &  $  -0.0215            \pm  0.0286  \pm  0.0017  $  \\
$(8.25,   9.50)$   &  $(2.70,  2.85)$  &  $  -0.0070            \pm  0.0153  \pm  0.0033  $  &  $  \phantom{-}0.0710  \pm  0.0415  \pm  0.0013  $  \\
$(8.25,   9.50)$   &  $(2.85,  3.00)$  &  $  -0.0228            \pm  0.0157  \pm  0.0034  $  &  $  \phantom{-}0.0123  \pm  0.0395  \pm  0.0010  $  \\
$(8.25,   9.50)$   &  $(3.00,  3.15)$  &  $  -0.0236            \pm  0.0164  \pm  0.0037  $  &  $  \phantom{-}0.0747  \pm  0.0411  \pm  0.0023  $  \\
$(8.25,   9.50)$   &  $(3.15,  3.30)$  &  $  -0.0252            \pm  0.0182  \pm  0.0042  $  &  $  -0.0533            \pm  0.0459  \pm  0.0025  $  \\
$(8.25,   9.50)$   &  $(3.30,  3.70)$  &  $  -0.0036            \pm  0.0141  \pm  0.0037  $  &  $  \phantom{-}0.0152  \pm  0.0299  \pm  0.0009  $  \\
$(8.25,   9.50)$   &  $(3.70,  4.50)$  &  $  -0.0293            \pm  0.0220  \pm  0.0037  $  &  $  -0.0063            \pm  0.0448  \pm  0.0034  $  \\
$(9.50,   10.75)$  &  $(2.10,  2.70)$  &  $  \phantom{-}0.0060  \pm  0.0109  \pm  0.0032  $  &  $  \phantom{-}0.0022  \pm  0.0324  \pm  0.0022  $  \\
$(9.50,   10.75)$  &  $(2.70,  2.85)$  &  $  -0.0011           \pm  0.0183  \pm  0.0036  $  &  $  \phantom{-}0.0429  \pm  0.0491  \pm  0.0050  $  \\
$(9.50,   10.75)$  &  $(2.85,  3.00)$  &  $  \phantom{-}0.0122  \pm  0.0182  \pm  0.0036  $  &  $  \phantom{-}0.0513  \pm  0.0509  \pm  0.0021  $  \\
$(9.50,   10.75)$  &  $(3.00,  3.15)$  &  $  \phantom{-}0.0067  \pm  0.0204  \pm  0.0037  $  &  $  -0.0898            \pm  0.0499  \pm  0.0059  $  \\
$(9.50,   10.75)$  &  $(3.15,  3.30)$  &  $  -0.0462            \pm  0.0233  \pm  0.0037  $  &  $  -0.0220            \pm  0.0494  \pm  0.0034  $  \\
$(9.50,   10.75)$  &  $(3.30,  3.70)$  &  $  -0.0290            \pm  0.0181  \pm  0.0037  $  &  $  -0.0204            \pm  0.0353  \pm  0.0013  $  \\
$(9.50,   10.75)$  &  $(3.70,  4.50)$  &  $  -0.0243            \pm  0.0273  \pm  0.0037  $  &  $  -0.0849            \pm  0.0509  \pm  0.0026  $  \\
\pt [\gevc] & $y$ & $A_{\rm P}(\Bp)_{\sqs = 8\, \tev}$ & $A_{\rm P}(\Bz)_{\sqs = 8\, \tev}$ \\
\hline
$(10.75,  12.00)$  &  $(2.10,  2.70)$  &  $  \phantom{-}0.0191  \pm  0.0128  \pm  0.0032  $  &  $  \phantom{-}0.0034  \pm  0.0355  \pm  0.0056  $  \\
$(10.75,  12.00)$  &  $(2.70,  2.85)$  &  $  -0.0562            \pm  0.0220  \pm  0.0034  $  &  $  -0.0193            \pm  0.0593  \pm  0.0026  $  \\
$(10.75,  12.00)$  &  $(2.85,  3.00)$  &  $  \phantom{-}0.0172  \pm  0.0233  \pm  0.0037  $  &  $  \phantom{-}0.0198  \pm  0.0628  \pm  0.0066  $  \\
$(10.75,  12.00)$  &  $(3.00,  3.15)$  &  $  -0.0080            \pm  0.0262  \pm  0.0044  $  &  $  -0.0056            \pm  0.0565  \pm  0.0012  $  \\
$(10.75,  12.00)$  &  $(3.15,  3.30)$  &  $  \phantom{-}0.0162  \pm  0.0282  \pm  0.0038  $  &  $  -0.0638            \pm  0.0582  \pm  0.0040  $  \\
$(10.75,  12.00)$  &  $(3.30,  3.70)$  &  $  -0.0393            \pm  0.0233  \pm  0.0037  $  &  $  \phantom{-}0.0205  \pm  0.0454  \pm  0.0083  $  \\
$(10.75,  12.00)$  &  $(3.70,  4.50)$  &  $  \phantom{-}0.0317  \pm  0.0353  \pm  0.0038  $  &  $  \phantom{-}0.0139  \pm  0.0709  \pm  0.0009  $  \\
$(12.00,  15.00)$  &  $(2.10,  2.70)$  &  $  \phantom{-}0.0067  \pm  0.0106  \pm  0.0032  $  &  $  -0.0364            \pm  0.0278  \pm  0.0010  $  \\
$(12.00,  15.00)$  &  $(2.70,  2.85)$  &  $  -0.0232            \pm  0.0195  \pm  0.0035 $  &  $  -0.0007            \pm  0.0525  \pm  0.0026  $  \\
$(12.00,  15.00)$  &  $(2.85,  3.00)$  &  $  \phantom{-}0.0171  \pm  0.0211  \pm  0.0047  $  &  $  \phantom{-}0.0255  \pm  0.0467  \pm  0.0010  $  \\
$(12.00,  15.00)$  &  $(3.00,  3.15)$  &  $  \phantom{-}0.0065  \pm  0.0241  \pm  0.0046  $  &  $  \phantom{-}0.0080  \pm  0.0521  \pm  0.0017  $  \\
$(12.00,  15.00)$  &  $(3.15,  3.30)$  &  $  -0.0101            \pm  0.0273  \pm  0.0038  $  &  $  -0.0019            \pm  0.0491  \pm  0.0021  $  \\
$(12.00,  15.00)$  &  $(3.30,  3.70)$  &  $  -0.0214            \pm  0.0219  \pm  0.0039  $  &  $  -0.0526            \pm  0.0373  \pm  0.0045  $  \\
$(12.00,  15.00)$  &  $(3.70,  4.50)$  &  $  -0.0511            \pm  0.0340  \pm  0.0038  $  &  $  -0.0494            \pm  0.0605  \pm  0.0027  $  \\
$(15.00,  30.00)$  &  $(2.10,  2.70)$  &  $  -0.0203            \pm  0.0115  \pm  0.0033  $  &  $  \phantom{-}0.0217  \pm  0.0267  \pm  0.0012  $  \\
$(15.00,  30.00)$  &  $(2.70,  2.85)$  &  $  -0.0340            \pm  0.0252  \pm  0.0036  $  &  $  -0.0204            \pm  0.0491  \pm  0.0038  $  \\
$(15.00,  30.00)$  &  $(2.85,  3.00)$  &  $  -0.0231            \pm  0.0277  \pm  0.0055  $  &  $  \phantom{-}0.0878  \pm  0.0520  \pm  0.0020  $  \\
$(15.00,  30.00)$  &  $(3.00,  3.15)$  &  $  \phantom{-}0.0347  \pm  0.0317  \pm  0.0037  $  &  $  \phantom{-}0.0120  \pm  0.0534  \pm  0.0016  $  \\
$(15.00,  30.00)$  &  $(3.15,  3.30)$  &  $  -0.0064            \pm  0.0379  \pm  0.0068  $  &  $  \phantom{-}0.0153  \pm  0.0626  \pm  0.0025  $  \\
$(15.00,  30.00)$  &  $(3.30,  3.70)$  &  $  -0.0221            \pm  0.0311  \pm  0.0042  $  &  $  -0.0647            \pm  0.0434  \pm  0.0013  $  \\
$(15.00,  30.00)$  &  $(3.70,  4.50)$  &  $  -0.0987            \pm  0.0496  \pm  0.0063  $  &  $  \phantom{-}0.0394  \pm  0.0777  \pm  0.0042  $  \\
\end{longtable}
 \end{center}

\begin{table}[!ht]
 \begin{center}
  \caption{Values of $A_{\rm P}(\Bs)$ and $A_{\rm P}(\Lb)$ in each kinematic bin for data collected in proton-proton collisions at centre-of-mass energy of  7\tev. The first uncertainties are statistical and the second systematic. }   
 \label{tab:resultsBsLb2011}
   \begin{tabular}{c|c|c|c}
\pt [\gevc] & $y$ & $A_{\rm P}(\Bs)_{\sqs = 7\, \tev}$ & $A_{\rm P}(\Lb)_{\sqs = 7\, \tev}$ \\
\hline
$(2.00,   7.00)$   &  $(2.10,  3.00)$  &  $  \phantom{-}0.0166  \pm  0.0632  \pm  0.0125  $   &  $  -0.0892 \pm 0.0508 \pm 0.0214 $  \\
$(2.00,   7.00)$   &  $(3.00,  3.30)$  &  $  \phantom{-}0.0311  \pm  0.0773  \pm  0.0151  $   &  $  \phantom{-}0.0507 \pm 0.0539 \pm 0.0208 $  \\
$(2.00,   7.00)$   &  $(3.30,  4.50)$  &  $  -0.0833            \pm  0.0558  \pm  0.0132  $          &  $  \phantom{-}0.0849 \pm 0.0401 \pm 0.0188$  \\
$(7.00,   9.50)$   &  $(2.10,  3.00)$  &  $  \phantom{-}0.0364  \pm  0.0479  \pm  0.0068  $   &  $  \phantom{-}0.1374 \pm 0.0697 \pm 0.0313 $  \\
$(7.00,   9.50)$   &  $(3.00,  3.30)$  &  $  \phantom{-}0.0206  \pm  0.0682  \pm  0.0127  $   &  $  \phantom{-}0.0138 \pm 0.0913 \pm 0.0298  $  \\
$(7.00,   9.50)$   &  $(3.30,  4.50)$  &  $  \phantom{-}0.0058  \pm  0.0584  \pm  0.0089  $   &  $  \phantom{-} 0.0466 \pm 0.0770 \pm 0.0347 $  \\
$(9.50,   12.00)$  &  $(2.10,  3.00)$  &  $  -0.0039            \pm  0.0456  \pm  0.0121  $         &  $  -0.0128 \pm 0.0985 \pm 0.0367 $  \\
$(9.50,   12.00)$  &  $(3.00,  3.30)$  &  $  \phantom{-}0.1095  \pm  0.0723  \pm  0.0179  $  &  $  -0.0848 \pm 0.1379 \pm 0.0452  $  \\
$(9.50,   12.00)$  &  $(3.30,  4.50)$  &  $  \phantom{-}0.1539  \pm  0.0722  \pm  0.0212  $  &  $  -0.1523 \pm 0.1414 \pm 0.0488  $  \\
$(12.00,  30.00)$  &  $(2.10,  3.00)$  &  $  -0.0271            \pm  0.0336  \pm  0.0061  $        &  $  -0.0720 \pm 0.1248 \pm 0.0465  $  \\
$(12.00,  30.00)$  &  $(3.00,  3.30)$  &  $  -0.0542            \pm  0.0612  \pm  0.0106  $        &  $  \phantom{-} 0.3291 \pm 0.2299 \pm 0.0918 $  \\
$(12.00,  30.00)$  &  $(3.30,  4.50)$  &  $  -0.0586            \pm  0.0648  \pm  0.0150  $        &  $  -0.0571 \pm 0.2162 \pm 0.0800 $  \\
\end{tabular}
 \end{center}
\end{table}

\begin{table}[!ht]
 \begin{center}
  \caption{Values of $A_{\rm P}(\Bs)$ and $A_{\rm P}(\Lb)$ in each kinematic bin for data collected in proton-proton collisions at centre-of-mass energy of 8\tev. The first uncertainties are statistical and the second systematic.}   
 \label{tab:resultsBsLb2012}
   \begin{tabular}{c|c|c|c}
\pt [\gevc] & $y$ & $A_{\rm P}(\Bs)_{\sqs = 8\, \tev}$ & $A_{\rm P}(\Lb)_{\sqs = 8\, \tev}$ \\
\hline
$(2.00,   7.00)$   &  $(2.10,  3.00)$  &  $  \phantom{-}0.0412  \pm  0.0416  \pm  0.0150  $  &  $ \phantom{-}0.0032 \pm 0.0318 \pm 0.0139   $  \\
$(2.00,   7.00)$   &  $(3.00,  3.30)$  &  $  -0.0241            \pm  0.0574  \pm  0.0079  $         &  $ \phantom{-}0.0929 \pm 0.0392 \pm 0.0171  $  \\
$(2.00,   7.00)$   &  $(3.30,  4.50)$  &  $  \phantom{-}0.0166  \pm  0.0391  \pm  0.0092  $  &  $  \phantom{-}0.0437 \pm 0.0284 \pm 0.0173  $  \\
$(7.00,   9.50)$   &  $(2.10,  3.00)$  &  $  \phantom{-}0.0482  \pm  0.0320  \pm  0.0067  $  &  $  \phantom{-}0.0069 \pm 0.0434 \pm 0.0169  $  \\
$(7.00,   9.50)$   &  $(3.00,  3.30)$  &  $  \phantom{-}0.0983  \pm  0.0470  \pm  0.0155  $  &  $  \phantom{-}0.0076 \pm 0.0589 \pm 0.0259  $  \\
$(7.00,   9.50)$   &  $(3.30,  4.50)$  &  $  -0.0430            \pm  0.0386  \pm  0.0079  $         &  $  \phantom{-}0.1053 \pm 0.0524 \pm 0.0252 $  \\
$(9.50,   12.00)$  &  $(2.10,  3.00)$  &  $  \phantom{-}0.0067  \pm  0.0303  \pm  0.0063  $ &  $  -0.0512 \pm 0.0594 \pm 0.0215 $  \\
$(9.50,   12.00)$  &  $(3.00,  3.30)$  &  $  -0.1283            \pm  0.0503  \pm  0.0171  $         &  $  \phantom{-}0.2355 \pm 0.0877 \pm 0.0399  $  \\
$(9.50,   12.00)$  &  $(3.30,  4.50)$  &  $  -0.0500            \pm  0.0460  \pm  0.0104  $         &  $  \phantom{-}0.1531 \pm 0.0838 \pm 0.0320 $  \\
$(12.00,  30.00)$  &  $(2.10,  3.00)$  &  $  -0.0012            \pm  0.0222  \pm  0.0050  $         &  $  \phantom{-}0.0453 \pm 0.0762 \pm 0.0300 $  \\
$(12.00,  30.00)$  &  $(3.00,  3.30)$  &  $  \phantom{-}0.0421  \pm  0.0416  \pm  0.0162  $  &  $  -0.0934 \pm 0.1377 \pm 0.0493  $  \\
$(12.00,  30.00)$  &  $(3.30,  4.50)$  &  $  \phantom{-}0.0537  \pm  0.0447  \pm  0.0124  $  &  $  \phantom{-}0.3173 \pm 0.1411 \pm 0.0655  $  \\
\end{tabular}
 \end{center}
\end{table}

\begin{table}
\caption{Values of the production asymmetries in bins of $\pt$, integrated over $y$, for \Bp and \Bz mesons for data collected in proton-proton collisions at the centre-of-mass energy of 7 \tev. The first uncertainties are statistical and the second systematic. The uncertainties among the bins are correlated due to the external inputs: $A_{\CP}(\Bp\to\jpsi\Kp)$ and $A_{\rm D}(\Kzb)$ for $A_{\rm P}(\Bp)$, and $|q/p|$ for $A_{\rm P}(\Bz)$.}
\label{tab:AP_BpB0_2011_pt}
\begin{center}
\begin{tabular}{c|c|c}
$\pt$ [GeV/$c$]& $A_{\rm P}(\Bp)_{\sqs = 7\, \tev}$ & $A_{\rm P}(\Bz)_{\sqs = 7\, \tev}$  \\
\hline
$(0.00,   2.00)   $  &  $  \phantom{-}0.0015  \pm  0.0067  \pm  0.0036  $  &  $  \phantom{-}0.0215  \pm  0.0297  \pm  0.0025  $  \\
$(2.00,   4.50)   $  &  $  -0.0050            \pm  0.0040  \pm  0.0037  $  &  $  \phantom{-}0.0123  \pm  0.0163  \pm  0.0078  $  \\
$(4.50,   7.00)   $  &  $  -0.0010            \pm  0.0045  \pm  0.0038  $  &  $  \phantom{-}0.0124  \pm  0.0150  \pm  0.0042  $  \\
$(7.00,   8.25)   $  &  $  \phantom{-}0.0083  \pm  0.0080  \pm  0.0041  $  &  $  -0.0440            \pm  0.0219  \pm  0.0012  $  \\
$(8.25,   9.50)   $  &  $  -0.0078            \pm  0.0096  \pm  0.0039  $  &  $  -0.0476            \pm  0.0248  \pm  0.0038  $  \\
$(9.50,   10.75)  $  &  $  -0.0220            \pm  0.0114  \pm  0.0044  $  &  $  \phantom{-}0.0155  \pm  0.0297  \pm  0.0056  $  \\
$(10.75,  12.00)  $  &  $  -0.0045            \pm  0.0138  \pm  0.0043  $  &  $  \phantom{-}0.0404  \pm  0.0357  \pm  0.0040  $  \\
$(12.00,  15.00)  $  &  $  \phantom{-}0.0107  \pm  0.0124  \pm  0.0053  $  &  $  -0.0050            \pm  0.0269  \pm  0.0035  $  \\
$(15.00,  30.00)  $  &  $  -0.0146            \pm  0.0150  \pm  0.0065  $  &  $  \phantom{-}0.0333  \pm  0.0298  \pm  0.0077  $  \\
\end{tabular}
\end{center}
\end{table}

\begin{table}
\caption{Values of the production asymmetries in bins of $y$, integrated over $\pt$, for \Bp and \Bz mesons for data collected in proton-proton collisions at the centre-of-mass energy of 7 \tev. The first uncertainties are statistical and the second systematic. The uncertainties among the bins are correlated due to the external inputs: $A_{\CP}(\Bp\to\jpsi\Kp)$ and $A_{\rm D}(\Kzb)$ for $A_{\rm P}(\Bp)$, and $|q/p|$ for $A_{\rm P}(\Bz)$. }
\label{tab:AP_BpB0_2011_eta}
\begin{center}
\begin{tabular}{c|c|c}
$y$& $A_{\rm P}(\Bp)_{\sqs = 7\, \tev}$ & $A_{\rm P}(\Bz)_{\sqs = 7\, \tev}$ \\
\hline
$(2.10,  2.70)  $  &  $  \phantom{-}0.0007  \pm  0.0047  \pm  0.0036  $  &  $  \phantom{-}0.0488  \pm  0.0205  \pm  0.0017  $  \\
$(2.70,  2.85)  $  &  $  -0.0131            \pm  0.0064  \pm  0.0036  $  &  $  -0.0366            \pm  0.0232  \pm  0.0027  $  \\
$(2.85,  3.00)  $  &  $  -0.0063            \pm  0.0061  \pm  0.0037  $  &  $  -0.0251            \pm  0.0213  \pm  0.0010  $  \\
$(3.00,  3.15)  $  &  $  -0.0125            \pm  0.0061  \pm  0.0039  $  &  $  -0.0478            \pm  0.0203  \pm  0.0017  $  \\
$(3.15,  3.30)  $  &  $  -0.0009            \pm  0.0063  \pm  0.0039  $  &  $  -0.0130            \pm  0.0203  \pm  0.0018  $  \\
$(3.30,  3.70)  $  &  $  -0.0060            \pm  0.0044  \pm  0.0043  $  &  $  -0.0143            \pm  0.0133  \pm  0.0017  $  \\
$(3.70,  4.50)  $  &  $  \phantom{-}0.0041  \pm  0.0062  \pm  0.0046  $  &  $  \phantom{-}0.0044  \pm  0.0173  \pm  0.0045  $  \\
\end{tabular}
\end{center}
\end{table}

\begin{table}
\caption{Values of the production asymmetries in bins of $\pt$, integrated over $y$, for \Bp and \Bz mesons for data collected in proton-proton collisions at the centre-of-mass energy of  8 \tev. The first uncertainties are statistical and the second systematic. The uncertainties among the bins are correlated due to the external inputs: $A_{\CP}(\Bp\to\jpsi\Kp)$ and $A_{\rm D}(\Kzb)$ for $A_{\rm P}(\Bp)$, and $|q/p|$ for $A_{\rm P}(\Bz)$.}
\label{tab:AP_BpB0_2012_pt}
\begin{center}
\begin{tabular}{c|c|c}
$\pt$ [GeV/$c$]& $A_{\rm P}(\Bp)_{\sqs = 8\, \tev}$ & $A_{\rm P}(\Bz)_{\sqs = 8\, \tev}$  \\
\hline
$(0.00,   2.00)   $  &  $  -0.0105            \pm  0.0045  \pm  0.0031  $  &  $  \phantom{-}0.0065  \pm  0.0230  \pm  0.0017  $  \\
$(2.00,   4.50)   $  &  $  -0.0033            \pm  0.0026  \pm  0.0031  $  &  $  -0.0188            \pm  0.0103  \pm  0.0009  $  \\
$(4.50,   7.00)   $  &  $  -0.0093            \pm  0.0029  \pm  0.0032  $  &  $  -0.0111            \pm  0.0092  \pm  0.0011  $  \\
$(7.00,   8.25)   $  &  $  -0.0094            \pm  0.0051  \pm  0.0033  $  &  $  -0.0192            \pm  0.0141  \pm  0.0015  $  \\
$(8.25,   9.50)   $  &  $  -0.0126            \pm  0.0061  \pm  0.0033  $  &  $  \phantom{-}0.0015  \pm  0.0155  \pm  0.0009  $  \\
$(9.50,   10.75)  $  &  $  -0.0073            \pm  0.0073  \pm  0.0034  $  &  $  -0.0156            \pm  0.0177  \pm  0.0013  $  \\
$(10.75,  12.00)  $  &  $  \phantom{-}0.0036  \pm  0.0090  \pm  0.0034  $  &  $  \phantom{-}0.0017  \pm  0.0210  \pm  0.0027  $  \\
$(12.00,  15.00)  $  &  $  -0.0082            \pm  0.0079  \pm  0.0035  $  &  $  -0.0270            \pm  0.0171  \pm  0.0009  $  \\
$(15.00,  30.00)  $  &  $  -0.0251            \pm  0.0095  \pm  0.0040  $  &  $  \phantom{-}0.0137  \pm  0.0177  \pm  0.0009  $  \\
\end{tabular}
\end{center}
\end{table}

\begin{table}
\caption{Values of the production asymmetries in bins of $y$, integrated over $\pt$, for \Bp and \Bz mesons for data collected in proton-proton collisions at the centre-of-mass energy of  8 \tev. The first uncertainties are statistical and the second systematic. The uncertainties among the bins are correlated due to the external inputs: $A_{\CP}(\Bp\to\jpsi\Kp)$ and $A_{\rm D}(\Kzb)$ for $A_{\rm P}(\Bp)$, and $|q/p|$ for $A_{\rm P}(\Bz)$. }
\label{tab:AP_BpB0_2012_eta}
\begin{center}
\begin{tabular}{c|c|c}
$y$& $A_{\rm P}(\Bp)_{\sqs = 8\, \tev}$ & $A_{\rm P}(\Bz)_{\sqs = 8\, \tev}$ \\
\hline
$(2.10,  2.70)  $  &  $  -0.0023            \pm  0.0029  \pm  0.0031  $  &  $  -0.0082            \pm  0.0128  \pm  0.0012  $  \\
$(2.70,  2.85)  $  &  $  -0.0080            \pm  0.0041  \pm  0.0031  $  &  $  -0.0237            \pm  0.0173  \pm  0.0009  $  \\
$(2.85,  3.00)  $  &  $  \phantom{-}0.0003  \pm  0.0040  \pm  0.0032  $  &  $  \phantom{-}0.0148  \pm  0.0159  \pm  0.0015  $  \\
$(3.00,  3.15)  $  &  $  -0.0038            \pm  0.0040  \pm  0.0032  $  &  $  -0.0140            \pm  0.0151  \pm  0.0009  $  \\
$(3.15,  3.30)  $  &  $  -0.0123            \pm  0.0042  \pm  0.0034  $  &  $  -0.0193            \pm  0.0158  \pm  0.0021  $  \\
$(3.30,  3.70)  $  &  $  -0.0138            \pm  0.0030  \pm  0.0034  $  &  $  -0.0029            \pm  0.0103  \pm  0.0010  $  \\
$(3.70,  4.50)  $  &  $  -0.0144            \pm  0.0042  \pm  0.0037  $  &  $  -0.0201            \pm  0.0137  \pm  0.0010  $  \\
\end{tabular}
\end{center}
\end{table}

\begin{table}
\caption{Values of the production asymmetries in bins of $\pt$, integrated over $y$, for the \Bs meson and the \Lb baryon for data collected in proton-proton collisions at the centre-of-mass energy of 7 \tev. The first uncertainties are statistical and the second systematic. The uncertainties among the bins are correlated due to the external inputs: $A_{\CP}(\Bp\to\jpsi\Kp)$, $A_{\rm D}(\Kzb)$, $|q/p|_{\Bz}$ and $|q/p|_{\Bs}$ for $A_{\rm P}(\Lb)$, and $|q/p|_{\Bs}$ for $A_{\rm P}(\Bs)$. }
\label{tab:AP_BsLb_2011_pt}
\begin{center}
\begin{tabular}{c|c|c}
$\pt$ [GeV/$c$]& $A_{\rm P}(\Bs)_{\sqs = 7\, \tev}$ & $A_{\rm P}(\Lb)_{\sqs = 7\, \tev}$ \\
\hline
$(2.0,   7.0)$   &  $-0.0166            \pm  0.0393  \pm  0.0082$          &  $  -0.0130 \pm 0.0311 \pm 0.0133  $  \\
$(7.0,   9.5)$   &  $\phantom{-}0.0247  \pm  0.0334  \pm  0.0050$   &  $  \phantom{-} 0.0948 \pm 0.0476 \pm 0.0211  $  \\
$(9.5,   12.0)$  &  $\phantom{-}0.0566  \pm  0.0349  \pm  0.0096$  &  $  -0.0596 \pm 0.0722 \pm 0.0262  $  \\
$(12.0,  30.0)$  &  $-0.0382            \pm  0.0273  \pm  0.0054$        &  $  -0.0146 \pm 0.0985 \pm 0.0369  $  \\
\end{tabular}
\end{center}
\end{table}

\begin{table}
\caption{Values of the production asymmetries in bins of $y$, integrated over $\pt$, for the \Bs meson and the \Lb baryon for data collected in proton-proton collisions at the centre-of-mass energy of 7 \tev. The first uncertainties are statistical and the second systematic. The uncertainties among the bins are correlated due to the external inputs: $A_{\CP}(\Bp\to\jpsi\Kp)$, $A_{\rm D}(\Kzb)$, $|q/p|_{\Bz}$ and $|q/p|_{\Bs}$ for $A_{\rm P}(\Lb)$, and $|q/p|_{\Bs}$ for $A_{\rm P}(\Bs)$.}
\label{tab:AP_BsLb_2011_eta}
\begin{center}
\begin{tabular}{c|c|c}
$y$& $A_{\rm P}(\Bs)_{\sqs = 7\, \tev}$ & $A_{\rm P}(\Lb)_{\sqs = 7\, \tev}$ \\
\hline
$(2.1,  3.0)$  &  $\phantom{-}0.0151  \pm  0.0445  \pm  0.0088$  &  $  -0.0511 \pm 0.0399 \pm 0.0168  $  \\
$(3.0,  3.3)$  &  $\phantom{-}0.0296  \pm  0.0566  \pm  0.0111$  &  $  \phantom{-}0.0514 \pm 0.0448 \pm 0.0171  $  \\
$(3.3,  4.5)$  &  $-0.0554            \pm  0.0432  \pm  0.0101$  &  $  \phantom{-}0.0638 \pm 0.0348 \pm 0.0160  $  \\
\end{tabular}
\end{center}
\end{table}

\begin{table}
\caption{Values of the production asymmetries in bins of $\pt$, integrated over $y$, for the \Bs meson and the \Lb baryon for data collected in proton-proton collisions at the centre-of-mass energy of  8 \tev. The first uncertainties are statistical and the second systematic. The uncertainties among the bins are correlated, due to the external inputs: $A_{\CP}(\Bp\to\jpsi\Kp)$, $A_{\rm D}(\Kzb)$, $|q/p|_{\Bz}$ and $|q/p|_{\Bs}$, for $A_{\rm P}(\Lb)$  and $|q/p|_{\Bs}$ for $A_{\rm P}(\Bs)$. }
\label{tab:AP_BsLb_2012_pt}
\begin{center}
\begin{tabular}{c|c|c}
$\pt$ [GeV/$c$]& $A_{\rm P}(\Bs)_{\sqs = 8\, \tev}$ & $A_{\rm P}(\Lb)_{\sqs = 8\, \tev}$ \\
\hline
$(2.0,   7.0)$   &  $\phantom{-}0.0235  \pm  0.0264  \pm  0.0083$  &  $  \phantom{-}0.0292 \pm 0.0200 \pm 0.0096  $  \\
$(7.0,   9.5)$   &  $\phantom{-}0.0257  \pm  0.0223  \pm  0.0049$  &  $  \phantom{-}0.0367 \pm 0.0302 \pm 0.0127  $  \\
$(9.5,   12.0)$  &  $-0.0286            \pm  0.0230  \pm  0.0053$  &  $  \phantom{-}0.0442 \pm 0.0437 \pm 0.0164  $  \\
$(12.0,  30.0)$  &  $\phantom{-}0.0187  \pm  0.0186  \pm  0.0049$  &  $  \phantom{-}0.0902 \pm 0.0612 \pm 0.0253  $  \\
\end{tabular}
\end{center}
\end{table}

\begin{table}
\caption{Values of the production asymmetries in bins of $y$, integrated over $\pt$, for the \Bs meson and the \Lb baryon for data collected in proton-proton collisions at the centre-of-mass energy of  8 \tev. The first uncertainties are statistical and the second systematic. The uncertainties among the bins are correlated, due to the external inputs: $A_{\CP}(\Bp\to\jpsi\Kp)$, $A_{\rm D}(\Kzb)$, $|q/p|_{\Bz}$ and $|q/p|_{\Bs}$, for $A_{\rm P}(\Lb)$  and $|q/p|_{\Bs}$ for $A_{\rm P}(\Bs)$. }
\label{tab:AP_BsLb_2012_eta}
\begin{center}
\begin{tabular}{c|c|c}
$y$& $A_{\rm P}(\Bs)_{\sqs = 8\, \tev}$ & $A_{\rm P}(\Lb)_{\sqs = 8\, \tev}$ \\
\hline
$(2.1,  3.0)$  &  $\phantom{-}0.0364  \pm  0.0290  \pm  0.0103$  &  $ \phantom{-}0.0028 \pm 0.0247 \pm 0.0107  $  \\
$(3.0,  3.3)$  &  $-0.0078            \pm  0.0413  \pm  0.0063$  &  $  \phantom{-}0.0792 \pm 0.0317 \pm 0.0138  $  \\
$(3.3,  4.5)$  &  $\phantom{-}0.0055  \pm  0.0298  \pm  0.0070$  &  $  \phantom{-}0.0682 \pm 0.0242 \pm 0.0142  $  \\
\end{tabular}
\end{center}
\end{table}

\clearpage
\addcontentsline{toc}{section}{References}
\setboolean{inbibliography}{true}
\bibliographystyle{LHCb}
\bibliography{main,LHCb-PAPER,LHCb-DP,additional}
\clearpage
\centerline{\large\bf LHCb collaboration}
\begin{flushleft}
\small
R.~Aaij$^{40}$,
B.~Adeva$^{39}$,
M.~Adinolfi$^{48}$,
Z.~Ajaltouni$^{5}$,
S.~Akar$^{59}$,
J.~Albrecht$^{10}$,
F.~Alessio$^{40}$,
M.~Alexander$^{53}$,
S.~Ali$^{43}$,
G.~Alkhazov$^{31}$
% ...

{\footnotesize \it
% ...
$ ^{5}$Clermont Universit{\'e}, Universit{\'e} Blaise Pascal, CNRS/IN2P3, LPC, Clermont-Ferrand, France\\
% ...
$ ^{10}$Fakult{\"a}t Physik, Technische Universit{\"a}t Dortmund, Dortmund, Germany\\
% ...
$ ^{31}$Petersburg Nuclear Physics Institute (PNPI), Gatchina, Russia\\
% ...
$ ^{39}$Universidad de Santiago de Compostela, Santiago de Compostela, Spain\\
$ ^{40}$European Organization for Nuclear Research (CERN), Geneva, Switzerland\\
% ...
$ ^{43}$Nikhef National Institute for Subatomic Physics, Amsterdam, The Netherlands\\
% ...
$ ^{48}$H.H. Wills Physics Laboratory, University of Bristol, Bristol, United Kingdom\\
% ...
$ ^{53}$School of Physics and Astronomy, University of Glasgow, Glasgow, United Kingdom\\
% ...
$ ^{59}$University of Cincinnati, Cincinnati, OH, United States\\
% ...
}
\end{flushleft}
\end{document}